\documentclass{article}

\usepackage{enumitem}
\usepackage{placeins}

\usepackage[preprint]{neurips_2026/neurips_2026}

\usepackage[utf8]{inputenc} 
\usepackage[T1]{fontenc}    
\usepackage{hyperref}       
\usepackage{url}            
\usepackage{booktabs}       
\usepackage{amsfonts}       
\usepackage{nicefrac}       
\usepackage{microtype}      
\usepackage{xcolor}         
\usepackage{graphicx}
\usepackage{subcaption}
\usepackage{amsmath}
\usepackage{tcolorbox}

\newcommand{\dd}{\mathrm d}

\newcommand{\cN}{\mathcal N}

\newcommand{\cQ}{\mathcal Q}
\newcommand{\cS}{\mathcal S}

\newcommand{\1}{\mathbf 1}
\newcommand{\bh}{\bm h}
\newcommand{\bg}{\bm g}
\newcommand{\bv}{\bm v}
\newcommand{\bD}{\bm\Delta}

\newcommand{\bxi}{\bm\xi}
\newcommand{\bchi}{\bm\chi}

\newcommand{\what}[1]{\widehat{#1}}

\usepackage[utf8]{inputenc} 
\usepackage{float}
\usepackage[T1]{fontenc}    
\usepackage{hyperref}       
\usepackage{url}            
\usepackage{booktabs}       
\usepackage{amsfonts}       
\usepackage{nicefrac}       
\usepackage{microtype}      
\usepackage{bm}
\usepackage{amsthm}
\usepackage{amsmath}
\usepackage{amssymb}
\usepackage{xcolor}

\def\h{\bm h}

\def\W{\bm W}

\definecolor{myblue}{rgb}{0.2,0.2,0.7} 
\hypersetup{colorlinks=true,allcolors=myblue}

\title{Spectral Dynamics in Deep Networks: Feature Learning, Outlier Escape, and Learning Rate Transfer}

\author{
Clarissa Lauditi$^{1}$,
Cengiz Pehlevan$^{1,2}$\thanks{Equal senior authors.},
Blake Bordelon$^{3,4}$\footnotemark[1]
\\
$^{1}$John A. Paulson School of Engineering and Applied Sciences, Harvard University\\
$^{2}$Kempner Institute for the Study of Natural and Artificial Intelligence, Harvard University\\
$^{3}$Center for Mathematical Sciences and Applications, Harvard University\\
$^{4}$Oden Institute for Computational Engineering and Sciences \& Dept. of Neuroscience, UT Austin\\
\texttt{\{clauditi,cpehlevan\}@seas.harvard.edu, blake\_bordelon@utexas.edu}
}

\begin{document}

\maketitle

\begin{abstract}
We study the evolution of hidden-weight spectra in wide neural networks trained by (stochastic) gradient descent. We develop a two-level dynamical mean-field theory (DMFT) that jointly tracks bulk and outlier spectral dynamics for spiked ensembles whose spike directions remain statistically dependent on the random bulk. We apply this framework to two settings: (1) infinite-width nonlinear networks in mean-field/$\mu$P scaling and (2) deep linear networks in the proportional high-dimensional limit, where width, input dimension, and sample size diverge with fixed ratios. Our theory predicts how outliers evolve with training time, width, output scale, and initialization variance. In deep linear networks, $\mu$P yields width-consistent outlier dynamics and hyperparameter transfer, including width-stable growth of the leading NTK mode toward the edge of stability (EoS). In contrast, NTK parameterization exhibits strongly width-dependent outlier dynamics, despite converging to a stable large-width limit. We show that this bulk+outlier picture is descriptive of simple tasks with small output channels, but that tasks involving large numbers of outputs (ImageNet classification or GPT language modeling) are better described by a restructuring of the spectral bulk. We develop a toy model with extensive output channels that recapitulates this phenomenon and show that edge of the spectrum still converges for sufficiently wide networks. 

\end{abstract}

\vspace{-10pt}
\section{Introduction}
\vspace{-10pt}

Understanding how large neural networks learn internal representations remains a central problem in deep learning theory. Motivated by both theoretical tractability and empirical observations about the benefits of larger neural network models \cite{kaplan2020scaling, hoffmann2022training, achiam2023gpt}, significant theoretical attention has been paid to various infinite width limits of neural networks in recent years \cite{jacot2018neural, lee2019wide, bahri2021explaining, bordelon2020spectrum, loureiro2021learning}. While some infinite width limits of neural networks generate a \textit{kernel behavior} where hidden representations are frozen over the course of training, the mean-field / $\mu$P scaling enables strong feature learning even at infinite width \cite{mei2019mean, geiger2020disentangling, yang2021tensor, bordelon2022self}. In this feature-learning regime (often termed the \emph{rich regime}), training substantially modifies hidden weights and representations rather than merely fitting a linear model on top of static random features \cite{ghorbani2020neural, vyas2022limitations, montanari2025dynamical}. 

A natural way to study the evolution of hidden features is through the spectrum of hidden weight matrices \cite{mahoney2022practice, hodgkinson2025models}. 
At initialization, these spectra are well described by random matrix theory: the empirical singular value distribution forms a Marchenko-Pastur bulk \cite{marvcenko1967distribution, potters2020first}. 
During training, structured updates can create isolated singular values that detach from the random bulk. These outliers encode low-dimensional directions shaped by learning, while the bulk continues to reflect high-dimensional randomness and finite-width effects. In random matrix theory, the emergence of isolated eigenvalues from the bulk as signal strength increases is the \textbf{BBP phase transition}~\cite{baik2005phase}.

This perspective is especially relevant for questions of width scaling and hyperparameter transfer. Feature learning parameterizations such as $\mu$P and CompleteP are designed so that training dynamics remain comparable across widths and depths, generating consistent optimal hyperparameters across model sizes \cite{yang2021tensor,yang2022tensor, dey2025don}. However, a complete theoretical account of hyperparameter transfer is still lacking in regimes where both (1) model losses vary significantly across widths and (2) optimal learning rates remain consistent in $\mu$P but not in other parameterizations, such as NTK-P. The edge-of-stability (EoS) literature~\cite{xing2018walksgd,Jastrzebski2020The,cohen2021gradient,Cohen2022AdaptiveGM,andreyev2025edgestochasticstabilityrevisiting} suggests that the maximum stable learning rate is controlled by sharpness or the NTK eigenvalue~\cite{ jiang2026understanding}. Therefore, width-stable learning-rate transfer requires width-stable dynamics of the corresponding top spectral outlier \cite{ghosh2025understanding}. This makes the evolution of the full spectrum, and in particular its leading isolated modes, a natural object for testing and explaining when hyperparameter transfer succeeds or fails. A recent proposal by Ghosh et al.~\cite{ghosh2025understanding} suggests that feature learning proceeds through effectively low-rank weight updates whose dominant directions stabilize before the loss converges. In this work, we provide an exact theoretical characterization of these spectral dynamics.
\vspace{-5pt}
\paragraph{BBP for $\mu$P?} In this work, we develop a theoretical framework for tracking the dynamics of singular values of hidden weight matrices in wide, randomly initialized deep networks. We show that learning induces spike-like directions in the weights, which are in general \emph{statistically coupled} to the random matrix from initialization. 
This places the problem outside the scope of standard spiked random matrix models, which typically assume independence between spikes and noise \cite{baik2005phase,benaych2012singular, perry2018optimality, capitaine2018limiting, barbier2023fundamental}. 
To address this, we introduce a two-level dynamical mean field theory (DMFT) formalism that jointly describes the training dynamics generating the structured directions and the spectral probe used to resolve bulk and outlier singular values. 
The result is a general framework for spiked matrix ensembles with statistically dependent spikes, together with a finite-dimensional criterion that predicts when outliers escape from the bulk. 

Our main contributions are:


\vspace{-8pt}
\begin{itemize}
    \item We develop a two-level DMFT framework for random matrices with finite-rank spikes statistically coupled to the random bulk. This yields a finite-dimensional condition for BBP-style outlier escape in trained weight spectra.
    \item For infinite-width nonlinear networks in mean-field/$\mu$P scaling, the theory yields layerwise predictions for spectral outliers and their BBP transitions as a function of learning richness. This finite-outlier picture accurately captures the main spectral features observed in wide convolutional ResNets trained on CIFAR-10 image classification.
    \item We characterize width-dependent spectral dynamics in deep linear networks and show that $\mu$P yields substantially more consistent BBP transition times and outlier dynamics across width than NTK parameterization, providing a mechanistic signature of successful transfer across model size.
    \item For tasks with extensive output dimension, including compute-optimal language-model pretraining and large-class vision classification, training can restructure the entire spectral bulk rather than produce only finitely many isolated outliers. We capture this departure from the finite-rank bulk-plus-outlier picture with a simple linear-network toy model with extensive output channels. This connects to the generalized BBP for extensive-rank spikes~\cite{forner2025bbp}, here treated through DMFT for training-induced spikes and arbitrarily deep NNs.
\end{itemize}

\vspace{-10pt}
\subsection{Related Works}

\vspace{-5pt}
\paragraph{Infinite Limits and Hyperparameter Transfer} The recent emphasis on neural scaling laws has made it increasingly important to understand which aspects of training are stable across model size and which are not \citep{kaplan2020scaling, hoffmann2022training}. On the theory side, infinite-width limits have provided a complementary lens: depending on parameterization, these limits can yield kernel descriptions, mean-field descriptions, or nontrivial feature-learning limits that remain analytically tractable \citep{jacot2018neural, mei2019mean, yang2021tensor, yang2022tensor}. Our work fits into this broader program, but focuses specifically on the evolution of hidden-layer weights and kernel spectra and on whether the signal-carrying spectral structure of training is preserved across width.
\vspace{-5pt}
\paragraph{Regimes of Training} Our paper is also closely related to the distinction between lazy/kernel and rich feature-learning regimes. In the lazy regime, parameter updates remain small and training is well approximated by linearization around initialization, leading to effectively kernel-based dynamics \citep{jacot2018neural, chizat2019lazy}. In rich regimes, by contrast, initialization scale and parameterization lead to substantial representation learning during training \citep{woodworth2020kernel, mei2019mean, yang2021tensor}. This distinction is central to maximal-update parameterizations such as $\mu$P, which aim to preserve nontrivial training dynamics and enable hyperparameter transfer across width \citep{yang2022tensor, vyas2023feature}. We contribute to this literature by showing that the evolution of the spectral bulk and isolated outliers provides a mechanistic diagnostic of both the lazy-to-rich transition and the width consistency underlying hyperparameter transfer. 
\vspace{-5pt}
\paragraph{Dynamical Mean Field Theory for Machine Learning}  Originating in spin-glass physics, dynamical mean field theory (DMFT), provides an effective description of high-dimensional disordered dynamical systems \cite{sompolinsky1981dynamic, de1978dynamics, bordelon2026disordered}. In machine learning, DMFT has been used to analyze random and trained recurrent networks \cite{crisanti2018path, clark2026structure}, SGD dynamics in generalized linear models \cite{mignacco2020dynamical, mignacco2022effective, gerbelot2022rigorous, bordelon2024dynamical}, feature learning, generalization, and memorization in two-layer networks \cite{montanari2025dynamical}, and infinite-width/depth training limits across architectures \cite{bordelon2022self, bordelon2023dynamics, bordelon2024infinite, bordelon2024depthwise, jiang2026hyperparameter, chaintron2026resnets}. We extend these methods to a \textit{two-level DMFT} for tracking singular/eigenvalue evolution in random matrices whose training-induced updates remain statistically coupled with their initial random components.
\vspace{-5pt}
\paragraph{Random Matrix Theory and the BBP Transition} At initialization, the singular-value bulk of wide random weight matrices follows the Marchenko--Pastur law \citep{marvcenko1967distribution}. More generally, spiked random matrix theory and the BBP transition describe when low-rank perturbations generate isolated eigenvalues outside a random bulk \citep{baik2005phase, benaych2012singular, baik2006eigenvalues, forner2025bbp}. These results provide the vocabulary of bulk edges, spikes, and outlier detachment, but in our setting the spikes are learned and remain statistically coupled to the random initialization. 
\vspace{-5pt}
\paragraph{Solvable Model of Weight Spectra and Scaling Laws}
Most relevant to us, recent efforts have characterized the weight spectra of trained two-layer networks with quadratic activations, revealing a rich phase portrait of possibilities for the final spectrum ranging from bulk+spikes to heavy tails \cite{defilippis2025scaling, erba2026nuclear}. The authors of these works related the spectral properties of the trained model to scaling law behavior and generalization. In a similar spirit, an analysis of the key/query matrices in an attention model trained on sequence tasks generated from a teacher attention model revealed analogous phase transitions and a link between key/query spectra and generalization \cite{boncoraglio2025single}. These works describe how final spectra depend on target structure and available data, while our work focuses on the evolution of spectra and spikes throughout the training trajectory for wide nonlinear networks of arbitrary depth. A recent work utilizes DMFT to examine the dynamics of the extensive width quadratic networks trained with random data \cite{martin2026high}, finding agreement with the static analyses at long times \cite{maillard2024bayes, erba2026nuclear,defilippis2025scaling}. 

\vspace{-5pt}
\paragraph{Weight Spectra in Trained Networks} Our work is complementary to empirical RMT analyses of trained networks, which emphasize heavy-tailed spectral structure and implicit self-regularization in fully trained models \citep{martin2020heavy, martin2021implicit,thamm2022random,wang2023spectral}. Recent theoretical works have also begun to examine the behavior of weight spectra in one-hidden layer networks after one step or under Bayesian inference \cite{dandi2023how, moniri2023theory, dandi2024random, cui2024asymptotics}. We develop a theory for how learned spectral structure emerges from random initialization in deep networks under multiple steps of (stochastic) gradient descent.
\vspace{-5pt}
\paragraph{Deep Linear Networks} Our work connects to the tradition of using deep linear networks as tractable models of nonlinear optimization dynamics \cite{saxe2013exact, kunin2024get, domine2024lazy}. Our work extends prior DMFT analyses that enable study of randomly initialized deep linear networks in a proportional scaling \cite{bordelon2025deep}. We extend these results with our two-level DMFT to track spectral outliers throughout training. 
\vspace{-8pt}
\section{Results}
\vspace{-8pt}
We first present a general characterization of the singular-value spectrum of random matrices with statistically coupled finite-rank structure. 
\vspace{-5pt}
\subsection{Singular Values of Random Matrices with Statistically Coupled Spikes}

Let $\bm W(S) \in \mathbb{R}^{N_1 \times N_0}$ be a random matrix initialized with mean-zero i.i.d. entries of variance $\sigma^2$ with $S$ spikes
\begin{align}\label{eq:weight_update_structure}
    \bm W(S)
    =
    \bm W(0)
    +
    \frac{1}{\sqrt{N_0}}
    \sum_{t \in [S]} \bm g(t) \bm\phi(t)^\top.
\end{align}
Our general result assumes that the spike vectors $\bm g(t)$ and $\bm\phi(t)$ have asymptotically independent entries, as occurs naturally in large-width limits of neural networks. Crucially, however, the spikes are not independent of the initialization. Instead, they are coupled to $\bm W(0)$ through their dependence on the dynamical random vectors $\bm\chi(t)=\frac{1}{\sqrt{N_0}}\bm W(0)\bm\phi(t)$ and $\bm\xi(t)
=\frac{\sqrt{N_0}}{N_1}\bm W(0)^\top \bm g(t)$, being
\begin{align}\label{eq:feature_structure}
   \text{Feature Structural Condition:} \quad \phi_i(t) = \phi_t\left( \{ \xi_i(s)\}_{s \in [S]} \right)   \ , \ g_i(t) = g_{t}\left( \{\chi_i(s)\}_{s\in [S]} \right)
\end{align}
where $\phi_t$ and $g_t$ are asymptotically deterministic (possibly nonlinear) functions that take in $S$ arguments $\{\chi_i(s)\}$ or $\{ \xi_i(s)\}_{s \in [S]}$ \footnote{In the cases of interest in this work, the dependence on the variables $\chi,\xi$ will be causal so $\phi(t) = \phi_t( \{ \xi(s) \}_{s < t} )$}. We will first review the limiting equations that describe the distribution of $\bm\phi(t), \bm g(t),\bm\chi(t),\bm\xi(t)$ which take the form of a dynamical mean field theory (DMFT). The DMFT correlation and response functions will govern the macroscopic behavior of the system in the high dimensional limit. 

\begin{tcolorbox}[colframe=myblue, opacityback=0.95, title=Lemma 1: Generic DMFT for the Asymptotic Limit Dynamics] 
Assume proportional scaling $N_0, N_1 \to \infty$ with $N_1/N_0 = \alpha$ in a system that satisfies Equations \eqref{eq:weight_update_structure} \&  \eqref{eq:feature_structure} for constant spikes $S = \mathcal{O}(1)$. Each of the vectors $\bm\phi(t), \bm g(t),\bm\chi(t),\bm\xi(t)$ have i.i.d. entries which can be viewed as \textit{single-site} stochastic processes which obey
\begin{align}
    &\chi(t) = u^\chi(t) + \frac{\sigma^2}{\alpha} \sum_{s} R^\phi(t,s) g(s) \ , \ u^\chi(t) \sim \mathcal{N}\left(0, \sigma^2 C^\phi(t,s) \right)
    \\
    &\xi(t) = u^\xi(t) + \sigma^2 \sum_{s} R^g(t,s) \phi(s) \ , \ u^\xi(t) \sim \mathcal{N}\left(0, \sigma^2 \alpha^{-1} C^g(t,s) \right)
\end{align}
and $\phi(t) = \phi_t( \{ \xi \} )$ and $g(t) = g_t(\{ \chi \})$ are functions of the above processes. The correlation functions $C^\phi,C^g$ and response functions $R^\phi, R^g$ are determined by averages $\left< \cdot \right>$ over the $\{ u^\chi, u^\xi \}$ Gaussian processes
\begin{align}
    &C^\phi(t,s) = \left< \phi(t) \phi(s) \right> \ , \ C^g(t,s) = \left< g(t) g(s) \right>
    \\
    &R^\phi(t,s) = \left< \frac{\partial \phi(t)}{\partial u^\xi(s) }\right> \ , \ R^g(t,s) = \left< \frac{\partial g(t)}{\partial u^\chi(s) }\right>.
\end{align}
\end{tcolorbox}

After solving the above DMFT equations for the correlation functions $C^\phi, C^g$ and response functions $R^\phi$ and $R^g$, one is now in a position to compute the singular values of the weight matrix $\bm W(S)$.  Below, we characterize the singular values in a large system limit with $S = \mathcal{O}_N(1)$ held fixed. 

\begin{tcolorbox}[colframe=myblue, opacityback=0.95, title=Result 1: Singular Values of Random Matrices with Statistically Coupled Spikes]\label{res:sing_value_thm} 

The limiting eigenvalue density $\rho_{\bm M}(\lambda)$ of $\bm M = \frac{1}{N_1} \bm W(S)^\top \bm W(S)$ has Bulk + Spikes form
\begin{align}
\rho_{\bm M}(\lambda) \to  \rho_{\text{MP}}(\lambda;\alpha,\sigma) + \rho_{\text{spike}}(\lambda)  \ , \quad   N_0 \rho_{\text{spike}}(\lambda) \to \sum_{z \in  O_S} \delta( \lambda - z) 
\end{align}
where $\rho_{\text{MP}}(\lambda;\alpha,\sigma)$ is the Marchenko-Pastur eigenvalue density with aspect ratio $\alpha$ and scale $\sigma$. The outlier set $O_S = \{ z \in \mathbb{R}_+ : \det \bm A(z) = 0\}$ are solutions to a zero-determinant condition for a matrix valued function $\bm A(z) \in \mathbb{R}^{4S \times 4S}$  
\begin{align}
    \bm A(z) = 
    \begin{bmatrix}
        \mathcal G(z)^{-1} \bm I  & \bm 0 & - (\sigma^2 \alpha^{-1} \bm R^\phi + \bm C^\phi ) & - \sigma^2 \bm C^\phi   
        \\
        \bm 0 & \mathcal G(z)^{-1} \bm I & - (\bm R^\phi)^\top & - \sigma^2 (\bm R^\phi)^\top
        \\
        -(\sigma^2 \bm R^g + \bm C^g) & -\sigma^2 \alpha^{-1} \bm C^g &  (1 - \frac{\sigma^2}{\alpha} \mathcal G(z)) \bm I & \bm 0
        \\
        -(\bm R^g)^\top & -\sigma^2 \alpha^{-1} (\bm R^g)^\top & \bm 0 & (1 - \frac{\sigma^2}{\alpha} \mathcal G(z)) \bm I 
    \end{bmatrix}
\end{align}
where $\mathcal G(z) = \frac{\alpha}{2 z \sigma^2}\left[ z +\sigma^2(\alpha^{-1}-1) - \sqrt{ [z +\sigma^2(\alpha^{-1}-1)]^2 - 4 z \alpha^{-1} \sigma^2 } \right]$ is the Stieltjes transform of the Wishart matrix \cite{potters2020first, atanasov2024scaling} and the matrices $\bm R^\phi, \bm R^g, \bm C^\phi, \bm C^g \in \mathbb{R}^{T\times T}$ are the correlation and response functions of the DMFT for the original system. 
\end{tcolorbox}
\vspace{-8pt}
\paragraph{Buildup of Spike(s)} By increasing spikes $S$, one can access the dynamics of the outliers. At initialization $S=0$, there are no solutions in the outlier set $O_S$. As updates to $\bm W$ accumulate over spikes $S$, multiple outliers (possibly up to $S$) can emerge from the bulk, similar to the classical BBP transition for spiked random matrices \cite{baik2005phase}. However, in practice, the structure of the limiting dynamics typically maintains a low dimensional spike subspace. $S$ is usually linear in timesteps $T$ of a dynamical system and we define the \textbf{BBP time} as the minimum value of $T$ where the set $O_S$ is non-empty (the time when an outlier exits the bulk). This time represents when a signal from the weight updates is sufficiently large to drive a detectable change in the maximum singular value of the matrix $\bm W(S)$. We analyze simple toy models to gain intuition in Apps \ref{appendix:classical_bbp} and \ref{app:anti_hebb_GOE}.

This result covers two deep-learning regimes studied below: infinite-width feature-learning networks at fixed batch size and training time (Section~\ref{sec:inf_nonlinear}), and randomly initialized deep linear networks in a proportional limit of input dimension, sample size, and width (Section~\ref{sec:linear_networks}).
    
\section{Infinite Width Limits of Feature Learning Networks}\label{sec:inf_nonlinear}
\vspace{-5pt}
Consider an $L$-hidden-layer fully connected network with width $N$, input dimension $D$, parameters
$\bm\theta=\mathrm{Vec}\{\bm W^\ell\}_{\ell=0}^L$, hidden states $\bm h_\mu^\ell$, and output $f_\mu$
\begin{align}
    f_\mu =
    \frac{1}{\gamma N}\bm w^L \cdot \phi(\bm h_\mu^L),
    \qquad
    \bm h_\mu^{\ell+1}
    =
    \frac{1}{\sqrt N}\bm W^\ell \phi(\bm h_\mu^\ell),
    \qquad
    \bm h_\mu^1
    =
    \frac{1}{\sqrt D}\bm W^0 \bm x_\mu .
\end{align}
Here $\phi$ is applied elementwise and $\gamma$ is an output multiplier controlling the richness of the dynamics: $\gamma \to 0$ yields a kernel/lazy limit, while the $1/(\gamma N)$ readout scaling corresponds to the mean-field/$\mu$P scaling and preserves feature learning as $N\to\infty$ \cite{chizat2019lazy, lee2019wide, geiger2020disentangling, yang2021tensor, bordelon2022self, bordelon2023dynamics}. The parameters are initialized i.i.d. with variance $\sigma^2$ and trained by SGD on minibatches $\mathcal B_t=\{\bm x_\mu\}_{\mu=1}^B$ : $\bm\theta(t+1) = \bm\theta(t) - \eta  \gamma^2 N \frac{\partial}{\partial \bm\theta(t)} \frac{1}{B} \sum_{\bm x \in \mathcal B_t}  \ell( f(\bm x)) $.
\vspace{-5pt}
\paragraph{Super-Wide Networks} We first assume that the width $N$ is diverging while the input dimension $D$, batchsize $B$ and number of timesteps $t$ are held fixed \footnote{This is actually a very high performing feature learning limit \cite{vyas2023feature, yang2021tuning, bordelon2024depthwise} and should not be confused with limiting steps $t$ compared to input dimension $D$ which can have an arbitrary relationship.}. These constraints enable exact concentration of the network's prediction and loss dynamics and also  perfect interpolation when applied on finite datasets \cite{jacot2018neural, bordelon2022self}. This regime has been shown to capture the behavior of wide convolutional neural networks on simple tasks like CIFAR-10, but it is overly optimistic about model performance in the scaling law regime where total data is large compared to model size \cite{vyas2023feature,hoffmann2022training, bordelon2024dynamical}. We will relax this constraint in Section \ref{sec:linear_networks}, where finite $\frac{B t}{N}$ and $\frac{D}{N}$ effects can be characterized in linear networks. 

\begin{tcolorbox}[colframe=myblue, opacityback=0.95, title=Lemma 2: DMFT For Infinite Width Neural Networks in $\mu$P Trained with (S)GD] 
Infinite width networks of any depth $L$ in $\mu$P trained with SGD for $O_N(1)$ steps on $O_N(1)$ batchsize satisfies the structural assumptions needed for our Bulk + Outlier theory with $S=BT$ spikes. The signal strength of the spikes are controlled by output multiplier $\gamma$ \cite{bordelon2022self}.  
\end{tcolorbox}

\begin{figure}
    \centering
    \begin{subfigure}{0.32\linewidth}
    \includegraphics[width=\linewidth]{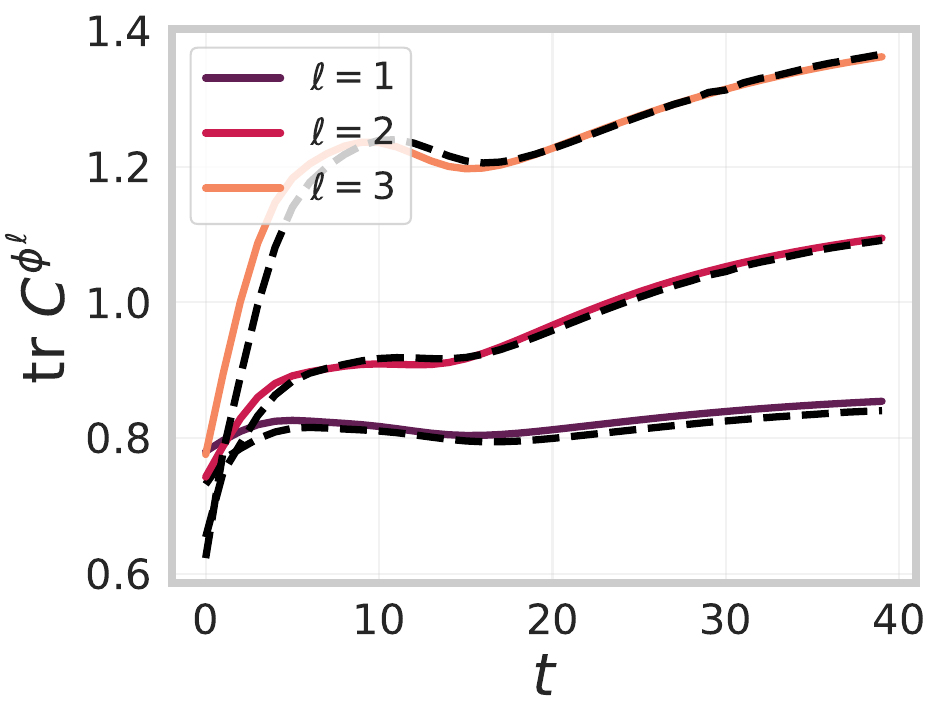}
    \caption{Feature Correlation Dynamics}
    \end{subfigure}
    \begin{subfigure}{0.32\linewidth}
    \includegraphics[width=\linewidth]{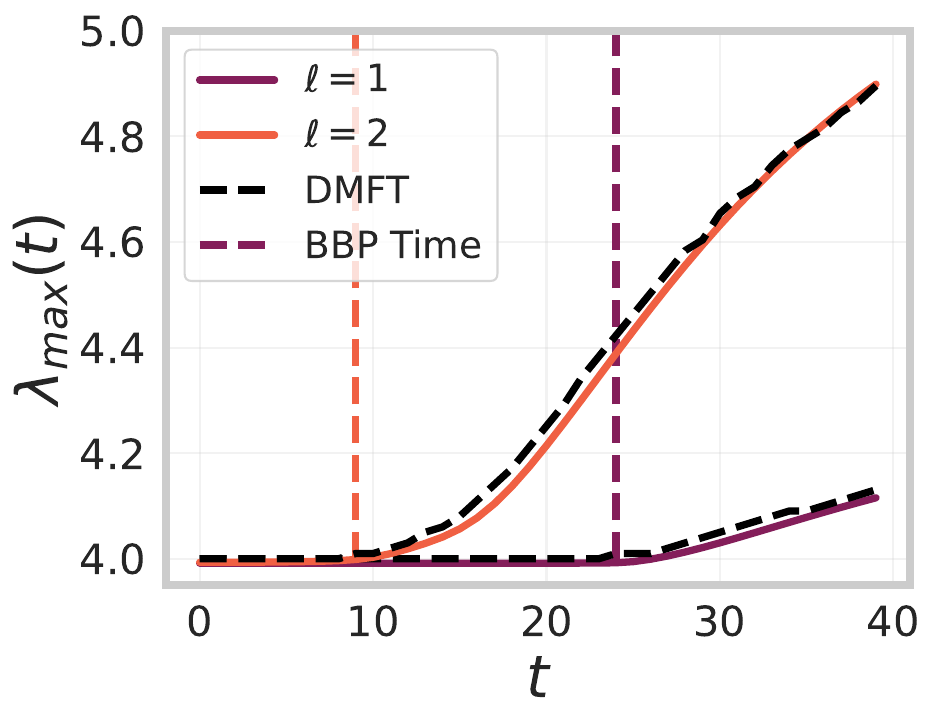}
    \caption{BBP Transition $\gamma = 2.5$}
    \end{subfigure}
    \begin{subfigure}{0.32\linewidth}
    \includegraphics[width=\linewidth]{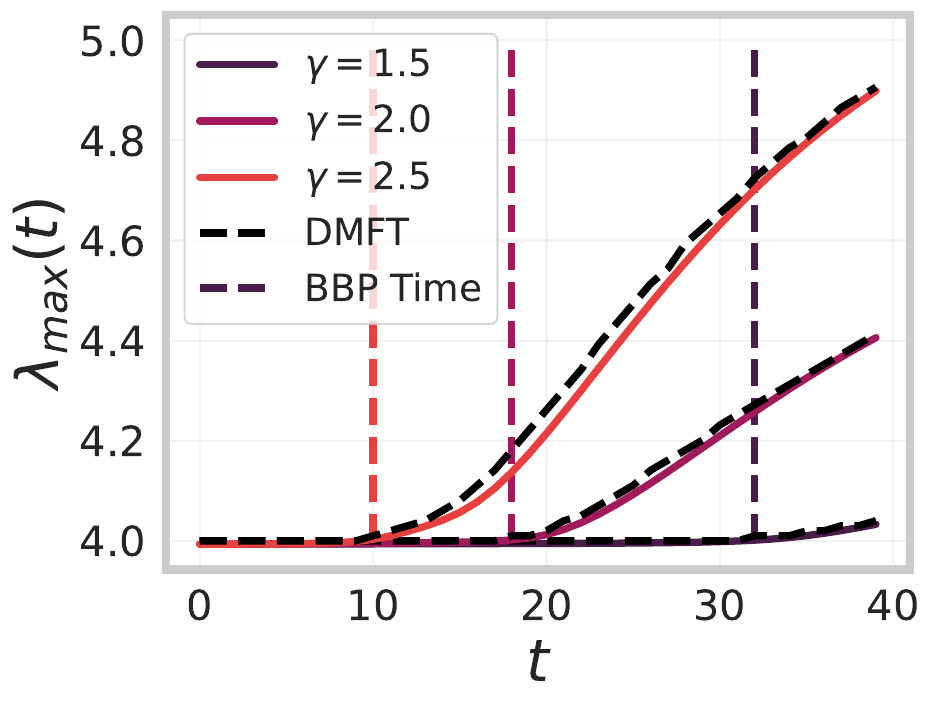}
    \caption{Last Layer $\ell=2$}
    \end{subfigure}
    \caption{Deep ($L=3$ hidden layers) nonlinear networks ($\phi=\tanh$) trained with gradient descent on a single-index polynomial target function with unit scale initialization $\sigma^2 = 1$. (a) The hidden feature kernel dynamics are predicted by the DMFT equations (dashed black lines). (b)-(c) The outlier dynamics of the hidden weights are accurately predicted by the $\bm A(z)$ matrix. }
    \label{fig:nonlin_tanh_DMFT}
\end{figure}

\begin{figure}
    \centering
    
    \begin{subfigure}{0.32\linewidth}
        \centering
        \includegraphics[width=\linewidth]{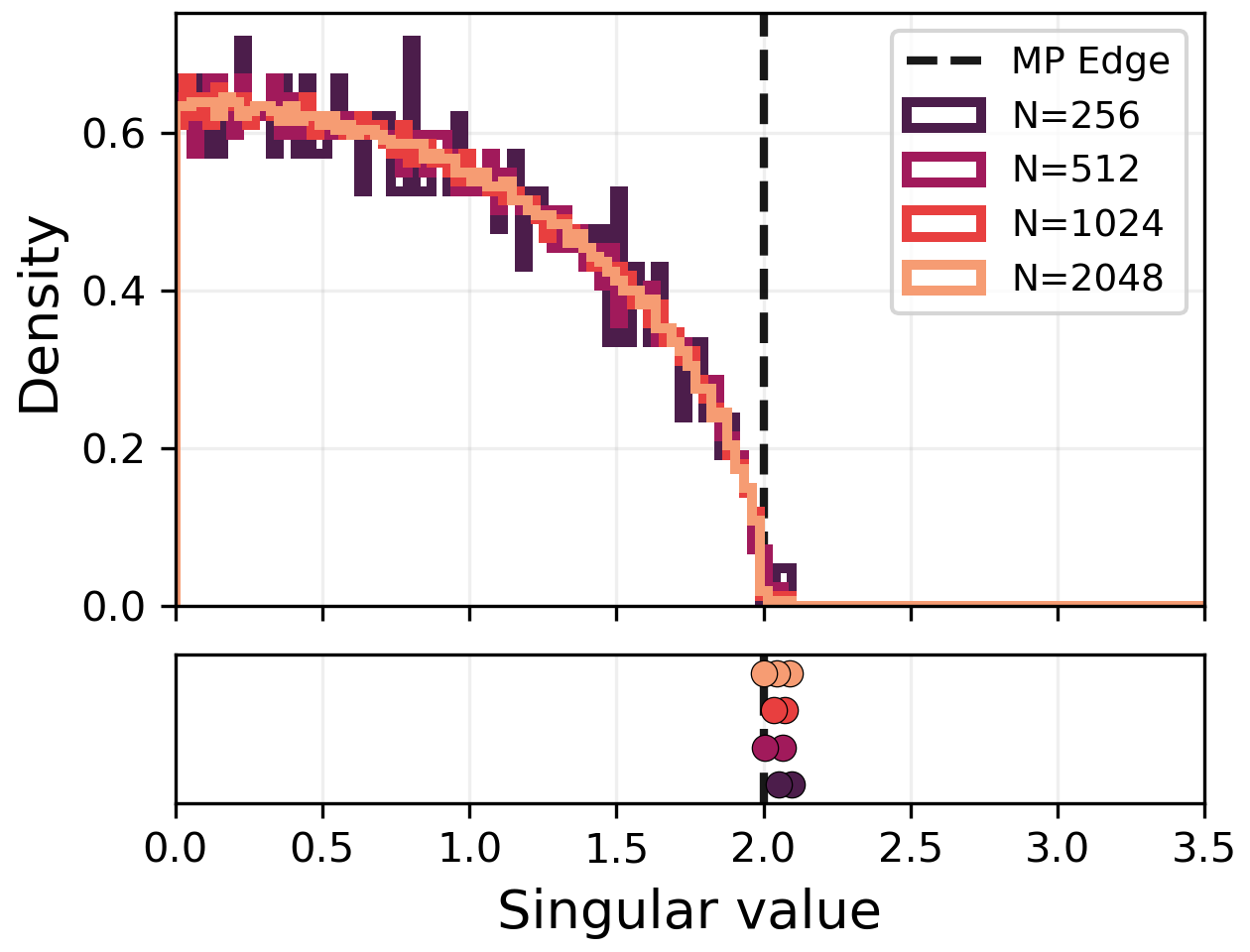}
        \caption{$T=100$}
    \end{subfigure}
    \hfill
    \begin{subfigure}{0.32\linewidth}
        \centering
        \includegraphics[width=\linewidth]{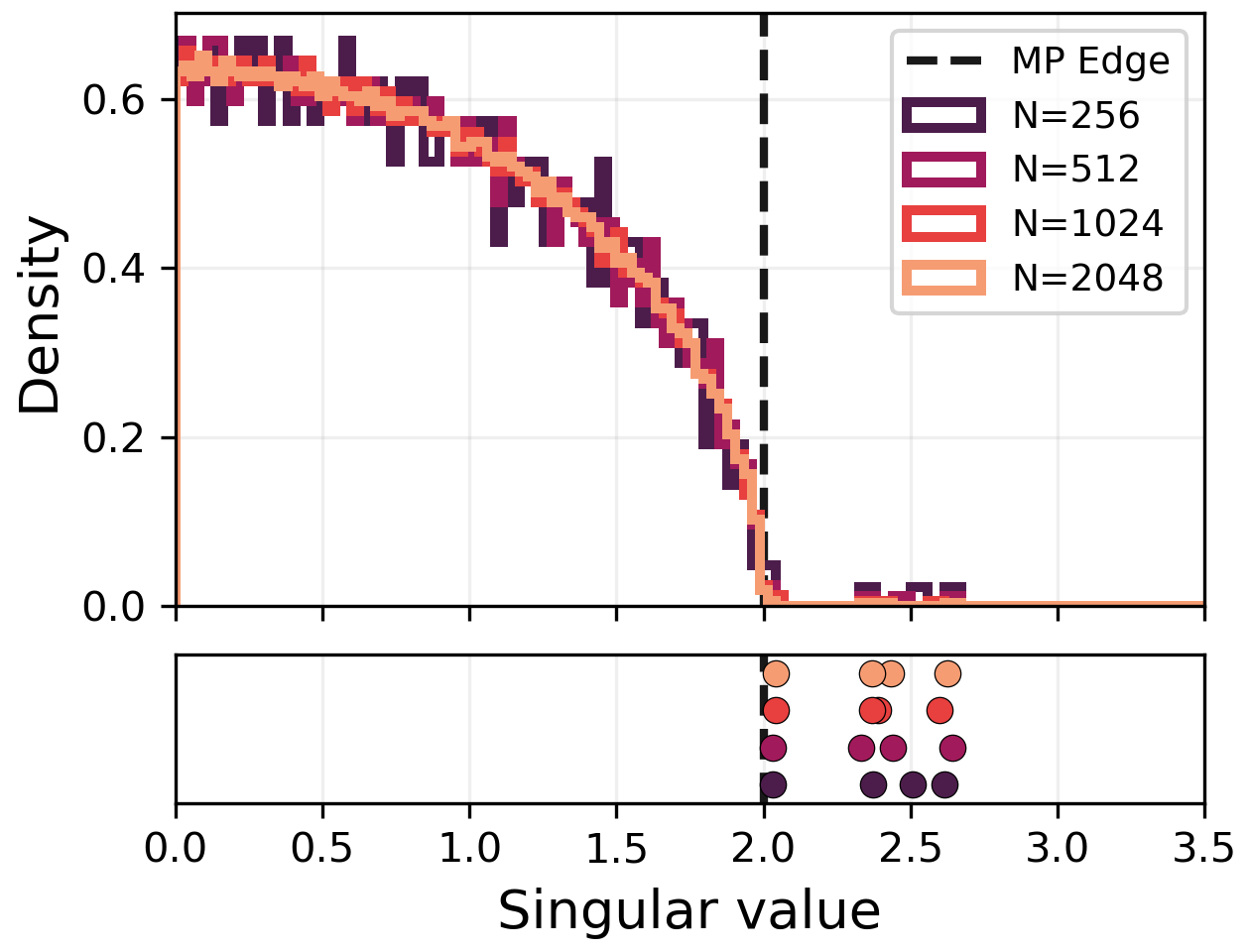}
        \caption{$T=250$}
    \end{subfigure}
    \hfill
    \begin{subfigure}{0.32\linewidth}
        \centering
        \includegraphics[width=\linewidth]{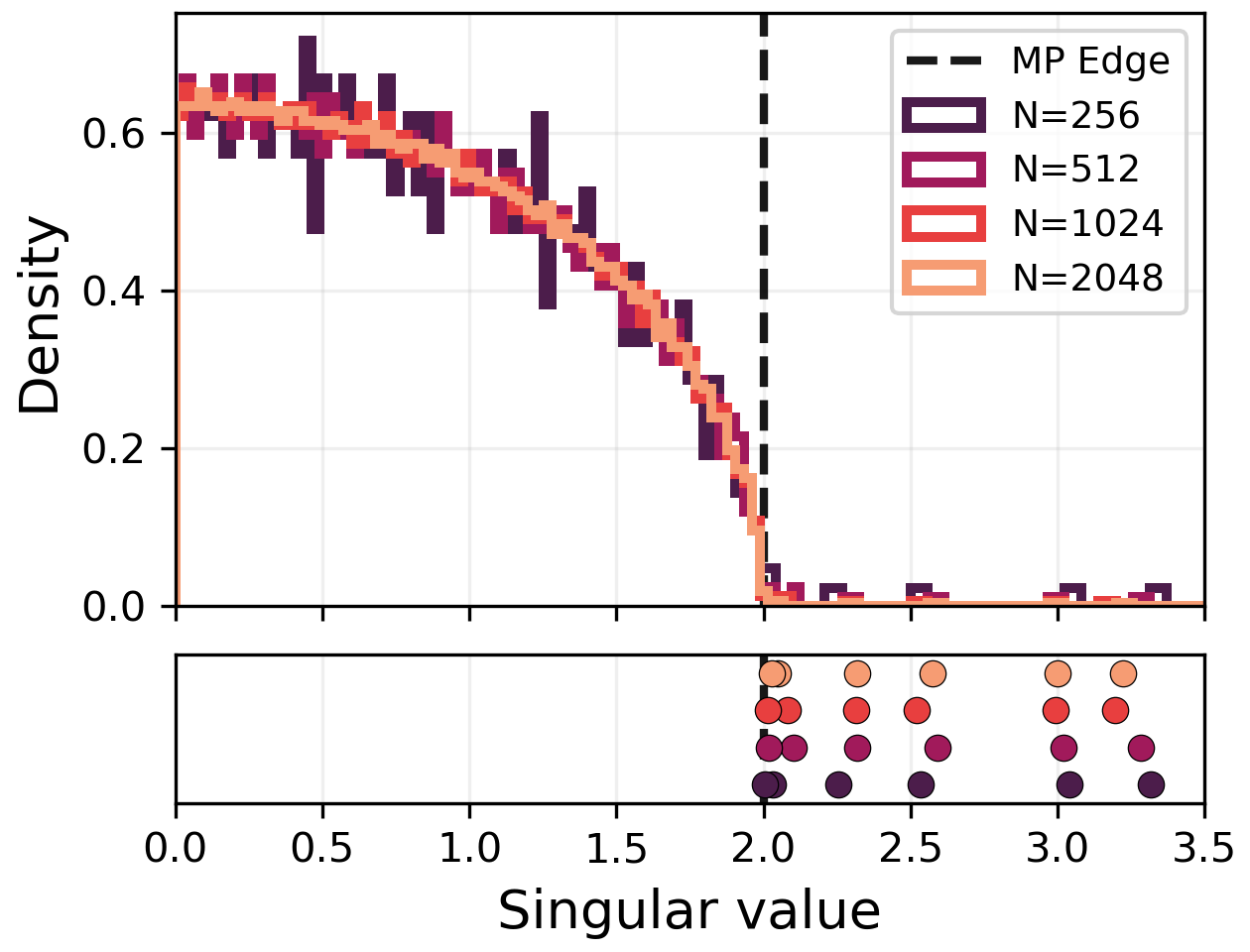}
        \caption{$T=1000$}
    \end{subfigure}

    \caption{Weight singular values of varying width $N$ ResNets (second convolution in the first residual block) on CIFAR-10 at different snapshots of training. Consistent with the infinite width theory, the MP bulk is unshifted, while the outliers undergo deterministic (and asymptotically width-independent) dynamics. Surprisingly even width $N=256$ models exhibit similar dynamics to $N=2048$. }
    \label{fig:mp_density_CNN}
\end{figure}

\paragraph{Spike Structure in Nonlinear Networks} In this setting, the spike variables are $\bm \phi_{\mu}^\ell(t) \equiv \phi(\bm h^\ell_\mu(t))$ and $\bm g_\mu^\ell(t) \equiv N \gamma \frac{\partial f_\mu(t) }{\partial \h^\ell_\mu(t)}$ are the forward and backward pass features respectively for all data in each minibatch $\bm x_\mu \in \mathcal B_t$. The dependence on the initial condition for weights occurs through 
\begin{align}
    &\bm\chi_\mu^{\ell+1}(t) = \frac{1}{\sqrt N} \bm W^\ell(0) \bm \phi_\mu^\ell(t) \ , \  \bm\xi^\ell_\mu(t) = \frac{1}{\sqrt N} \bm W^{\ell}(0)^\top \bm g^{\ell+1}_{\mu}(t)
    \\
    &\bm\phi^\ell_\mu(t) \sim_{N \to \infty} \phi_{\mu, t}^\ell\left( \{ \bm\chi^{\ell}_\nu(s), \bm\xi^\ell_\nu(s)  \}_{\nu;s<t} ; \gamma \right) \ , \ \bm g^\ell_\mu(t) \sim_{N \to \infty} g_{\mu, t}^\ell\left( \{ \bm\chi^{\ell}_\nu(s), \bm\xi^\ell_\nu(s)  \}_{\nu;s<t} ; \gamma \right)  
\end{align}
where the functions $\phi^\ell_{\mu,t}$ and $g^\ell_{\mu,t}$ only act \emph{elementwise} over the neurons. Further, each neuron statistically decouples as an i.i.d. random variable whose distribution can be identified from the DMFT. We give the explicit formulas for $\bm\phi^\ell_\mu(t), \bm g^\ell_\mu(t)$ spikes in the Appendix \ref{app:nonlinear_inner_dmft_eqn}, which allows Result 1 to be applied to nonlinear networks (see as well~\cite{bordelon2022self, bordelon2023dynamics, bordelon2024depthwise}). In Figure \ref{fig:nonlin_tanh_DMFT} we illustrate our theory of the spike dynamics for a depth $4$ ($L=3$ hidden layer) $\tanh$ network trained on a single index model $y = q(\bm\beta \cdot \bm x)$ where $\bm x \in \mathbb{R}^{10}$ and $q(z) = z + \frac{1}{2} z^2$. The DMFT predicts the feature correlation functions, as well as the maximum eigenvalue dynamics through Result 1. 
\vspace{-5pt}
\paragraph{Bulk+Outlier Picture Holds in $\mu$P CNNs on CIFAR-10}
DMFT can in principle be applied to architectures such as CNNs, recurrent networks, and transformers \cite{bordelon2022self, clark2026structure, bordelon2024infinite}, but solving the exact outlier equations on realistic tasks is often computationally hard. We therefore test the bulk+outlier picture empirically in $\mu$P CNNs trained on CIFAR-10. As shown in Figure~\ref{fig:mp_density_CNN}, across large channel widths $N$, a finite number of spikes detach from an otherwise stable bulk after thousands of training steps, consistent with a BBP-like transition in the \textit{super-wide} regime.

\section{Linear Networks That Incorporate Width / Data Effects }\label{sec:linear_networks}
\vspace{-5pt}
In the previous section, we saw that the theory for nonlinear networks could accurately predict outlier dynamics for \emph{super-wide networks}. However, this regime is incapable of characterizing variations in training dynamics across network widths.
\vspace{-5pt}
\paragraph{Proportionally Wide Linear Networks}
    To access regimes where width can vary in a meaningful way that impacts performance and dynamics, we next turn to a setting where we can characterize networks that are wide but where width, input dimension, and data are comparable. This setting is challenging for general architectures and data distributions, but can be analyzed in closed form for \emph{random and high dimensional data} in deep linear networks (for details see~\cite{bordelon2025deep} or Appendix~\ref{app:linear_two_level_dmft}).  

\begin{tcolorbox}[colframe=myblue, opacityback=0.95, title=Lemma 3: DMFT For Proportional Width Linear Networks in $\mu$P Trained with (S)GD ] 
Consider a randomly initialized width-$N$ linear network with $C=O(1)$ output channels and linear targets
$y_c(\bm x)=D^{-1/2}\bm\beta_c\cdot \bm x+\epsilon_c$. We study either full-batch GD on $P$ random inputs in $D$ dimensions or online SGD with batch size $B$, in the proportional limits
\begin{align}
  \begin{cases}
      P,N,D \to \infty, \quad \alpha=P/D, \quad \nu=N/D, \quad S = CT,
      & \text{full-batch GD},\\
      B,N,D \to \infty, \quad \alpha=B/D, \quad \nu=N/D, \quad S = CT,
      & \text{online SGD}.
  \end{cases}
\end{align}
(S)GD satisfies the conditions of Result~1, and the loss and weight dynamics are characterized by self-averaging correlation and response $\{C^\phi,C^g,R^\phi,R^g\}$ for each hidden layer $\ell\in[L]$.
\end{tcolorbox}

\begin{figure}[!t]
    \centering
    \begin{subfigure}{0.32\linewidth}
        \centering
        \includegraphics[width=\linewidth]{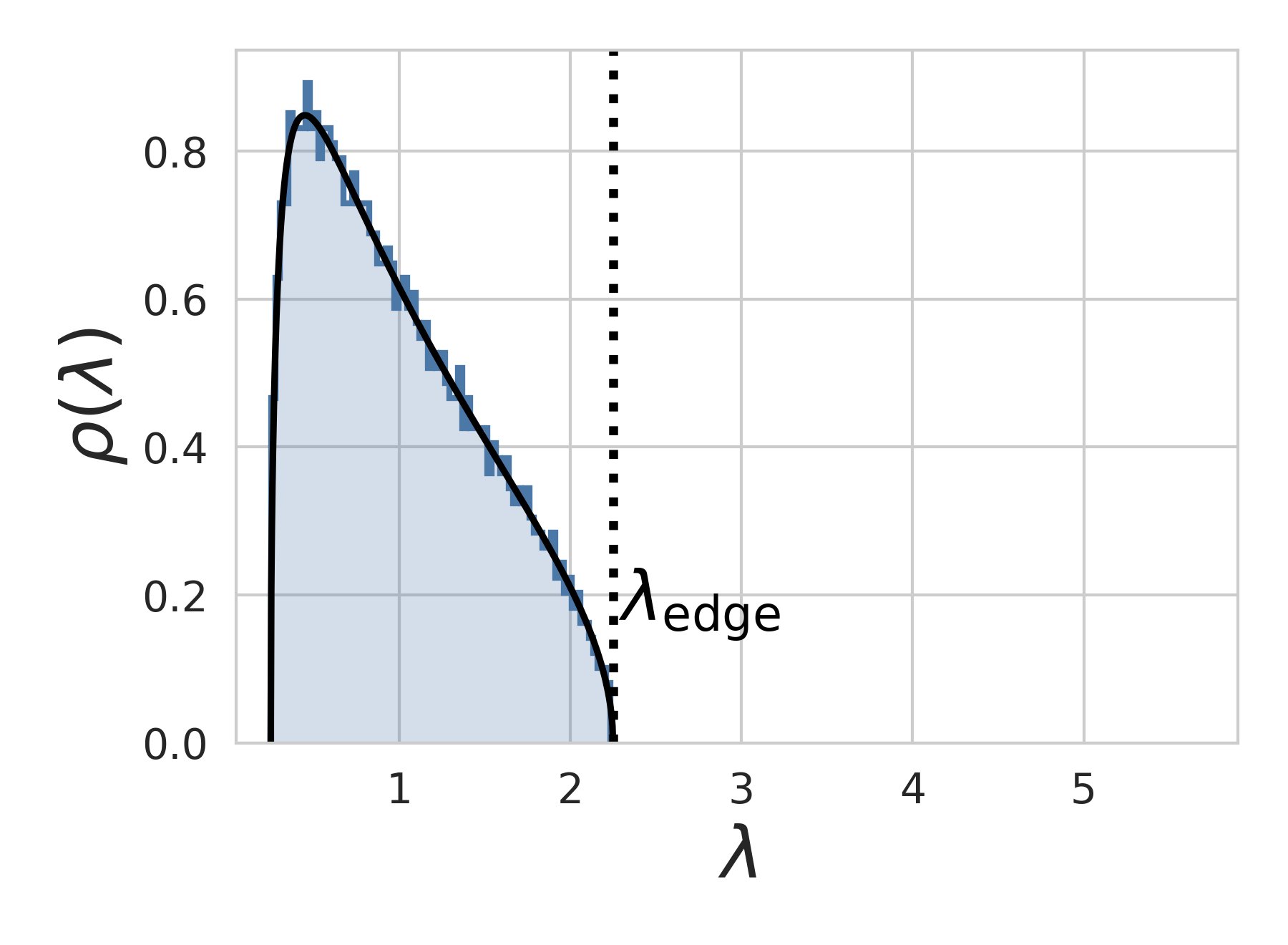}
        \caption{$T=0$}
    \end{subfigure}
    \hfill
    \begin{subfigure}{0.32\linewidth}
        \centering
        \includegraphics[width=\linewidth]{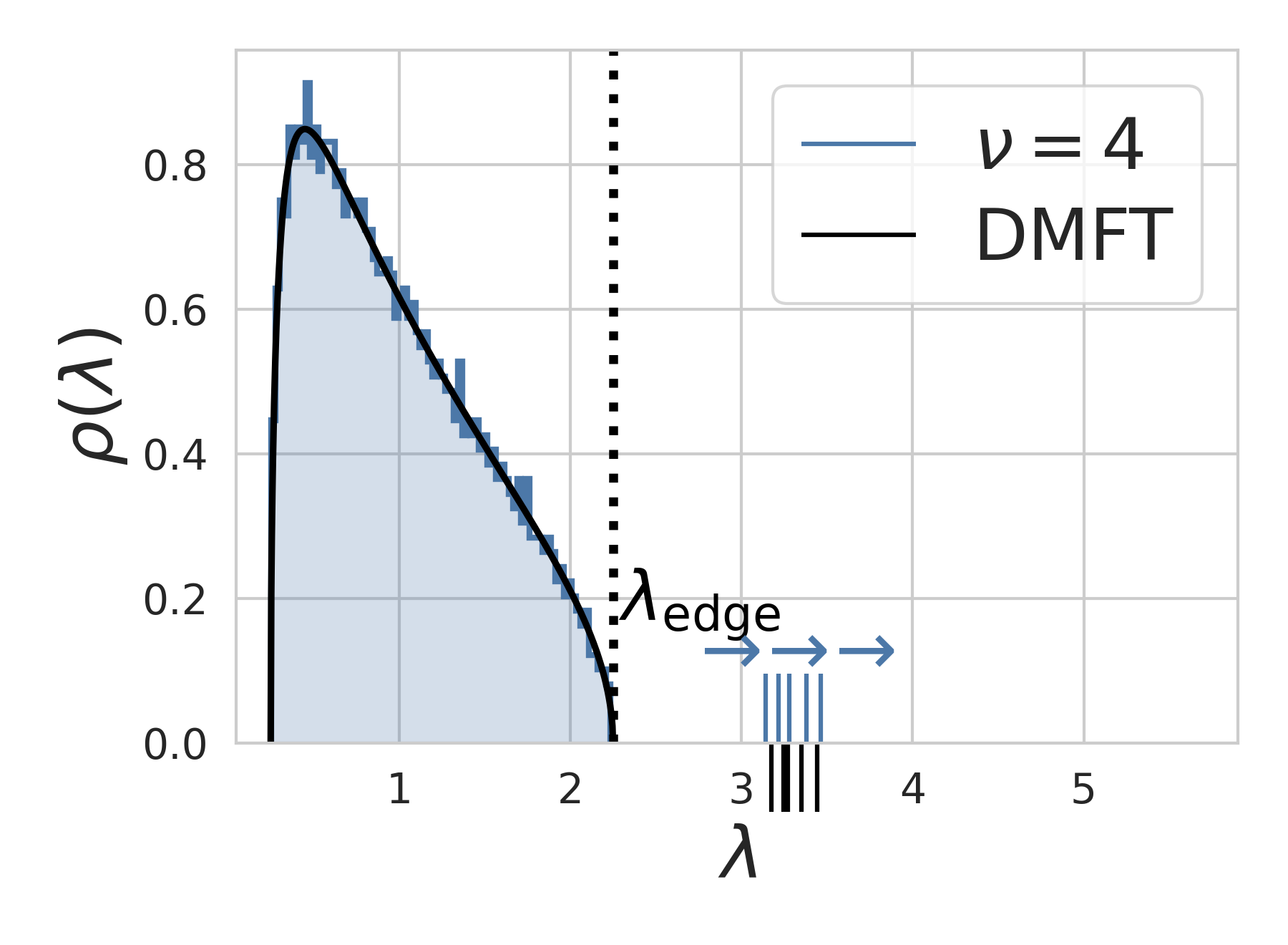}
        \caption{$T=5$}
    \end{subfigure}
    \hfill
    \begin{subfigure}{0.32\linewidth}
        \centering
        \includegraphics[width=\linewidth]{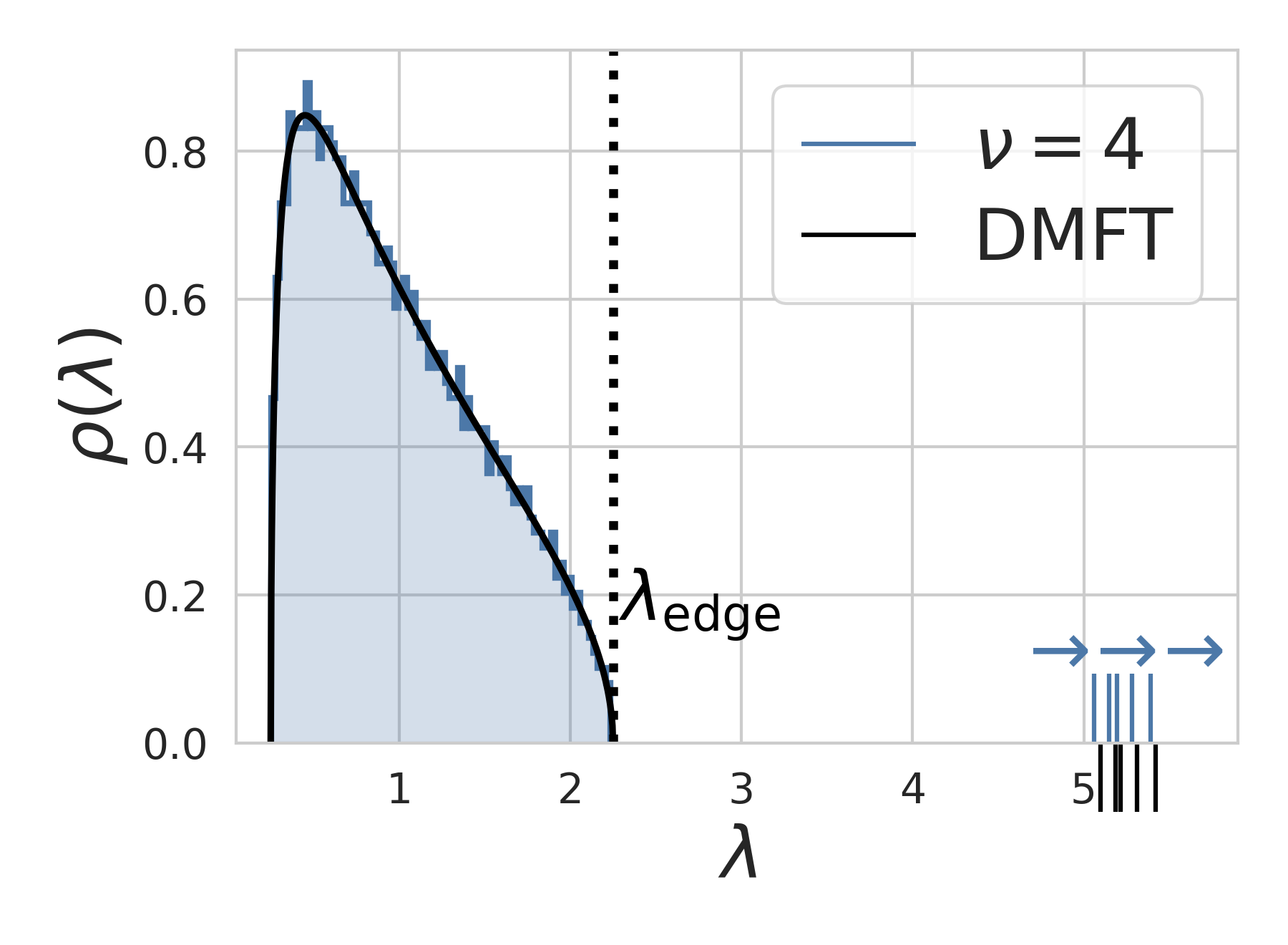}
        \caption{$T=50$}
    \end{subfigure}
    \begin{subfigure}{0.32\linewidth}
        \centering
        \includegraphics[width=\linewidth]{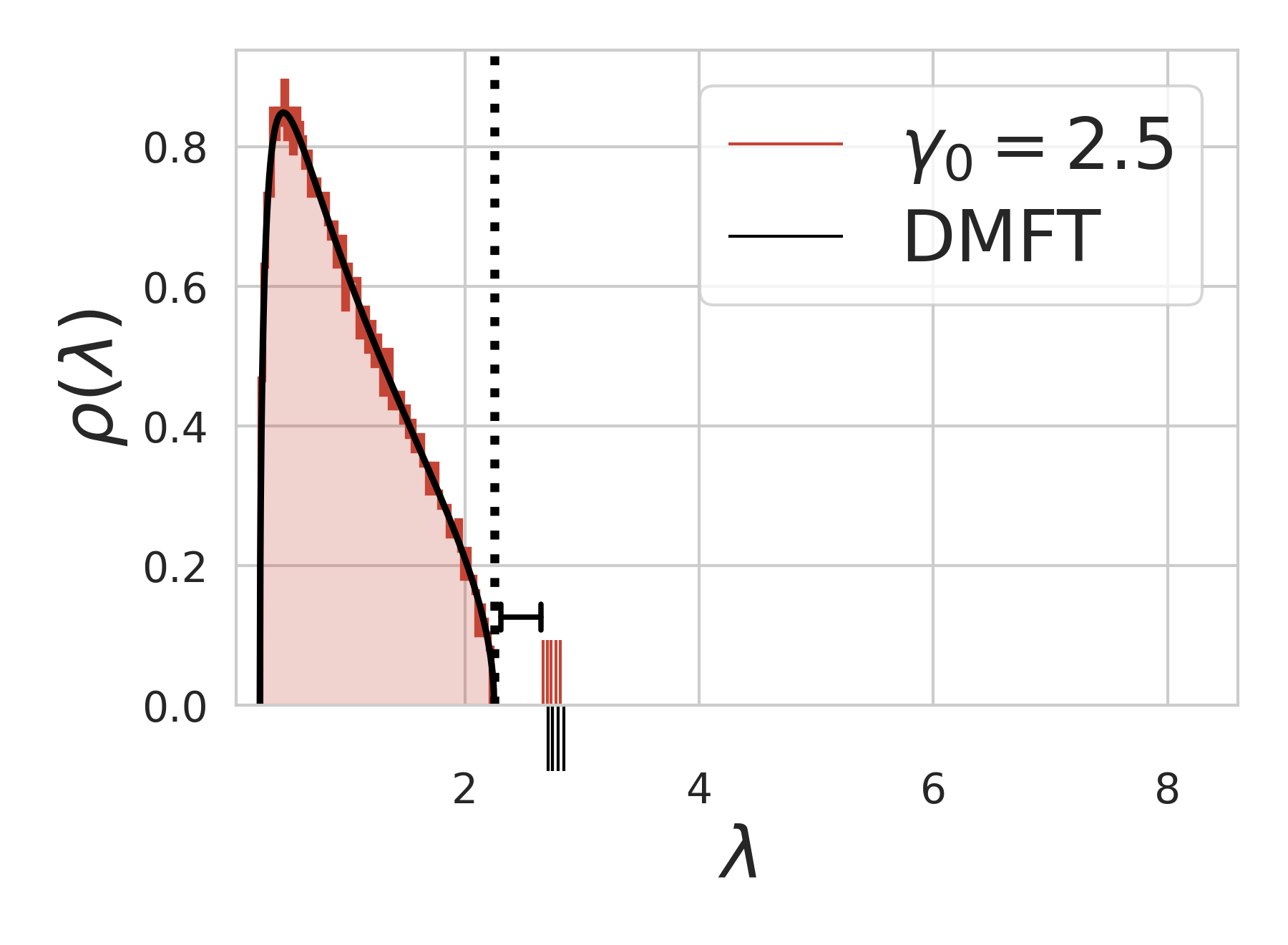}
        \caption{$T=100, \nu=4$}
    \end{subfigure}
    \hfill
    \begin{subfigure}{0.32\linewidth}
        \centering
        \includegraphics[width=\linewidth]{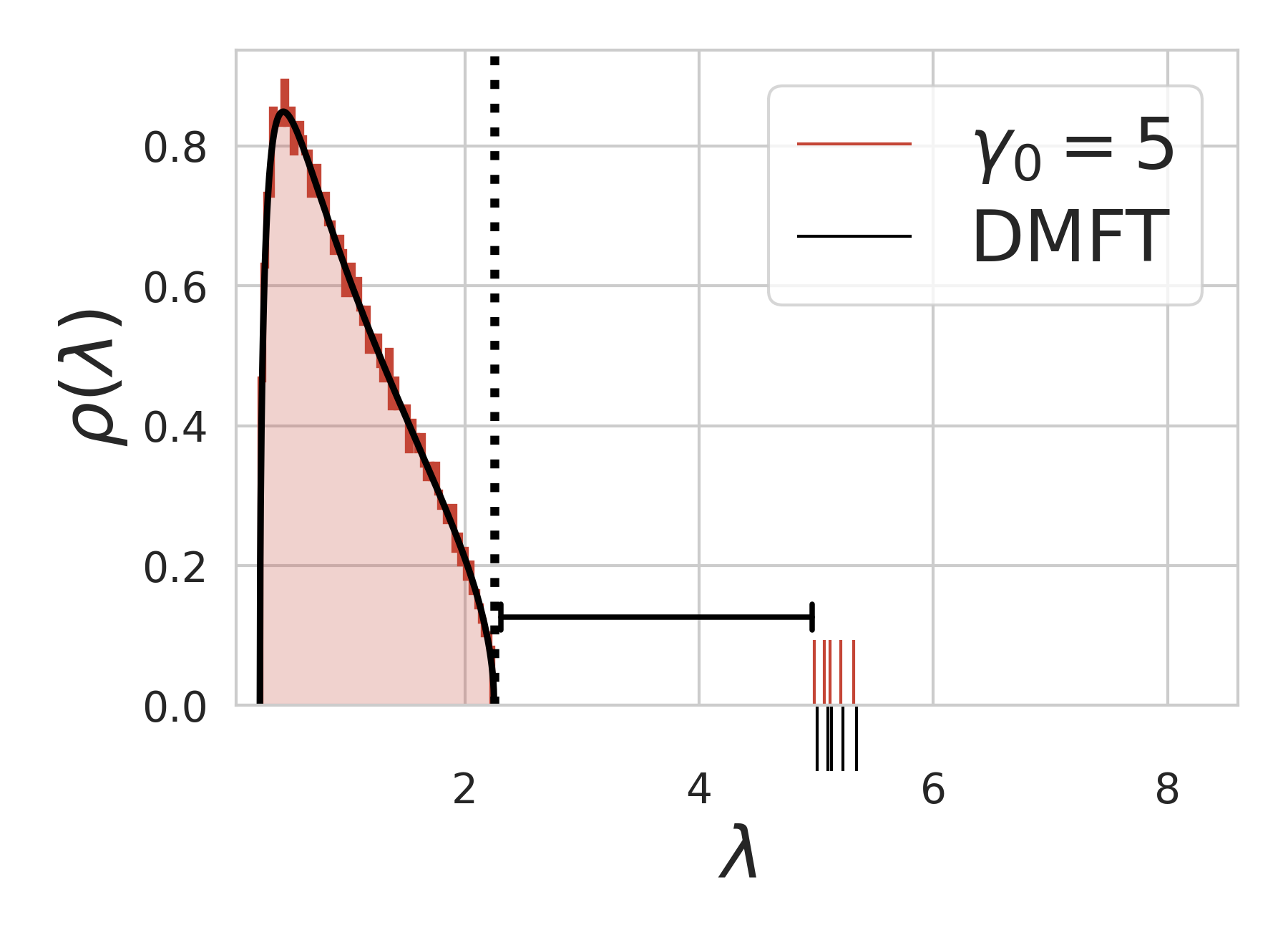}
        \caption{$T=100, \nu=4$}
    \end{subfigure}
    \hfill
    \begin{subfigure}{0.32\linewidth}
        \centering
        \includegraphics[width=\linewidth]{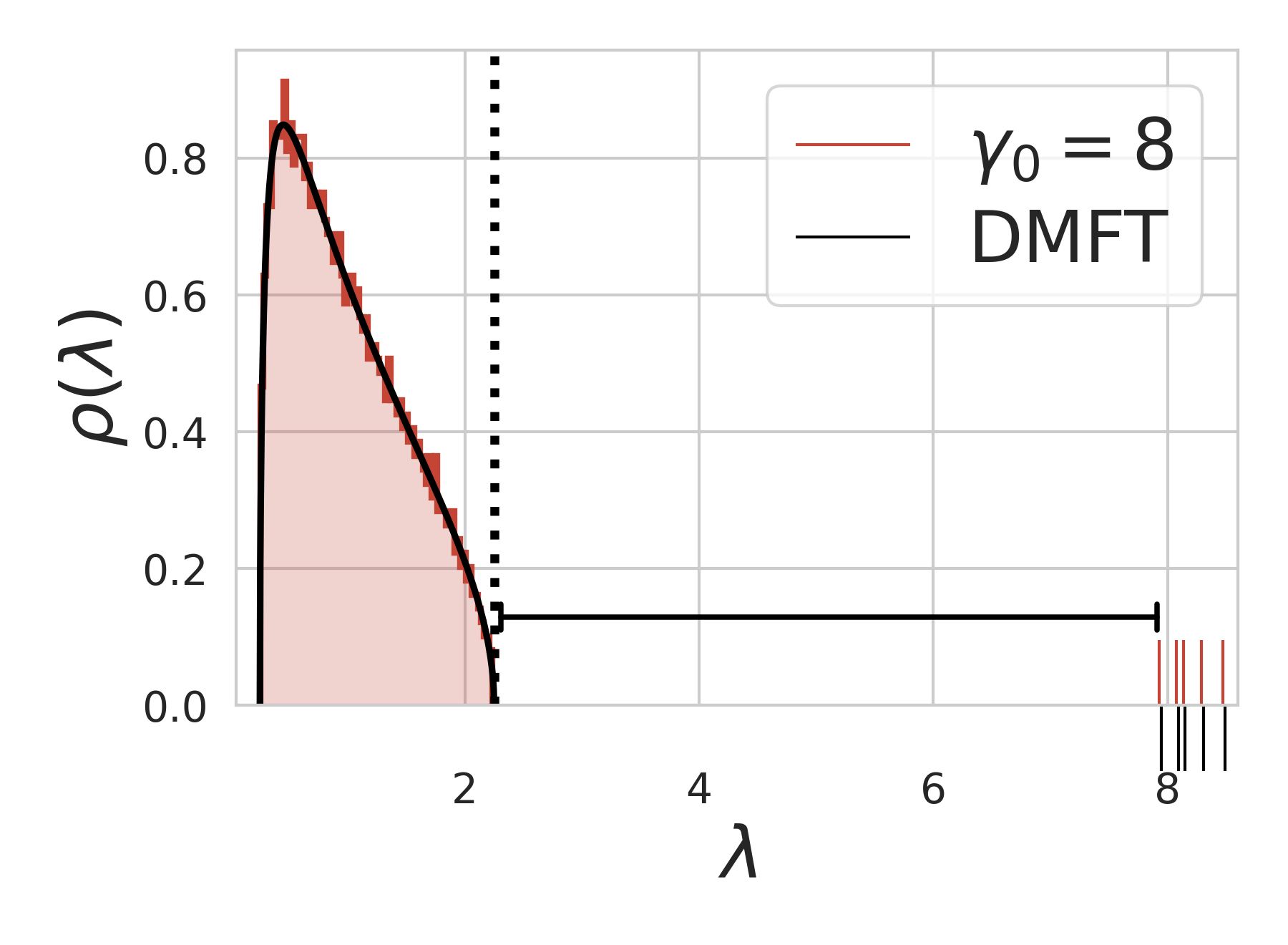}
        \caption{$T=100, \nu=4$}
    \end{subfigure}
    \caption{Spectral density $\rho (\lambda)$ of the weight covariance of a $L=2$ linear network. (a) At initialization ($T=0$), the spectrum follows a Marchenko–Pastur (MP) distribution. (b)-(c) As training progresses under GD, the bulk remains unchanged, while a small number of eigenvalues detach from the bulk and evolve as outliers. (d)-(f) The richness $\gamma_0$ enhances the scale of the outliers. }
    \label{fig:mp_density_linear_vary_T}
\end{figure}
\begin{figure*}
    \centering
    \begin{subfigure}[b]{0.65\linewidth}
\includegraphics[width=\linewidth]{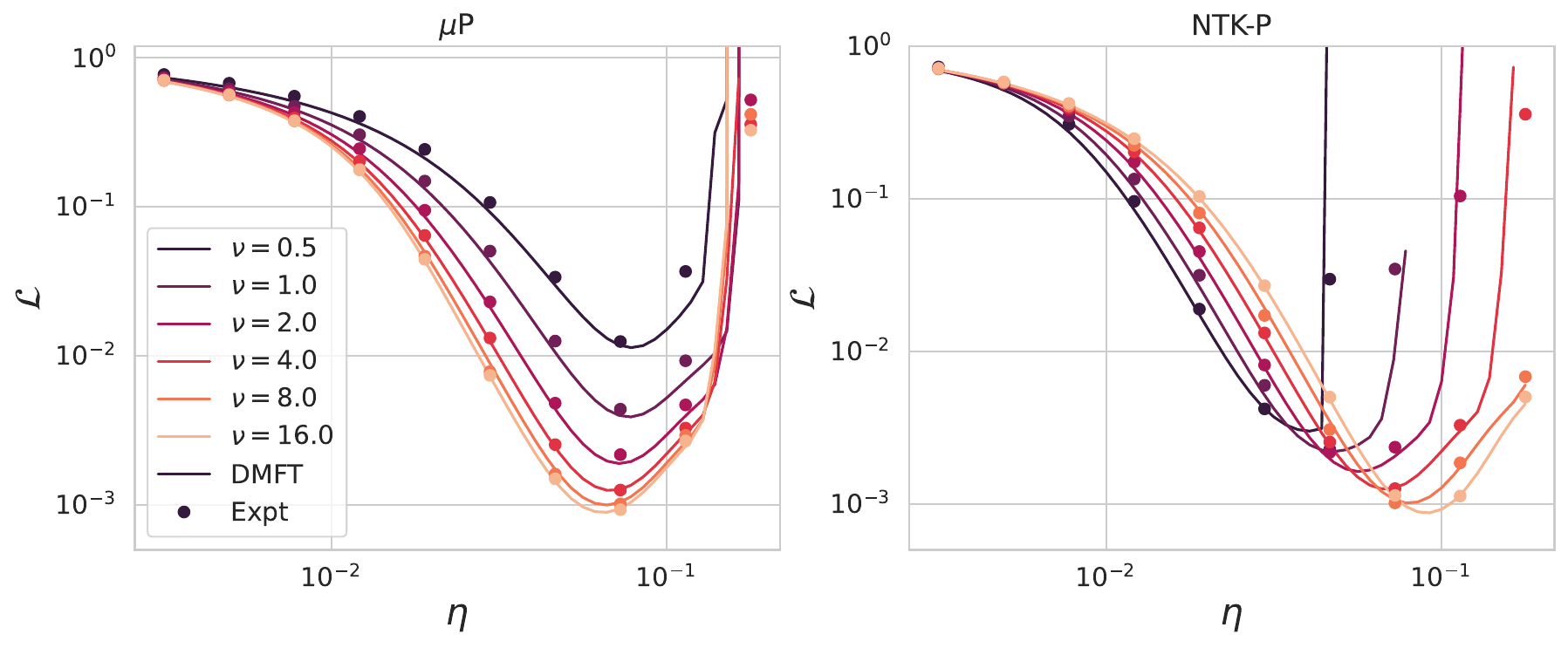}
        \caption{Learning rate transfer under SGD with DMFT Prediction }
    \end{subfigure}
    \begin{subfigure}[b]{0.33\linewidth}
        \includegraphics[width=\linewidth]{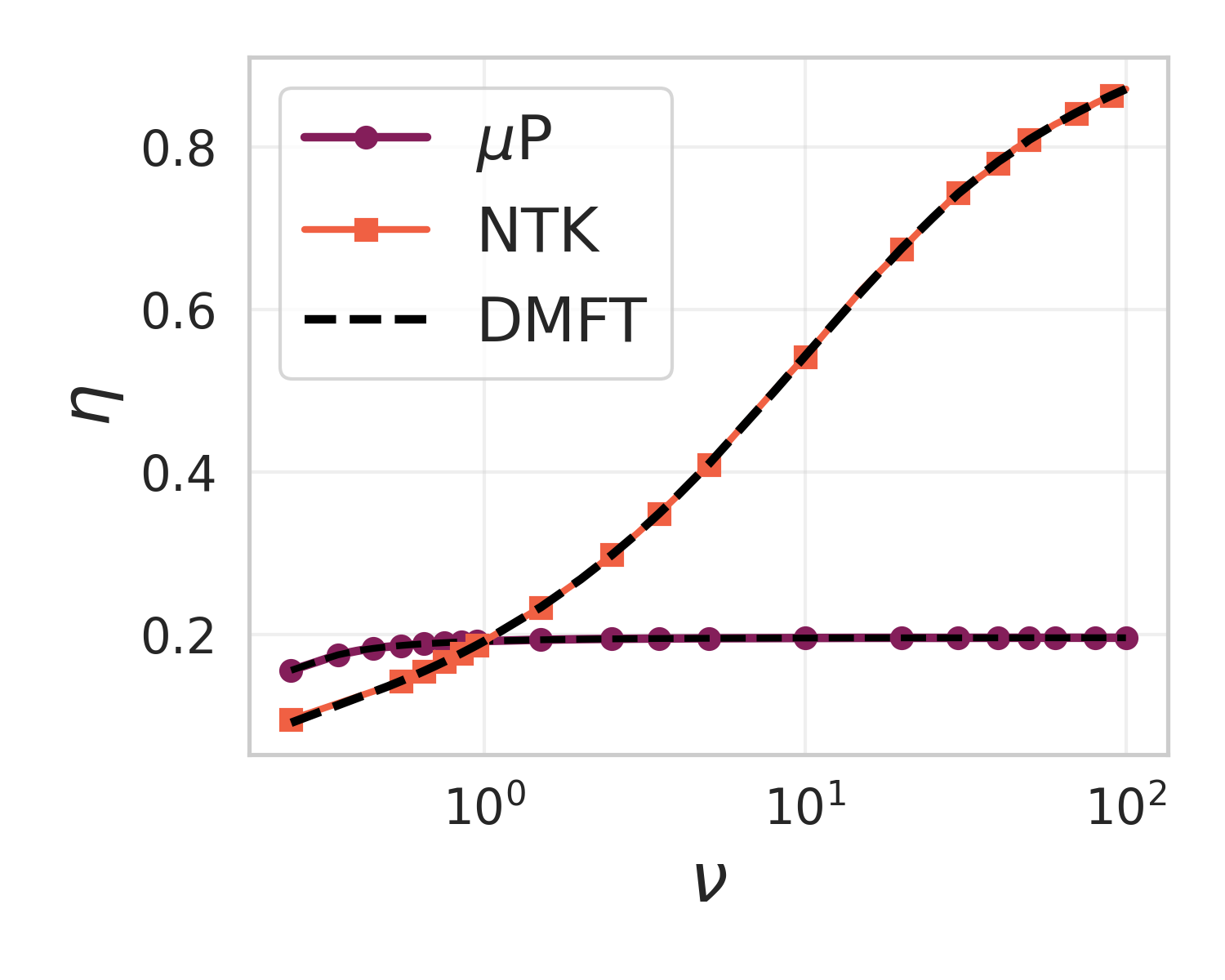}
        \caption{$\eta_{\text{max}}$}
    \end{subfigure}
    \caption{ (a) DMFT can predict the success and failure of learning rate transfer across widths $\nu$ in the proportionally wide linear networks. (b) The maximum stable learning rate as a function of $\nu$ is weakly varying in $\mu$P and is predicted by the NTK top eigenvalue (predicted from our outlier theory). }
    \label{fig:lr_transfer}
    \end{figure*}
\begin{figure*}
    \centering
    \begin{subfigure}[t]{0.32\linewidth}
        \centering
        \includegraphics[width=\linewidth]{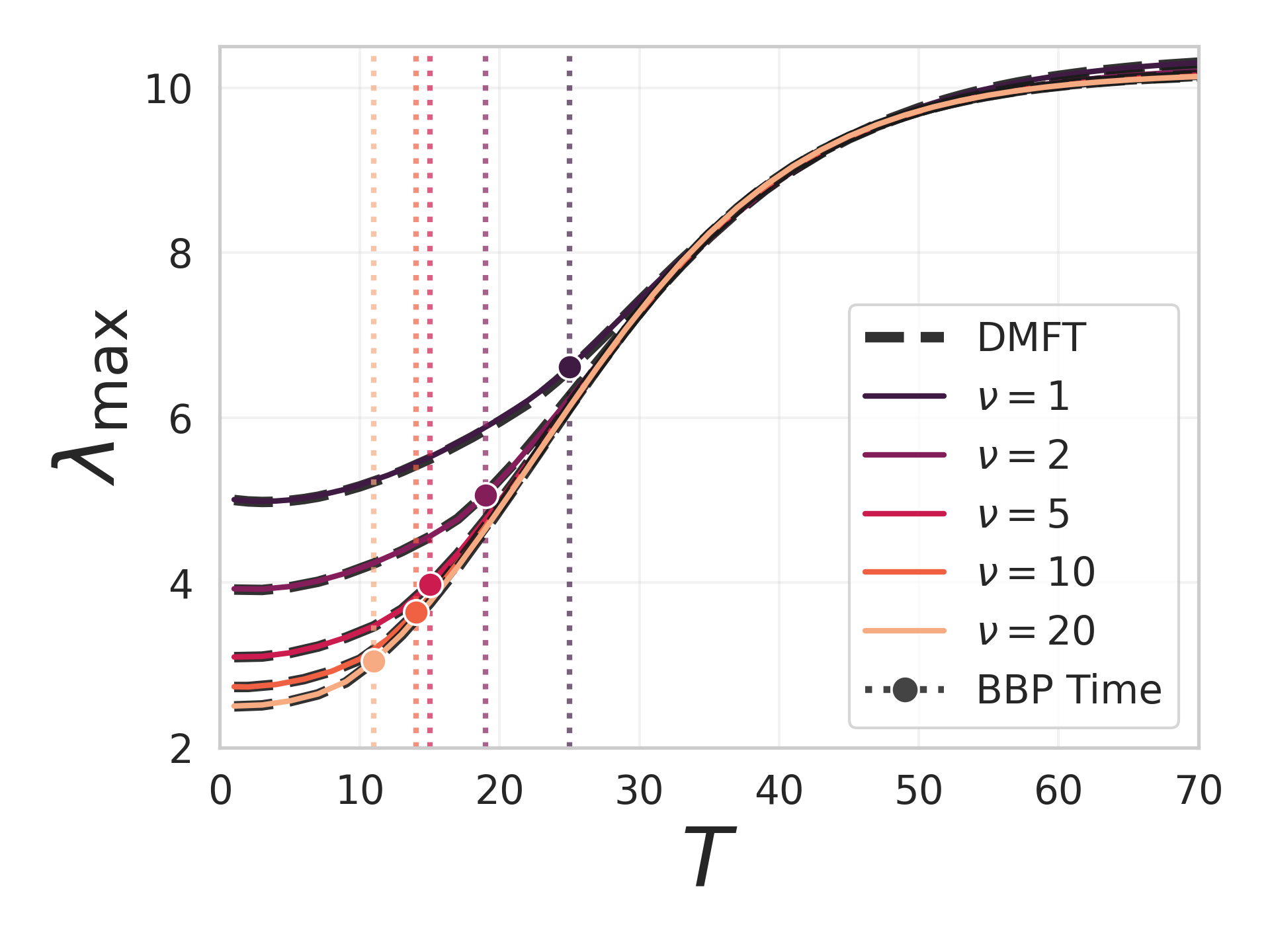}
        \caption{$\mu$P}
        \label{fig:lambda_T_mup}
    \end{subfigure}
    \begin{subfigure}[t]{0.34\linewidth}
        \centering
        \includegraphics[width=\linewidth]{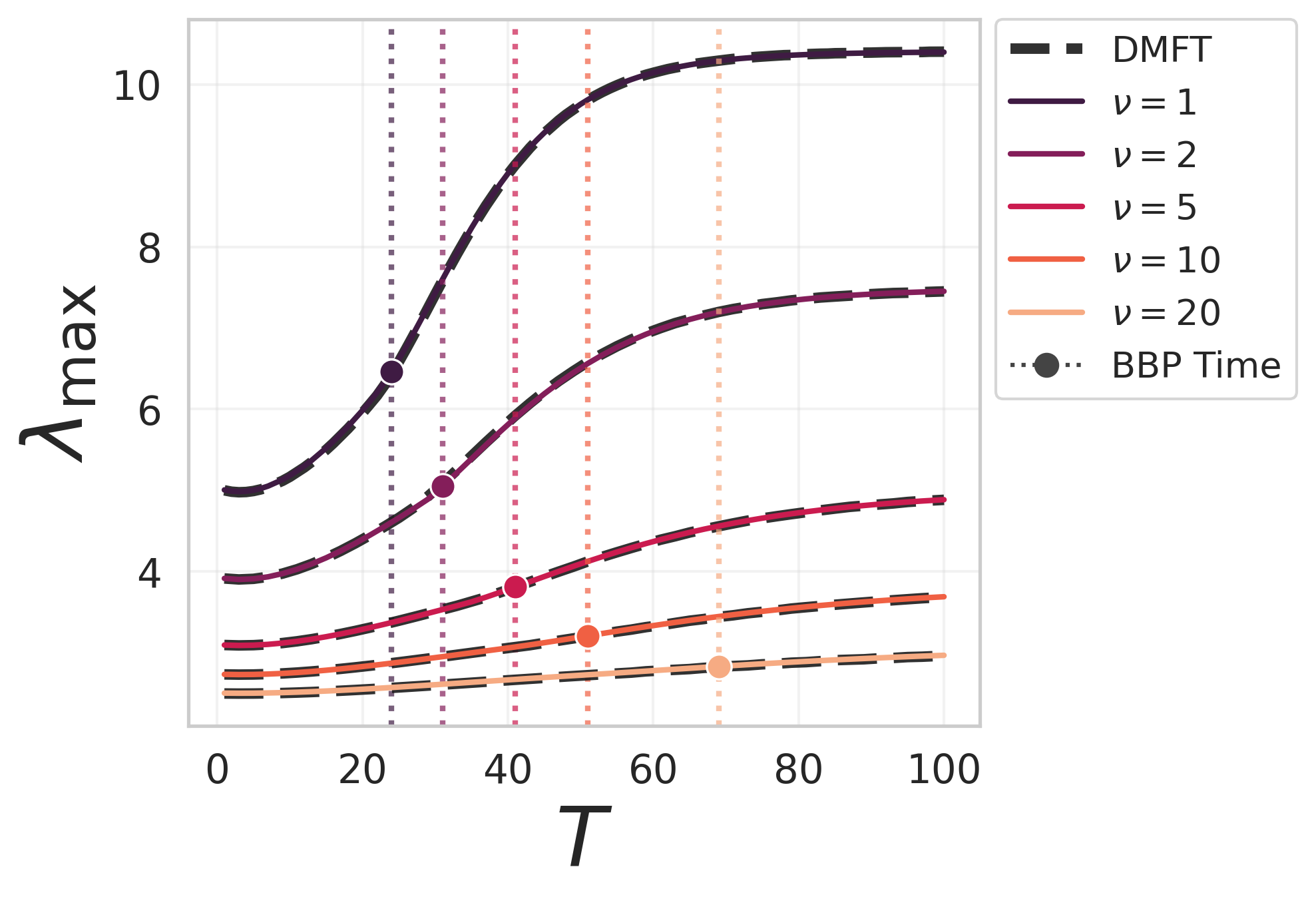}
        \caption{NTK-P}
        \label{fig:lambda_T_ntkp}
    \end{subfigure}
    \begin{subfigure}[t]{0.32\linewidth}
        \centering
        \includegraphics[width=\linewidth]{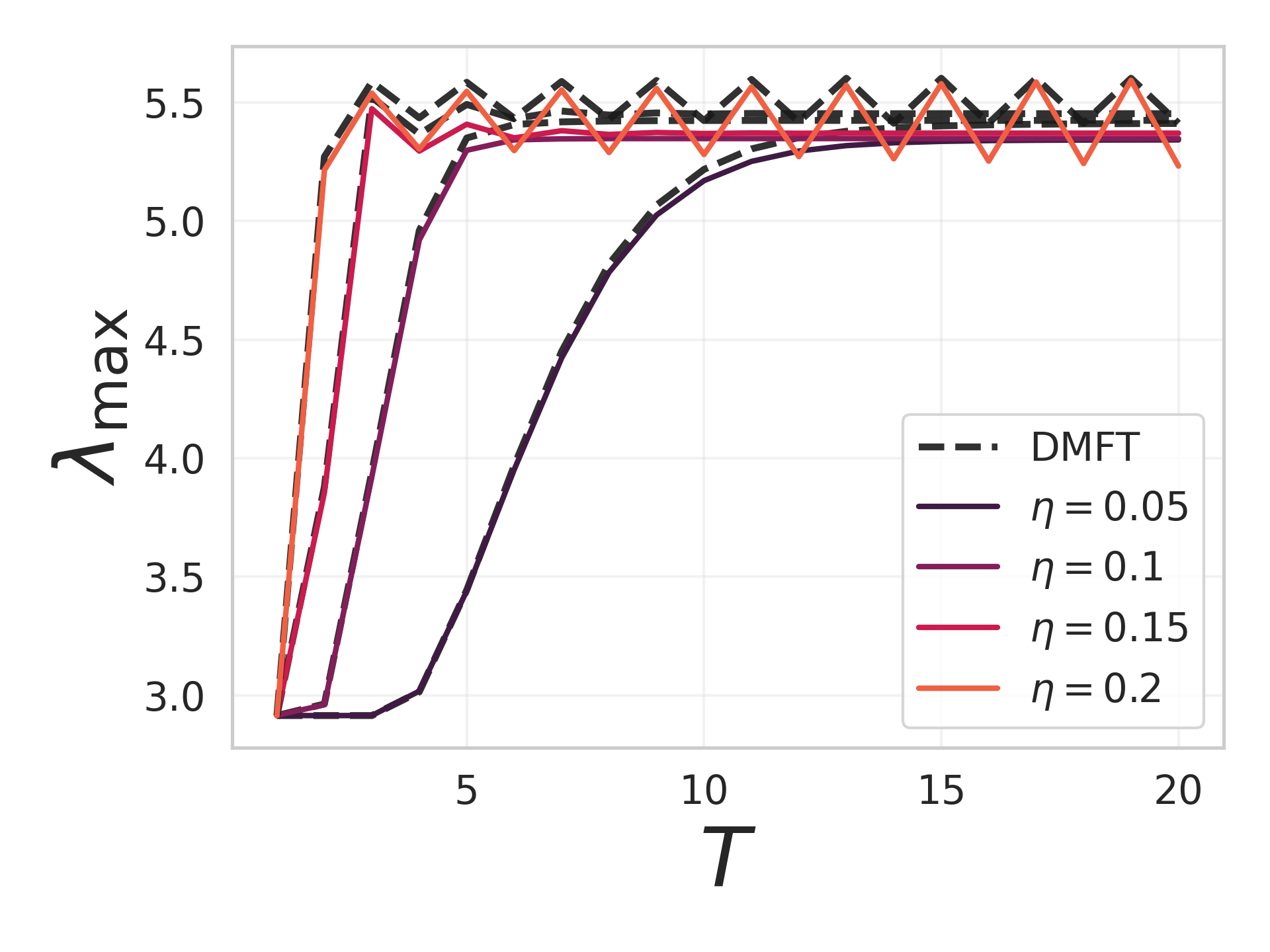}
        \caption{Weight covariance eigenvalue}
        \label{fig:lambda_eta}
    \end{subfigure}
    \caption{Progressive sharpening of the top NTK mode in a deep ($L=2$) linear network. (a,b) Evolution of the top NTK eigenvalue $\lambda_{\max}$ for different aspect ratios $\nu$ and parameterizations. In $\mu$P, the post-BBP dynamics is nearly $\nu$-independent, whereas in NTK-P it remains strongly $\nu$-dependent and approaches a lazy large-$\nu$ limit. (c) $\lambda_{\max}(T)$ for different learning rates $\eta$: larger $\eta$ speeds up sharpening until edge-of-stability (EoS).}
    \label{fig:ntk_sharpening_combined}
\end{figure*}

The weight updates are formed as sums over both time steps $t \in [T]$ and class indices $c \in [C]$
\begin{align}
    &\bm W^\ell(T) = \bm W^\ell(0) + \frac{\eta \gamma_0}{\sqrt N} \sum_{t=1}^T \sum_{c=1}^C \bm g^{\ell+1}_c(t) \bm \phi^{\ell}_c(t)^\top. 
\end{align}
The DMFT equations provide the necessary description of the random (Gaussian in this case) fields $\bm g^\ell_{c}(t) \sim g^{\ell}_{c,t}( \{ \bm\chi^\ell_{c'}(s) , \bm\xi^{\ell}_{c'}(s) \}_{s<t}   )$ and $\bm\phi^{\ell}_{c}(t) \sim \phi^{\ell}_{c,t}( \{ \bm\chi^\ell_{c'}(s) , \bm\xi^{\ell}_{c'}(s) \}_{s<t}   )$, where $\sim $ indicates equivalence in the joint scaling limit and the functions $\phi_{c,t}^\ell(\cdot)$ and $g^\ell_{c,t}(\cdot)$ act elementwise over the vectors. In Figure \ref{fig:mp_density_linear_vary_T} we illustrate the emergence of $C$ outlier eigenvalues from the bulk as $T$ increases. The behavior of these outliers is well predicted by the theory. Similarly, Figure  \ref{fig:mp_density_linear_vary_T} illustrates at a fixed number of steps $T$, the impact of the output multiplier $\gamma_0$ on the strength of the outlier spike. Figures~\ref{fig:lr_transfer} and~\ref{fig:ntk_sharpening_combined} connect these outlier dynamics to learning-rate transfer by showing that the maximal stable learning rate is controlled by the top NTK mode: in $\mu$P, this is nearly width-stable, whereas in NTK-P it remains strongly $\nu$-dependent.

\section{Extensive Output Channels: Beyond Constant Bulk + Dynamic Spikes}
\begin{figure}
    \centering     
\begin{subfigure}[b]{0.37\linewidth}
        \includegraphics[width=\linewidth]{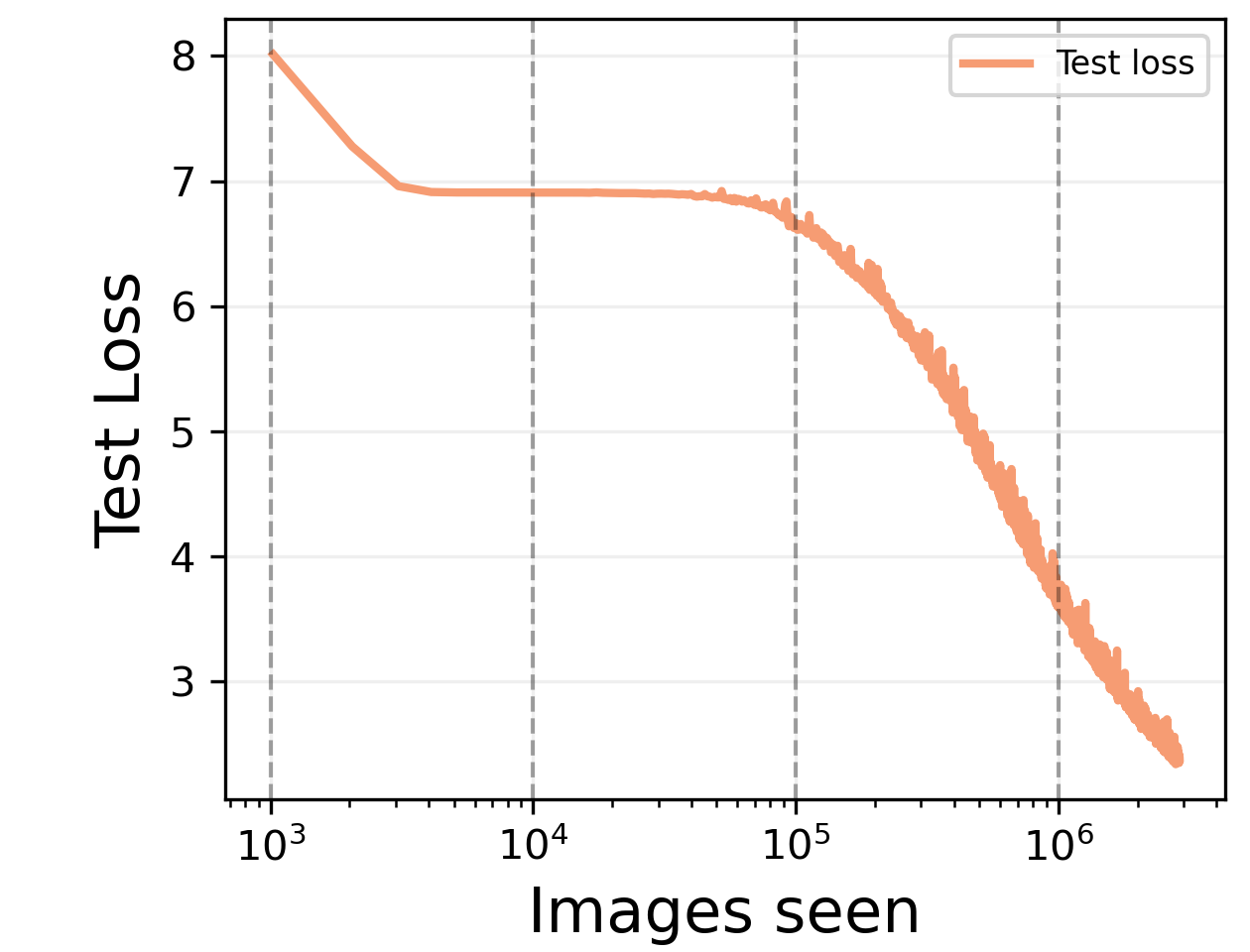}
        \caption{Loss Dynamics}
    \end{subfigure}
\begin{subfigure}[b]{0.37\linewidth}
        \includegraphics[width=\linewidth]{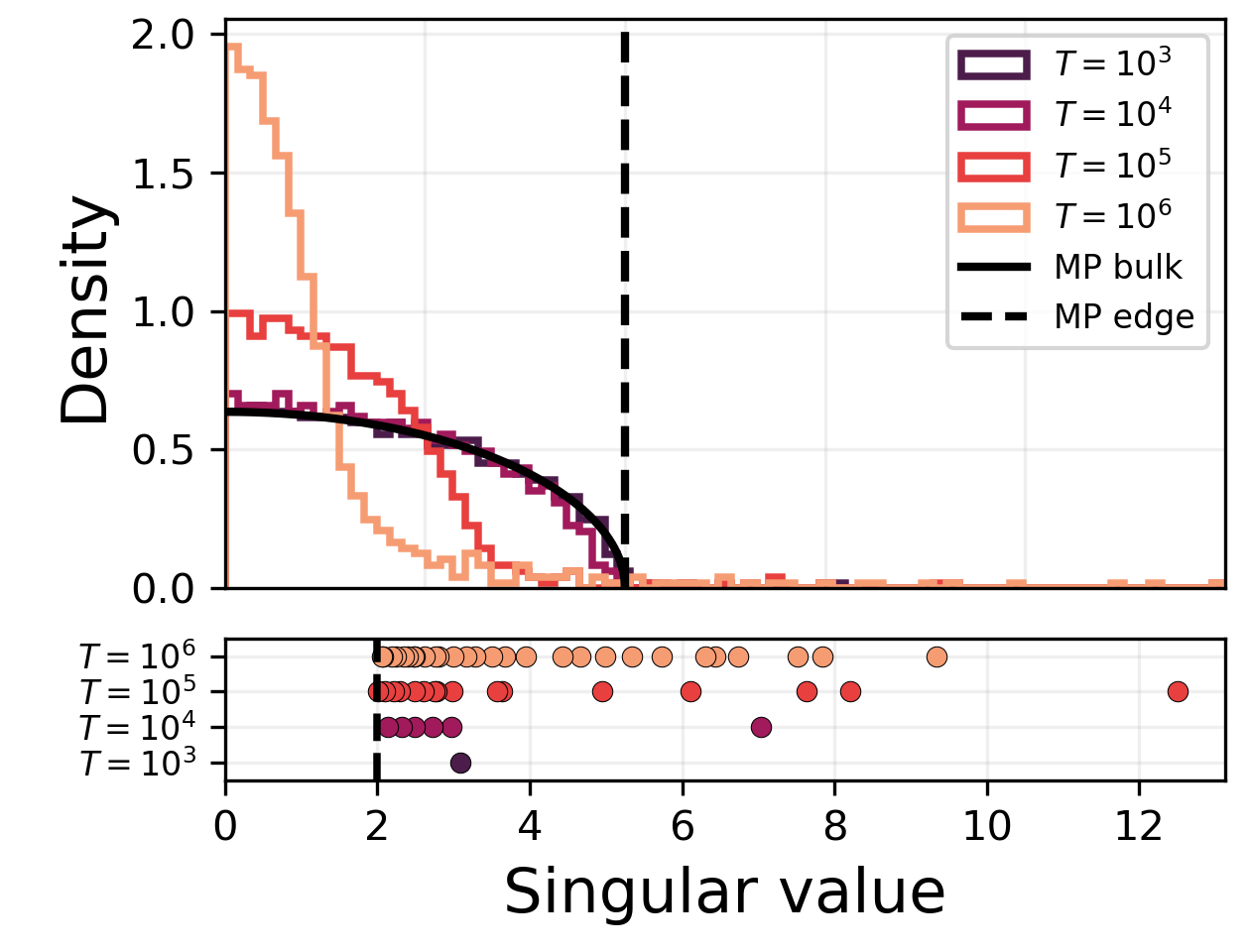}
        \caption{Weight Spectra vs $T$}
    \end{subfigure}
    \caption{Width-256 $\mu$P ResNet18 on ImageNet. Test loss drops around $10^5$ images seen, while the MP-normalized spectrum of the first hidden-layer ResNet block deforms: the bulk shifts left and a right tail emerges beyond the MP edge.}
    \label{fig:resnet_imagenet}
\end{figure}

\begin{figure}
    \centering     
\begin{subfigure}[b]{0.37\linewidth}
        \includegraphics[width=\linewidth]{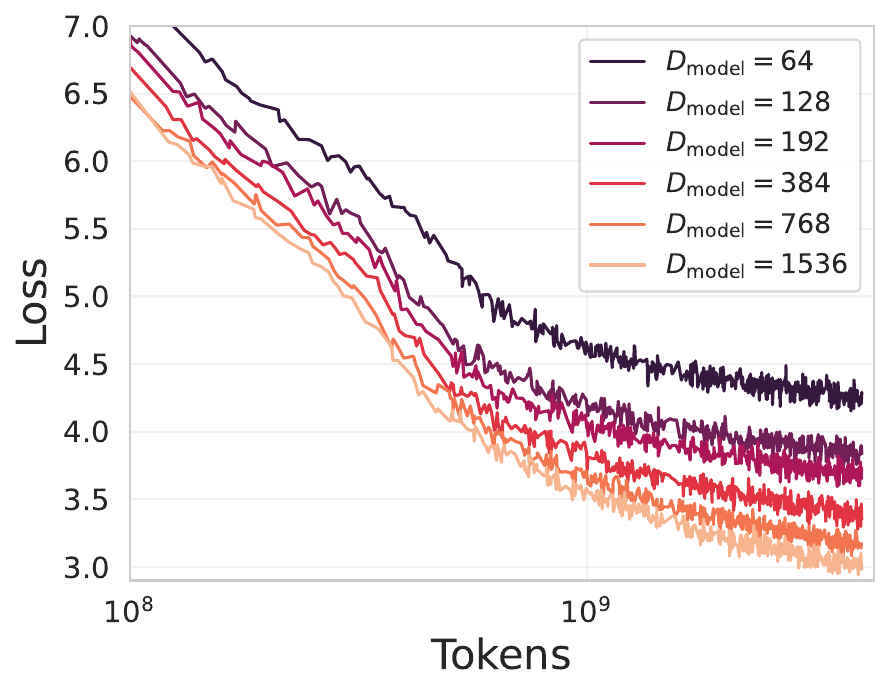}
        \caption{Loss Dynamics Across Widths}
    \end{subfigure}
\begin{subfigure}[b]{0.37\linewidth}
        \includegraphics[width=\linewidth]{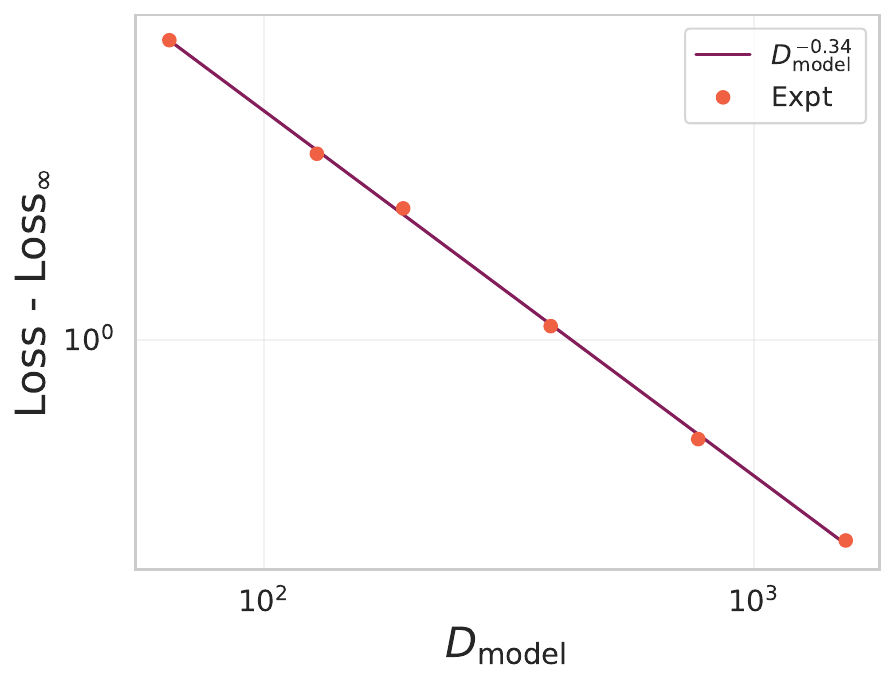}
        \caption{Non-trivial Scaling Exponent}
    \end{subfigure}
\begin{subfigure}[b]{0.85\linewidth}
\includegraphics[width=\linewidth]{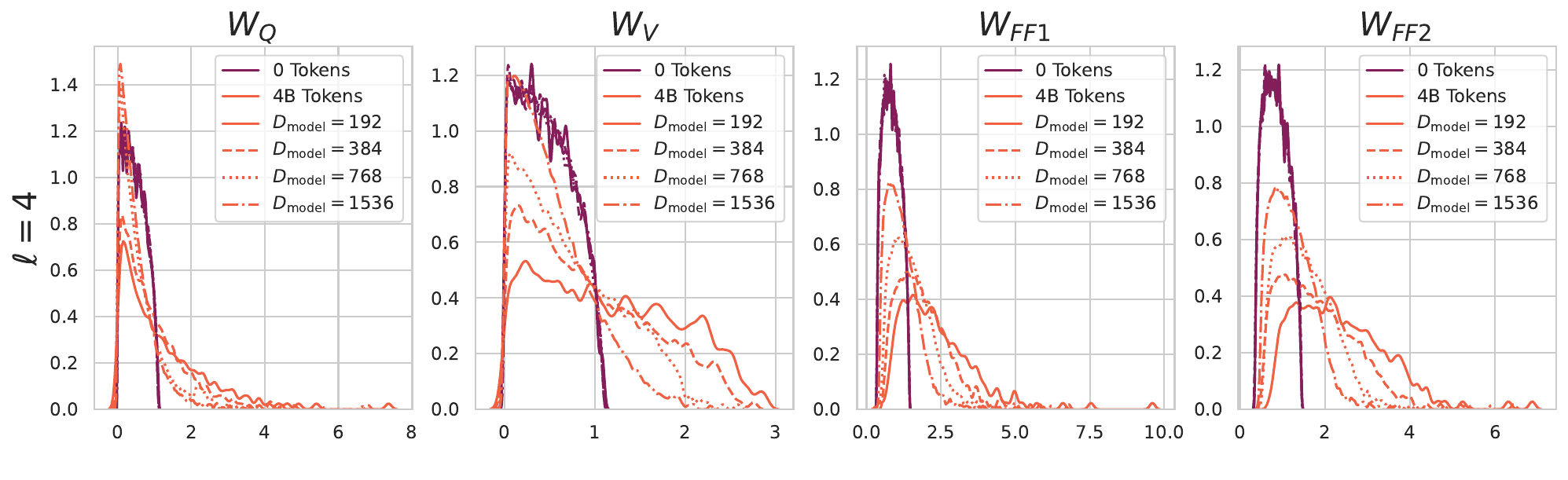}
\caption{Weight Spectra Before and After Training}
\end{subfigure}
    \caption{Losses and spectral densities of weights before and after language model pretraining in $\mu$P on $4$B tokens on the C4 dataset in models of varying widths $D_{\text{model}} = 64 \times \text{heads}$. The models range from $H=3$ heads to $H=24$ heads with $L=6$ layers and spectra are plotted before and after training. The spectral densities are not well described by a constant bulk with finite set of outliers, falling outside the scope of Result 1.  }
    \label{fig:language_models}
\end{figure}

Although Result 1 captures the finite-outlier regime observed in simple models and CIFAR-10 CNNs, larger-output tasks exhibit a different spectral phenomenology. We first observe this departure in an ImageNet-scale ResNet18, where training modifies the bulk of the hidden-weight spectrum rather than producing only a few isolated outliers (Fig.~\ref{fig:resnet_imagenet}). The same effect appears more prominently in GPT-style language models trained on next-token prediction on the C4 dataset (Fig.~\ref{fig:language_models}), where the output vocabulary is much larger than the hidden width. These observations motivate a large-output-channel regime in which training restructures an extensive part of the spectrum.


\vspace{-5pt}
\subsection{Candidate Theory: Large Output Channels} 
\vspace{-5pt}
 Motivated by the fact that the output vocabulary size for language models is often larger than the width (in Fig.~\ref{fig:language_models} the vocabulary is $\sim 50$k and the width for $H=12$ is $N=768$), we consider an extension of our linear network theory where the output channels $C$ is of the same order as width $N$.

\begin{tcolorbox}[colframe=myblue, opacityback=0.95, title=Result 2: Linear Network with Extensive Classes at One Step] 
Consider randomly initialized width-$N$ linear network with jointly diverging output dimension $C$ and input dimension $D$. For $\bm M^\ell=\frac{1}{N}\bm W^\ell(T)^\top \bm W^\ell(T)$, we study the joint limit
\begin{equation}
N,P,D,C\to\infty,\qquad \alpha=P/D,\quad \nu=N/D,\quad \chi=C/D,
\end{equation}
in which the singular-value density converges as $\rho_{\bm M^\ell}(s)\to \rho^\ell(s;\alpha,\nu,\chi,\gamma_0)$.
For $\gamma_0,\chi>0$, this limiting bulk density is generally not Marchenko--Pastur. In the large $\nu, \alpha \to \infty$ limit, the leading finite-$\nu$ correction to the top eigenvalue for a two-layer model approaches
\begin{equation}
    \lambda_{\rm max}  = \lambda_{\rm max}^{\infty} +  \frac{1}{\nu} \  \lambda_{\rm max}^\infty
    \left(\sqrt{\chi} +\frac{\chi}{\gamma_0^2(1+\sqrt{\chi})} \right) +
    \mathcal O(\nu^{-2})
\end{equation}
so, in the extensive-output regime, width-consistent $\lambda_{\rm max}$ can still occur if correction is small.
\end{tcolorbox}
In the population limit $\alpha\to\infty$, one GD step (see Appendix~\ref{app:extensive_outputs} for details) gives
\begin{equation}
    \bm W^{\ell}(1)
    =
    \bm W^{\ell}(0)
    +
    \frac{\eta\gamma_0}{\sqrt{CD}}
    \bm G^{\ell+1}(0)\bm B^\star \bm H^\ell(0)^\top.
\end{equation}
Here $\bm H^\ell\in\mathbb R^{N\times D}$ is the forward feature map, $\bm G^{\ell+1}\in\mathbb R^{N\times C}$ is the back-propagated readout, and $\bm B^\star\in\mathbb R^{C\times D}$ is the target matrix. At initialization, these matrices are statistically independent and asymptotically free~\cite{potters2020first}, so the one-step update is a product of independent random matrices. Extending this analysis to full training requires tracking the correlations that develop across layers, which we leave for future work. 
\begin{figure*}
\centering
\begin{subfigure}[t]{0.32\textwidth}
    \centering
    \includegraphics[width=\linewidth]{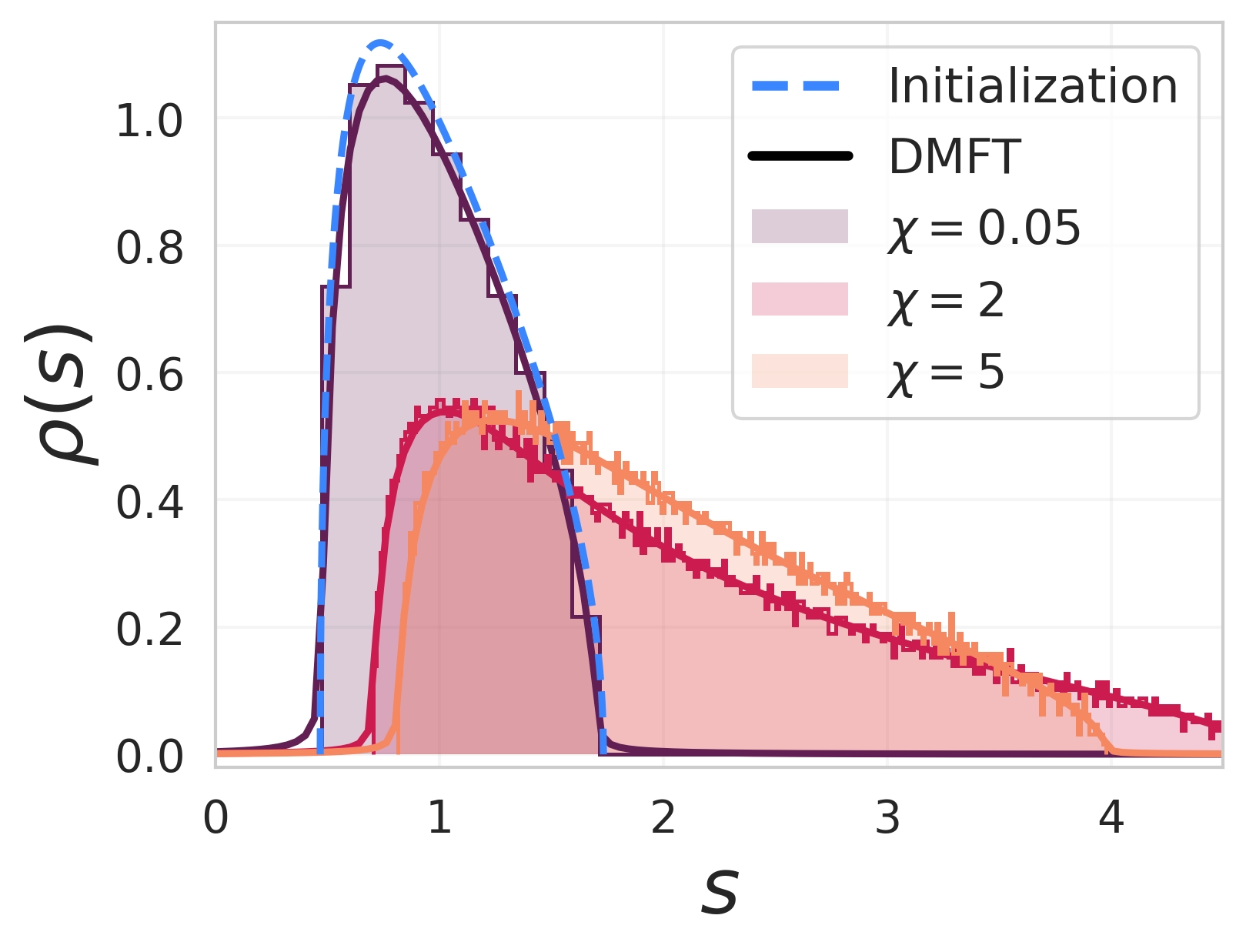}
    \caption{$\gamma_0 = 1$}
\end{subfigure}
\hfill
\begin{subfigure}[t]{0.32\textwidth}
    \centering
    \includegraphics[width=\linewidth]{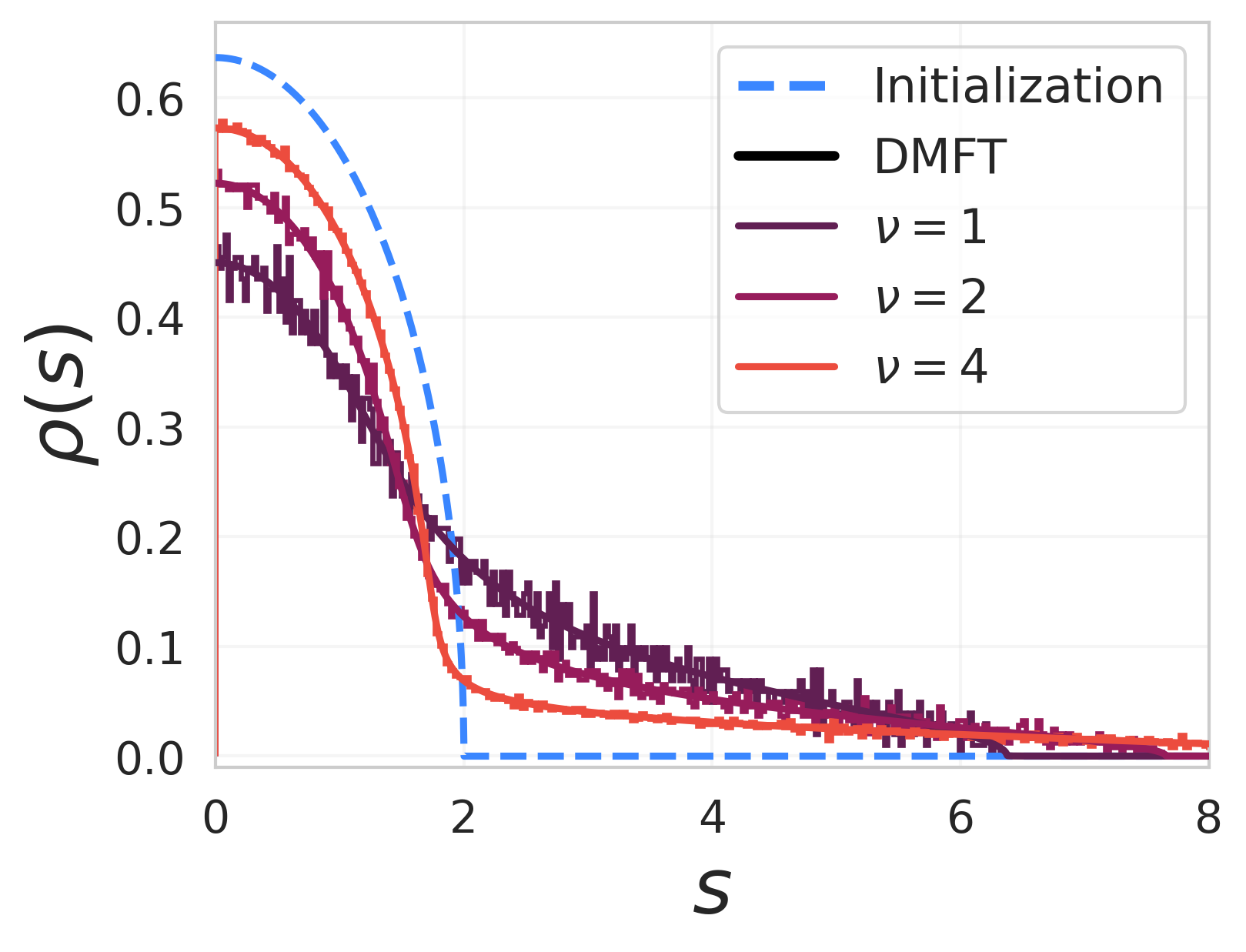}
    \caption{$\gamma_0 = 2$}
\end{subfigure}
\hfill
\begin{subfigure}[t]{0.32\textwidth}
    \centering
    \includegraphics[width=\linewidth]{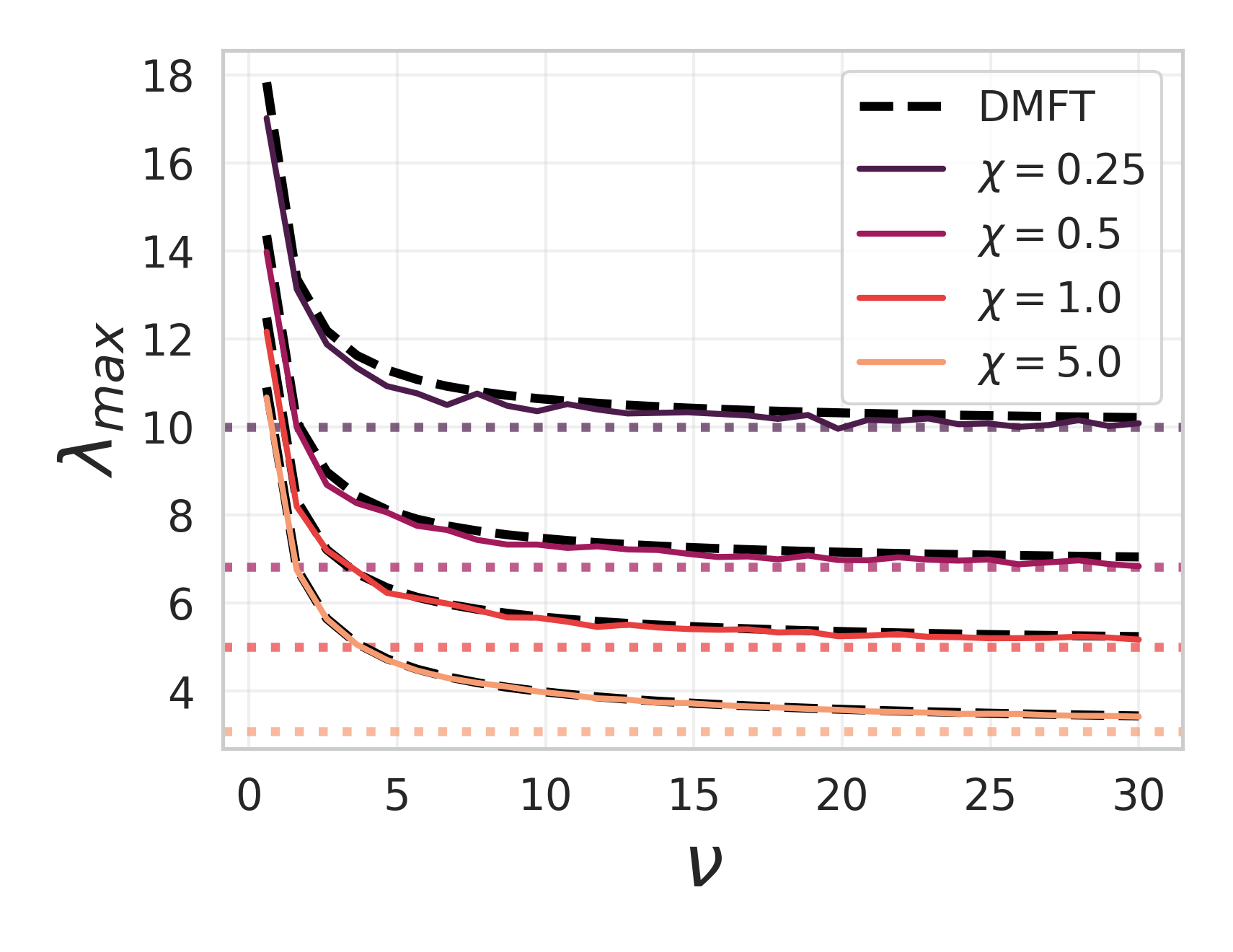}
    \caption{$\gamma_0 = 1$}
\end{subfigure}

\caption{
Large-output spectral densities.
(a) Two-layer bulk shifts away from MP as $\chi$ increases.
(b) Three-layer hidden-weight spectrum sharpens with width $\nu$.
(c) $\lambda_{\max}$ stabilizes across width and approaches the shifted-MP plateau
$1+\gamma_0^2(1+\chi^{-1/2})^2$.
}
\label{fig:largeC_threepanel}
\end{figure*}
We illustrate the spectra of this model across $\chi$, $\gamma_0$ and $\nu$ in Fig.~\ref{fig:largeC_threepanel}. 

\section{Discussion}
We developed a method to compute the dynamics of spectral outliers for spiked matrix ensembles where the spikes have a causal dependence on the random component of the matrix. This method enabled us to characterize the outlier dynamics of weights in (1) super wide nonlinear networks and (2) proportional width linear networks. We showed that this theory describes some regimes of network training, such as CNNs on CIFAR-10 but fails to capture others such as language model pretraining, for which we propose a simple toy model to account for an extensive number of classes. Several recent works have proposed alternative solvable models of extensive rank adaptations in fully trained quadratic networks \cite{maillard2024bayes, erba2026nuclear} and student-teacher attention networks \cite{boncoraglio2025single}. Further, similar DMFT treatment in these solvable models has been fruitfully applied \cite{martin2026high}. Our two-level DMFT framework could potentially be useful to examine the evolution of the spectrum and outliers in these and related extensive rank models. 

\vspace{-8pt}
\paragraph{Limitations} 
Currently the theory only applies to either (1) nonlinear networks in the super-wide regime where the constant bulk + spikes always holds or (2) a proportionally wide regime for linear networks where the outliers are also finite when learning a finite set of output channels. Further, the outlier computation scales as $\mathcal O(S^3)$ in the number of spike directions since the zero determinant condition. Lastly, our analysis of the extensive class linear network is currently restricted to a single step of GD. More detailed analysis of the dynamics of the spectrum is left for future work.

\subsection*{Acknowledgements}
The authors would like to thank Enrico Malatesta, Denny Wu, and Nikhil Ghosh for inspiring conversations about weight and kernel spectra in $\mu$P networks. We also thank Jacob Zavatone-Veth, Alex Atanasov, Alex Meterez, Adam Lee, Mary Letey, Itay Lavie, David Clark, Hamza Chaudhry, Pierfrancesco Beneventano, Francesco Mori, Francesca Mignacco, Jamie Simon, Daniel Kunin, Bruno Loureiro, Florent Krzakala, and Emanuele Troiani for useful discussions and feedback on this work.

C.L. is supported by DARPA grants DIAL-FP-038 and AIQ-HR00112520041. C.P. is supported by an NSF CAREER Award (IIS-2239780), DARPA grants DIAL-FP-038 and AIQ-HR00112520041, the Simons Collaboration on the Physics of Learning and Neural Computation, and the William F. Milton Fund from Harvard University. This work has been made possible in part by a gift from the Chan Zuckerberg Initiative Foundation to establish the Kempner Institute for the Study of Natural and Artificial Intelligence. B.B. is supported by the Center of Mathematical Sciences and Applications at Harvard University. B.B. acknowledges the Texas Advanced Computing Center (TACC) at The University of Texas at Austin for providing resources that have contributed to the research results reported within this paper, specifically allocations DMS26007 and DMS26010 on the Lonestar6-GPU system.

\bibliographystyle{unsrt}
\bibliography{neurips_2026/references}


\appendix
\section{Classical BBP Transition: Wigner + Rank-One Spike}\label{appendix:classical_bbp}
\begin{figure}[H]
    \centering
    \includegraphics[width=0.5\linewidth]{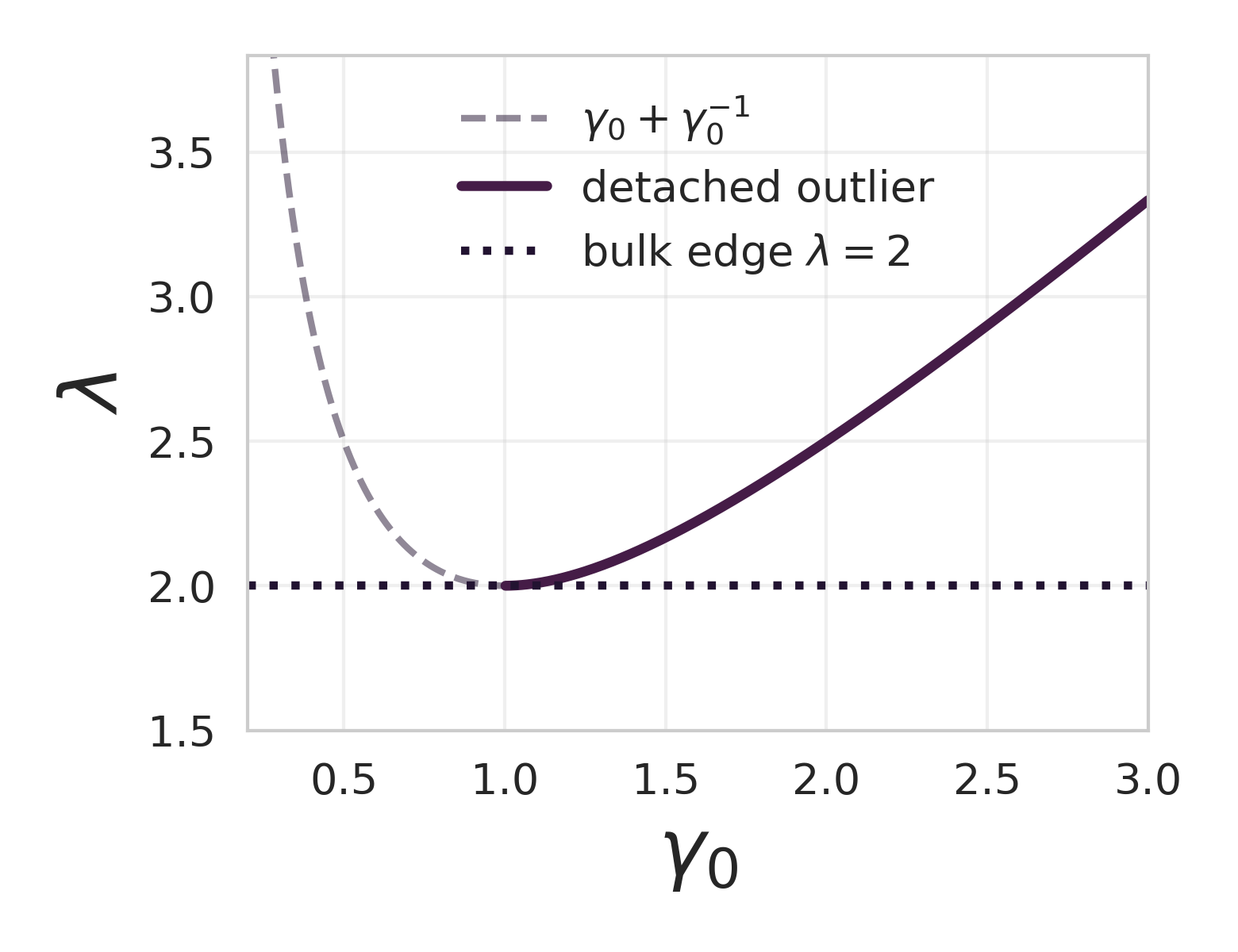}
    \caption{Classical BBP transition in the rank-one spiked Wigner model. The outlier location at $\lambda = \gamma_0 + \gamma_0^{-1}$ detaches from the Wigner bulk edge $\lambda =2$ for $\gamma_0>1$.}
    \label{fig:placeholder}
\end{figure}
As a warm-up, we consider the standard rank-one spiked Wigner model
\begin{equation}
    \bm M= \bm M_0 + \frac{\gamma_0}{D}\bm v\bm v^\top,
    \qquad
    \|\bm v\|^2=D
\end{equation}
where $\bm M_0$ is a GOE matrix. The goal is to recover the classical BBP outlier condition using the same response-function method that we later apply to training-induced spikes. This example illustrates the main idea in the simplest setting: the normalized trace response gives the random-matrix bulk, while the projection of the linear probe onto the spike direction detects isolated eigenvalues.

We introduce an auxiliary spectral-time dynamics
\begin{align}
    \partial_\tau \bm h(\tau)
    &=
    -\bm M\bm h(\tau)
    \nonumber \\
    &=
    -\bm M_0\bm h(\tau)
    -
    \gamma_0 C_{vh}(\tau)\bm v
\end{align}
where
\begin{equation}
    C_{vh}(\tau)
    =
    \frac{1}{D}\bm v\cdot \bm h(\tau).
\end{equation}
The trace response of this linear system is
\begin{equation}
    R_h(\tau,\tau')
    =
    -\frac{i}{D}
    \bm h(\tau)\cdot \hat{\bm h}(\tau')
\end{equation}
and its Fourier transform gives the resolvent of the Wigner bulk.

Averaging over the GOE disorder gives the usual single-site DMFT equation
\begin{equation}
    \partial_\tau h(\tau)
    =
    u_h(\tau)
    +
    \int d\tau'\,
    R_h(\tau,\tau')h(\tau')
    -
    \gamma_0 C_{vh}(\tau)v
\end{equation}
with Gaussian noise
\begin{equation}
    u_h
    \sim
    \mathcal N(0,C_h).
\end{equation}
In Fourier space
\begin{equation}
    h(\omega)
    =
    \left[
    i\omega
    -
    R_h(\omega)
    \right]^{-1}
    \left(
    u_h(\omega)
    -
    \gamma_0 C_{vh}(\omega)v
    \right).
\end{equation}
The bulk response is therefore the GOE resolvent
\begin{equation}
    R_h(\omega)
    =
    \frac{
    i\omega
    -
    \sqrt{(i\omega)^2-4}
    }{2}
\end{equation}
which captures the continuous Wigner bulk.

To detect outliers, we project the probe dynamics onto the spike direction. The projected correlation $C_{vh}(\omega)$ satisfies
\begin{equation}
    C_{vh}(\omega)
    =
    -
    \gamma_0
    \left[
    i\omega
    -
    R_h(\omega)
    \right]^{-1}
    C_{vh}(\omega).
\end{equation}
Equivalently, allowing for an initial overlap $C_0$,
\begin{equation}
    C_{vh}(\omega)
    =
    \frac{C_0}
    {
    i\omega
    +
    \gamma_0
    -
    R_h(\omega)
    }.
\end{equation}
Thus isolated eigenvalues correspond to poles of the projected resolvent, i.e. to zeros of
\begin{equation}
    i\omega
    +
    \gamma_0
    -
    R_h(\omega)
    =
    0.
\end{equation}
Analytically continuing to the real spectral axis with $i\omega=-\lambda-i0^+$, the outlier condition becomes
\begin{equation}
    -\lambda
    +
    \gamma_0
    =
    \frac{
    -\lambda
    +
    \sqrt{\lambda^2-4}
    }{2}.
\end{equation}
Solving gives the classical spiked-Wigner outlier location
\begin{equation}
    \lambda_{\rm out}
    =
    \gamma_0
    +
    \frac{1}{\gamma_0}.
\end{equation}
This solution lies outside the Wigner bulk only when $\gamma_0>1$, giving the usual BBP threshold. Thus the toy model demonstrates the two ingredients used throughout the paper: the trace response determines the bulk, while the spike-probe correlation reveals outliers through their poles.

\section{Anti-Hebbian Wigner dynamics}\label{app:anti_hebb_GOE}
\begin{figure}[H]
    \centering
    \includegraphics[width=0.85\linewidth]{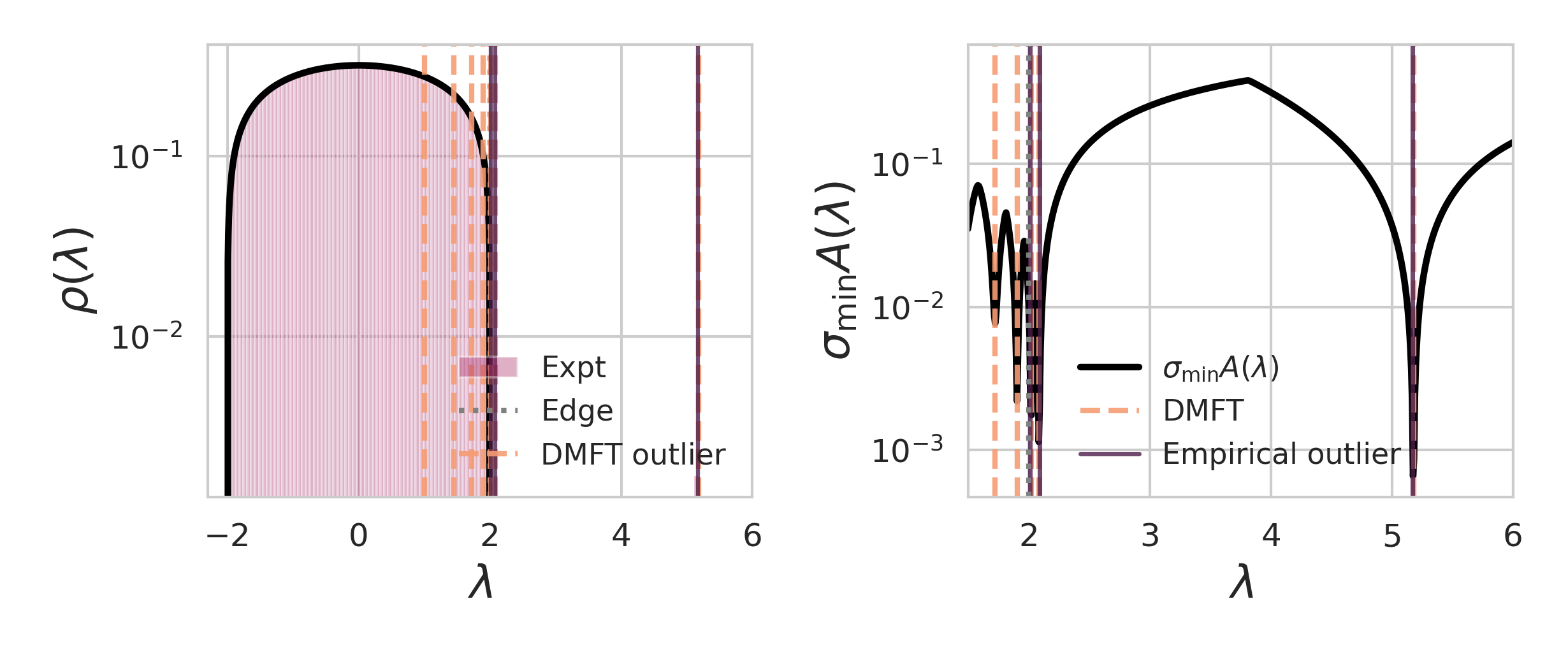}
    \caption{Anti-Hebbian Wigner toy model with dynamically generated spike directions, shown after $T=200$ discretized physical-time steps with step size $\Delta t=0.05$. Left: empirical eigenvalue density of $\bm M(T)$ compared with the GOE semicircle bulk; vertical lines mark the empirical outliers and the DMFT predictions from the projected resolvent condition. Right: minimum singular value of the finite-dimensional matrix $\bm A(\lambda)$, whose zeros give the outlier condition. The dips align with the empirical detached eigenvalues, showing that the two-level DMFT captures outliers generated by spike directions statistically coupled to the random bulk.
}
    \label{fig:placeholder}
\end{figure}
We next consider a toy model in which the spike directions are not planted independently of the random matrix, but are generated dynamically by the random matrix itself. This makes explicit the statistical dependence between the random bulk and the learned spike directions, which is the key difficulty in trained neural networks. Consider $D$ degrees of freedom $v_i (t)$ with $i \in \{D\}$, evolving under a time-dependent symmetric coupling matrix $\bm M(t)$
\begin{equation}
    \partial_t \bm v(t)=-\bm M(t)\bm v(t).
\end{equation}
The coupling matrix $\bm M (t)$ is decomposed into a quenched random part and a plastic part
\begin{equation}
    \bm M(t) = \bm M_0 + \bm A(t)
\end{equation}
where $\bm M_0$ is drawn from the GOE ensemble $\langle(\bm{M}_{0})_{ij}\rangle=0$ and $\langle(\bm{M}_{0})_{ij}(\bm{M}_{0})_{kl}\rangle=\frac{1}{D}(\delta_{ik}\delta_{jl}+\delta_{il}\delta_{jk})$, while the plastic component is generated by the trajectory itself through the local anti-Hebbian rule (similarly to~\cite{PhysRevX.14.021001})
\begin{equation}
     \partial_t \bm A(t)= \frac{\gamma_0}{D}\bm v(t) \bm v(t)^\top.
\end{equation}
At a fixed snapshot of time $t$, we want to characterize the spectrum of $\bm M (t)$. This has two parts: the $\mathcal{O}(D)$-dimensional bulk, which is described by the normalized resolvent $G_t(z) = \frac{1}{D}\text{Tr}(z-\bm M(t))^{-1}$, and a $\text{rank}(t)$ possible isolated eigenvalues generated by the finite-rank deformation. Given $t = \mathcal{O}(1)$, the contribution of these eigenvalues to the normalized trace has weights $\mathcal{O}(D^{-1})$, and disappears from the large $D$ limit of $G_t(z)$. To see the outliers, we must resolve the finite-dimensional subspace spanned by the learned directions $\{\bm v(s):0\leq s\leq t\}$. 

As before, we thus introduce a fictitious \textit{spectral} time $\tau$ and a linear probe $\bm h (\tau)$ evolving under the random matrix
\begin{equation}\label{eq::probe_dyns}
    \partial_\tau \bm h (\tau) = -\bm{M}_{0}\bm{h}(\tau)-\gamma_{0}\int_{0}^{t}dt'C_{vh}(t',\tau)\bm{v}(t') +\bm j (\tau).
\end{equation}
The two objects we wish to track are
\begin{equation}
    C_{vh}(t,\tau) = \frac{1}{D}\bm v (t)\cdot \bm h (\tau), \quad R_h (\tau,\tau') = \frac{1}{D} \text{Tr} \frac{\partial \bm h (\tau)}{\partial \bm j(\tau')^{\top}}
\end{equation}
the cross correlation $C_{vh}(t,\tau)$ between the structured direction $\bm v (t)$ and the linear probe $\bm h (\tau)$, which has the role of a resolvent projected onto the learned subspace, and $R_h(\tau,\tau')$ which is the trace response of the spectral probe, hence the Stieltjes transform of the empirical spectrum. In the large $D$ limit, $R_h(\tau,\tau')$ encodes the bulk, while the poles of $C_{vh} (t,\tau)$ give the outliers. 

We thus end up with a two-level dynamical mean field theory (DMFT) description. The first level describes the training dynamics of $\bm v (t)$ evolving under the physical time $t$
\begin{equation}
    \partial_{t}\bm{v}(t)=-\bm{M}_{0}\bm{v}(t)-\gamma_{0}\int_{0}^{t}C_{v}(t,t')\bm{v}(t')dt
\end{equation}
which generates the spike subspace, while the second level refers to the linear system of Eq.~\eqref{eq::probe_dyns} which probes the spectrum of the random matrix $\bm M (t)$.

\subsection{Average over the quenched disorder}
Computing the average over the GOE $\bm{M}_{0}$ matrix

\begin{equation}
    \begin{split}
        &\Bigg<\exp\Bigg(i\text{Tr}\bm{M}_{0}\Bigg(\int dt\bm{v}(t)\hat{\bm{v}}(t)^{\top}+\int d\tau\bm{h}(\tau)\hat{\bm{h}}(\tau)^{\top}\Bigg)\Bigg>_{\bm{M}_{0}}=\\
        &\exp\Bigg(-\frac{1}{2}\int dtdt'C_{v}(t,t')\hat{\bm{v}}(t)\cdot\hat{\bm{v}}(t')+\frac{D}{2}\int dtdt'R_{v}(t,t')R_{v}(t',t)\Bigg)\\
        &\times\exp\Bigg(-\frac{1}{2}\int d\tau d\tau'C_{h}(\tau,\tau')\hat{\bm{h}}(\tau)\cdot\hat{\bm{h}}(\tau')+\frac{D}{2}\int d\tau d\tau'R_{h}(\tau,\tau')R_{h}(\tau',\tau)\Bigg)\\
        &\times\exp\Bigg(-\int dtd\tau C_{vh}(t,\tau)\hat{\bm{v}}(t)\cdot\hat{\bm{h}}(\tau)+D\int dtd\tau R_{hv}(\tau,t)R_{vh}(t,\tau)\Bigg)
    \end{split}
\end{equation}
with definitions
\begin{align}
    &C_{v}(t,t')	=\frac{1}{D}\bm{v}(t)\cdot\bm{v}(t')\\
    &R_{v}(t,t')	=-\frac{i}{D}\bm{v}(t)\cdot\hat{\bm{v}}(t')\\
    &C_{h}(\tau,\tau')	=\frac{1}{D}\bm{h}(\tau)\cdot\bm{h}(\tau')\\
    &R_{h}(\tau,\tau')	=-\frac{i}{D}\bm{h}(\tau)\cdot\hat{\bm{h}}(\tau')\\
    &C_{vh}(t,\tau)	=\frac{1}{D}\bm{v}(t)\cdot\bm{h}(\tau)\\
    &R_{hv}(\tau,t)	=-\frac{i}{D}\bm{h}(\tau)\cdot\hat{\bm{v}}(t)
\end{align}
that we can enforce through Fourier representation of the Dirac delta function
\begin{align}
    1&\equiv\int\frac{dR_{v}(t,t')d\hat{R}_{v}(t,t')}{2\pi}\exp\Bigg(-D\int dtdt'R_{v}(t,t')\hat{R}_{v}(t,t')-i\int dtdt'\hat{R}_{v}(t,t')\bm{v}(t)\cdot\hat{\bm{v}}(t')\Bigg)\\1&\equiv\int\frac{dR_{h}(\tau,\tau')d\hat{R}_{h}(\tau,\tau')}{2\pi}\exp\Bigg(-D\int d\tau d\tau'R_{h}(\tau,\tau')\hat{R}_{h}(\tau,\tau')-i\int d\tau d\tau'\hat{R}_{h}(\tau,\tau')\bm{h}(\tau)\cdot\hat{\bm{h}}(\tau')\Bigg)\\1&\equiv\int\frac{dR_{hv}(\tau,t)d\hat{R}_{hv}(\tau,t)}{2\pi}\exp\Bigg(-D\int d\tau dtR_{hv}(\tau,t)\hat{R}_{hv}(\tau,t)-i\int d\tau dt\hat{R}_{hv}(\tau,t)\bm{h}(\tau)\cdot\hat{\bm{v}}(t)\Bigg)\\1&\equiv\int\frac{dR_{vh}(t,\tau)d\hat{R}_{vh}(t,\tau)}{2\pi}\exp\Bigg(-D\int dtd\tau R_{vh}(t,\tau)\hat{R}_{vh}(t,\tau)-i\int dtd\tau\hat{R}_{vh}(t,\tau)\bm{v}(t)\cdot\hat{\bm{h}}(\tau)\Bigg).
\end{align}
After averaging over the GOE disorder, its effect is replaced by a Gaussian dynamical noise and by retarded self-interactions. At the saddle point, the physical training trajectory obeys the usual single-site DMFT equation
\begin{equation}
     \partial_{t}v(t)=u_{v}(t)-\gamma_{0}\int_{0}^{t}dt'C_{v}(t',t)v(t')+\int dt'R_{v}(t,t')v(t')
\end{equation}
where $u_v $ is the Gaussian process with covariance $\langle u_v (t) u_v(t')\rangle = C_v (t,t')$. The order parameters $C_v, R_v$ are fixed self-consistently from this training-level DMFT. Once this first DMFT problem has been solved, the spectral problem can be simplified as well as
\begin{equation}
    \partial_{\tau}h=u_{h}(\tau)-\gamma_{0}\int_{0}^{t}dt'C_{vh}(t',\tau)v(t')+\int d\tau'R_{h}(\tau,\tau')h(\tau')+\int dt' R_{hv}(\tau,t')v(t')
\end{equation}
with joint Gaussian noise $\left(\begin{array}{c}
u_{v}\\
u_{h}
\end{array}\right)\sim\mathcal{N}\left(0,\left(\begin{array}{cc}
C_{v} & C_{vh}\\
C_{vh}^{\top} & C_{h}
\end{array}\right)\right)$. Now, the bulk response of the probe is still the GOE resolvent. Therefore in Fourier space
\begin{equation}
    R_h=\frac{1}{2}\Big[i\omega-\sqrt{(i\omega)^{2}-4}\Big]
\end{equation}
with the physical branch fixed by $R_h (\omega) \sim \frac{1}{i\omega}$ at large $|\omega|$. The only spectral information not contained in the normalized trace is carried by the projected correlations
\begin{equation}
    C_{vh}(t,\omega) = \frac{1}{D}\bm v (t)\cdot \bm h (\omega), \quad R_{vh} (t,\omega) =- \frac{i}{D} \bm h (\omega) \cdot \hat{\bm v}(t). 
\end{equation}
Taking the Fourier transform of the probe equation gives
 \begin{equation}
     h(\omega)=\Big[i\omega-R_{h}(\omega)\Big]^{-1}\Big(h_{0}+u_{h}(\omega)+\int_{0}^{t}dt'\Big(R_{hv}(\omega,t')-\gamma_{0}C_{vh}(t',\omega)\Big)v(t')\Big)
 \end{equation}
 from which 
\begin{align}
    C_{vh}(t'',\omega)&=R_{h}(\omega)\Big(\langle v(t'')h_{0}\rangle+\langle v(t'')u_{h}(\omega)\rangle+\int_{0}^{t}dt'\Big(R_{hv}(\omega,t')-\gamma_{0}C_{vh}(t',\omega)\Big)C_{v}(t'',t')\Big)\\
    R_{hv}(\omega,t'')&=R_{h}(\omega)\Big(\int_{t''}^{t}dt'\Big(R_{hv}(\omega,t')-\gamma_{0}C_{vh}(t',\omega)\Big)R_{v}(t',t'')\Big)
\end{align}
given the initial condition $\bm c_0 =\frac{1}{D}\bm v (t'') \cdot \bm h_0 $ and
\begin{align}
    \langle v(t'')u_{h}(\omega)\rangle=\int_{0}^{t''}dsC_{vh}(s,\omega)R_{v}(t'',s).
\end{align}
After discretizing the physical time $t$, these two equations become a closed $2T \times 2T$ system
\begin{align}
    \underbrace{\left(\begin{array}{cc}
\bm{I}-R_{h}(\omega)(\bm{R}_{v}\odot\bm{\Theta})+\gamma_{0}\Delta tR_{h}(\omega)\bm{C}_{v} & -\Delta tR_{h}(\omega)\bm{C}_{v}\\
\gamma_{0}R_{h}(\omega)(\bm{R}_{v}\odot\bm{\Theta})^{\top} & \bm{I}-R_{h}(\omega)(\bm{R}_{v}\odot\bm{\Theta})^{\top}
\end{array}\right)}_{\mathcal{A}(\omega)}\left(\begin{array}{c}
\bm{C}_{vh}(\omega)\\
\bm{R}_{hv}(\omega)
\end{array}\right)=\left(\begin{array}{c}
R_{h}(\omega)\bm{c}_{0}\\
0
\end{array}\right).
\end{align}
Here $\odot$ denotes elementwise multiplication. The outliers correspond to poles of the projected resolvent, equivalently to the zeros of
\begin{equation}
    \det \bm{\mathcal A}(\omega)
    =
    0.
\end{equation}

This toy model makes the role of the two-level construction explicit: the first DMFT level generates the statistically dependent spike directions, while the second level probes the spectrum of the resulting random matrix. The classical BBP pole condition of Appendix~\ref{appendix:classical_bbp} is recovered when the spike direction is fixed.

\section{Derivation of Outlier Dynamics for Singular Values of Matrices with Statistically Dependent Spikes}

\paragraph{Coupled Dynamics between System and Probe} We consider an evolution equation for a probe vector $\bm\psi_0(\tau)$ as $\partial_{\tau} \bm\psi(\tau) = \bm M(t) \bm \psi(\tau)$. We decompose these dynamics into two components that are each linear in $\bm W(0)$ 
\begin{align}
\textbf{Probe Dynamics}: \quad
 &\bm\psi_1(\tau) = \frac{1}{\sqrt{N_0}} \bm W(0) \bm\psi(\tau) + \sum_{s} c_0(s) \bm g(s)
    \\
    &\partial_{\tau} \bm\psi_0(\tau) = \frac{\sqrt{N_0}}{N_1} \bm W(0)^\top \bm \psi_1(\tau) + \sum_{s} c_1(s) \bm \phi(s) 
\end{align}
where we introduced the overlaps
\begin{align}
    \bm c_0(\tau, s) \equiv \frac{1}{N_0} \bm\psi_0(\tau) \cdot \bm\phi(s) \ , \ \bm c_1(\tau, s) \equiv \frac{1}{N_1} \bm\psi_1(\tau) \cdot \bm g(s) 
\end{align}
These equations are combined with the defining equations for $\bm\chi(t)$ and $\bm\xi(t)$
\begin{align}
    \textbf{System Dynamics}: \quad &\bm \chi(t) \equiv \frac{1}{\sqrt{N_0}} \bm W(0) \bm \phi(t) \ , \ \bm\xi(t) \equiv \frac{\sqrt{N_0}}{N_1} \bm W(0)^\top \bm g(t) 
    \\
    &\bm\phi(t) = \phi_t( \{ \bm\chi(s) , \bm\xi(s) \} )  \ , \ \bm g(t) = g_t( \{ \bm\chi(s) , \bm\xi(s) \} ) 
\end{align}
where $\phi_t$ and $g_t$ act elementwise over the entries of the vectors. We assume that the entries of $\bm W(0)$ are iid random variables with variance $\sigma^2$. For concreteness, we can assume Gaussian entries
\begin{align}
    W_{ij}(0) \sim \mathcal N(0, \sigma^2)  .
\end{align}
This condition defines a distribution over the random vectors $\bm\psi_0, \bm\psi_1, \bm\chi, \bm\xi$. We now aim to characterize this distribution using a formal path integral approach. 

\paragraph{Martin--Siggia--Rose path integral.}
We set \(N=N_0\) and \(\alpha=N_1/N_0\). The probe and system variables obey
\begin{align}
    \bm\psi_1(\tau)
    &=
    \frac{1}{\sqrt N}\bm W(0)\bm\psi_0(\tau)
    +
    \sum_s c_0(\tau,s)\bm g(s),
    \label{eq:probe-psi1-constraint}
    \\
    \partial_\tau \bm\psi_0(\tau)
    &=
    \frac{\sqrt N}{N_1}\bm W(0)^\top \bm\psi_1(\tau)
    +
    \sum_s c_1(\tau,s)\bm\phi(s)
    +
    \bm j_\psi(\tau),
    \label{eq:probe-psi0-constraint}
    \\
    \bm\chi(t)
    &=
    \frac{1}{\sqrt N}\bm W(0)\bm\phi(t),
    \label{eq:chi-constraint}
    \\
    \bm\xi(t)
    &=
    \frac{\sqrt N}{N_1}\bm W(0)^\top \bm g(t).
    \label{eq:xi-constraint}
\end{align}
Here
\[
    c_0(\tau,s)
    =
    \frac{1}{N}\bm\psi_0(\tau)\cdot\bm\phi(s),
    \qquad
    c_1(\tau,s)
    =
    \frac{1}{N_1}\bm\psi_1(\tau)\cdot\bm g(s).
\]
The MSR generating functional is
\begin{align}
    Z[\bm j_\psi,\bm j_\chi,\bm j_\xi]
    =
    \left\langle
    \exp\left(
    \int d\tau\, \bm j_\psi(\tau)\cdot\bm\psi_0(\tau)
    +
    \sum_t \bm j_\chi(t)\cdot\bm\chi(t)
    +
    \sum_t \bm j_\xi(t)\cdot\bm\xi(t)
    \right)
    \right\rangle_{\bm W(0)}.
\end{align}
Equivalently, imposing
\eqref{eq:probe-psi1-constraint}--\eqref{eq:xi-constraint} with MSR response fields gives
\begin{align}
    Z
    =
    \int
    &\mathcal D\bm\psi_0\mathcal D\bm\psi_1
    \mathcal D\bm\chi\mathcal D\bm\xi
    \mathcal D\hat{\bm\psi}_0\mathcal D\hat{\bm\psi}_1
    \mathcal D\hat{\bm\chi}\mathcal D\hat{\bm\xi}\,
    \exp\{S_{\rm MSR}\},
\end{align}
where
\begin{align}
    S_{\rm MSR}
    =
    &i\int d\tau\,\hat{\bm\psi}_1(\tau)\cdot
    \left[
    \bm\psi_1(\tau)-\sum_s c_0(\tau,s)\bm g(s)
    \right]
    \\
    &+
    i\int d\tau\,\hat{\bm\psi}_0(\tau)\cdot
    \left[
    \partial_\tau\bm\psi_0(\tau)
    -
    \sum_s c_1(\tau,s)\bm\phi(s)
    -
    \bm j_\psi(\tau)
    \right]
    \nonumber\\
    +
    i\sum_t\hat{\bm\chi}(t)\cdot\bm\chi(t)
    &+
    i\sum_t\hat{\bm\xi}(t)\cdot\bm\xi(t)
    -
    i\sum_{ij}W_{ij}(0)\mathcal X_{ij}.
\end{align}
The source conjugate to \(W_{ij}(0)\) is
\begin{align}
    \mathcal X_{ij}
    =
    \frac{1}{\sqrt N}
    \left[
    \int d\tau\,\hat\psi_{1,i}(\tau)\psi_{0,j}(\tau)
    +
    \sum_t \hat\chi_i(t)\phi_j(t)
    \right]
    +
    \frac{\sqrt N}{N_1}
    \left[
    \int d\tau\,\psi_{1,i}(\tau)\hat\psi_{0,j}(\tau)
    +
    \sum_t g_i(t)\hat\xi_j(t)
    \right].
    \label{eq:Xij-source}
\end{align}

\paragraph{Averaging over the initialization.}
Since \(W_{ij}(0)\sim\mathcal N(0,\sigma^2)\),
\begin{align}
    \left\langle
    \exp\left(
    -i\sum_{ij}W_{ij}(0)\mathcal X_{ij}
    \right)
    \right\rangle_{\bm W(0)}
    =
    \exp\left(
    -\frac{\sigma^2}{2}\sum_{ij}\mathcal X_{ij}^2
    \right).
\end{align}
Thus
\[
    Z
    =
    \int \mathcal D(\cdots)
    \exp\{S_0-G\},
    \qquad
    G
    =
    \frac{\sigma^2}{2}\sum_{ij}\mathcal X_{ij}^2.
\]
Expanding \(G\) generates all quadratic contractions between the row-side variables
\[
    \hat{\bm\psi}_1,\quad \bm\psi_1,\quad \hat{\bm\chi},\quad \bm g
\]
and the column-side variables
\[
    \bm\psi_0,\quad \hat{\bm\psi}_0,\quad \bm\phi,\quad \hat{\bm\xi}.
\]
Explicitly,
\begin{align}
    G
    =
    \frac{\sigma^2}{2}\sum_{ij}
    \Bigg\{
    &
    \frac{1}{N}
    \int d\tau d\tau'\,
    \hat\psi_{1,i}(\tau)\hat\psi_{1,i}(\tau')
    \psi_{0,j}(\tau)\psi_{0,j}(\tau')
    \nonumber\\
    &+
    \frac{1}{\alpha^2N}
    \int d\tau d\tau'\,
    \psi_{1,i}(\tau)\psi_{1,i}(\tau')
    \hat\psi_{0,j}(\tau)\hat\psi_{0,j}(\tau')
    \nonumber\\
    &+
    \frac{1}{N}
    \sum_{t,s}
    \hat\chi_i(t)\hat\chi_i(s)
    \phi_j(t)\phi_j(s)
    \nonumber\\
    &+
    \frac{1}{\alpha^2N}
    \sum_{t,s}
    g_i(t)g_i(s)
    \hat\xi_j(t)\hat\xi_j(s)
    \nonumber\\
    &+
    \frac{2}{\alpha N}
    \int d\tau d\tau'\,
    \hat\psi_{1,i}(\tau)\psi_{1,i}(\tau')
    \psi_{0,j}(\tau)\hat\psi_{0,j}(\tau')
    \nonumber\\
    &+
    \frac{2}{N}
    \int d\tau\sum_t
    \hat\psi_{1,i}(\tau)\hat\chi_i(t)
    \psi_{0,j}(\tau)\phi_j(t)
    \nonumber\\
    &+
    \frac{2}{\alpha N}
    \int d\tau\sum_t
    \hat\psi_{1,i}(\tau)g_i(t)
    \psi_{0,j}(\tau)\hat\xi_j(t)
    \nonumber\\
    &+
    \frac{2}{\alpha N}
    \int d\tau\sum_t
    \psi_{1,i}(\tau)\hat\chi_i(t)
    \hat\psi_{0,j}(\tau)\phi_j(t)
    \nonumber\\
    &+
    \frac{2}{\alpha^2N}
    \int d\tau\sum_t
    \psi_{1,i}(\tau)g_i(t)
    \hat\psi_{0,j}(\tau)\hat\xi_j(t)
    \nonumber\\
    &+
    \frac{2}{\alpha N}
    \sum_{t,s}
    \hat\chi_i(t)g_i(s)
    \phi_j(t)\hat\xi_j(s)
    \Bigg\}.
    \label{eq:G-expanded-full}
\end{align}

\paragraph{Order parameters.}
The contractions in \eqref{eq:G-expanded-full} are expressed in terms of the following
macroscopic order parameters. For the probe,
\begin{align}
    C^{\psi_0}(\tau,\tau')
    &=
    \frac{1}{N}
    \bm\psi_0(\tau)\cdot\bm\psi_0(\tau'),
    &
    C^{\psi_1}(\tau,\tau')
    &=
    \frac{1}{N_1}
    \bm\psi_1(\tau)\cdot\bm\psi_1(\tau'),
    \\
    R^{\psi_0}(\tau,\tau')
    &=
    -\frac{i}{N}
    \bm\psi_0(\tau)\cdot\hat{\bm\psi}_0(\tau'),
    &
    R^{\psi_1}(\tau,\tau')
    &=
    -\frac{i}{N_1}
    \bm\psi_1(\tau)\cdot\hat{\bm\psi}_1(\tau').
\end{align}
For the training-level system,
\begin{align}
    C^\phi(t,s)
    &=
    \frac{1}{N}
    \bm\phi(t)\cdot\bm\phi(s),
    &
    C^g(t,s)
    &=
    \frac{1}{N_1}
    \bm g(t)\cdot\bm g(s),
    \\
    R^\phi(t,s)
    &=
    -\frac{i}{N}
    \bm\phi(t)\cdot\hat{\bm\xi}(s),
    &
    R^g(t,s)
    &=
    -\frac{i}{N_1}
    \bm g(t)\cdot\hat{\bm\chi}(s).
\end{align}
For the mixed probe--system quantities,
\begin{align}
    c_0(\tau,t)
    &=
    \frac{1}{N}
    \bm\psi_0(\tau)\cdot\bm\phi(t),
    &
    c_1(\tau,t)
    &=
    \frac{1}{N_1}
    \bm\psi_1(\tau)\cdot\bm g(t),
    \\
    \rho_0(\tau,t)
    &=
    -\frac{i}{N}
    \bm\psi_0(\tau)\cdot\hat{\bm\xi}(t),
    &
    \rho_1(\tau,t)
    &=
    -\frac{i}{N_1}
    \bm\psi_1(\tau)\cdot\hat{\bm\chi}(t).
\end{align}

\paragraph{Conjugate order parameters.}
Each order parameter is enforced by a Fourier representation of a delta function.
For example,
\begin{align}
    1
    =
    \int dC^{\psi_0}d\widehat C^{\psi_0}
    \exp\left\{
    -N
    \int d\tau d\tau'\,
    \widehat C^{\psi_0}(\tau,\tau')
    \left[
    C^{\psi_0}(\tau,\tau')
    -
    \frac{1}{N}
    \bm\psi_0(\tau)\cdot\bm\psi_0(\tau')
    \right]
    \right\},
\end{align}
and
\begin{align}
    1
    =
    \int dR^{\psi_0}d\widehat R^{\psi_0}
    \exp\left\{
    -N
    \int d\tau d\tau'\,
    \widehat R^{\psi_0}(\tau,\tau')
    \left[
    R^{\psi_0}(\tau,\tau')
    +
    \frac{i}{N}
    \bm\psi_0(\tau)\cdot\hat{\bm\psi}_0(\tau')
    \right]
    \right\}.
\end{align}
Similarly, we insert identities for
\[
    C^{\psi_1},C^\phi,C^g,
    R^{\psi_1},R^\phi,R^g,
    c_0,c_1,\rho_0,\rho_1.
\]
We collect all order parameters and conjugate fields into
\[
    Q
    =
    \left\{
    C,R,c,\rho,
    \widehat C,\widehat R,\widehat c,\widehat\rho
    \right\}.
\]

\paragraph{Large-\(N\) effective action.}
After inserting the order-parameter identities, the microscopic variables factorize over
row and column indices. Therefore the generating functional has the large-deviation form
\begin{align}
    Z
    =
    \int dQ\,
    \exp\left\{
    -N S_{\rm eff}[Q]
    \right\},
    \label{eq:Z-large-dev-Q}
\end{align}
where
\begin{align}
    S_{\rm eff}[Q]
    =
    S_{\rm op}[Q]
    -
    \log Z_0[Q]
    -
    \alpha \log Z_1[Q].
    \label{eq:Seff-factorized}
\end{align}
Here \(Z_0[Q]\) is the single-site path integral for a representative column-side site
\((\psi_0,\phi,\xi)\), \(Z_1[Q]\) is the single-site path integral for a representative
row-side site \((\psi_1,g,\chi)\), and \(S_{\rm op}\) contains the order-parameter and
conjugate-order-parameter terms.

In the proportional limit, \(N,N_1\to\infty\) with \(N_1/N=\alpha\), the integral
\eqref{eq:Z-large-dev-Q} is evaluated by saddle point:
\[
    \frac{\delta S_{\rm eff}}{\delta Q}=0.
\]
Variation with respect to conjugate order parameters recovers the definitions of
\(C,R,c,\rho\). Variation with respect to the order parameters fixes the conjugates,
or equivalently the covariances and retarded self-interactions of the effective
single-site processes.

\paragraph{Saddle point in \((\tau,t)\)-space.}
At the saddle point, the disorder-averaged theory is equivalent to the following
self-consistent stochastic process:
\begin{align}
    \partial_\tau \psi_0(\tau)
    &=
    u^{\psi_0}(\tau)
    +
    \sigma^2
    \int d\tau'\,
    R^{\psi_1}(\tau,\tau')\psi_0(\tau')
    +
    \sigma^2
    \sum_s \rho_1(\tau,s)\phi(s)
    +
    \sum_s c_1(\tau,s)\phi(s)
    +
    j_\psi(\tau),
    \label{eq:saddle-psi0-tau}
    \\
    \psi_1(\tau)
    &=
    u^{\psi_1}(\tau)
    +
    \frac{\sigma^2}{\alpha}
    \int d\tau'\,
    R^{\psi_0}(\tau,\tau')\psi_1(\tau')
    +
    \frac{\sigma^2}{\alpha}
    \sum_s \rho_0(\tau,s)g(s)
    +
    \sum_s c_0(\tau,s)g(s),
    \label{eq:saddle-psi1-tau}
    \\
    \chi(t)
    &=
    u^\chi(t)
    +
    \frac{\sigma^2}{\alpha}
    \sum_s R^\phi(t,s)g(s),
    \label{eq:saddle-chi-t}
    \\
    \xi(t)
    &=
    u^\xi(t)
    +
    \sigma^2
    \sum_s R^g(t,s)\phi(s).
    \label{eq:saddle-xi-t}
\end{align}
The functions \(\phi(t)\) and \(g(t)\) are then obtained from the elementwise structural
relations
\[
    \phi(t)=\phi_t(\{\chi(s),\xi(s)\}_{s<T}),
    \qquad
    g(t)=g_t(\{\chi(s),\xi(s)\}_{s<T}).
\]
The Gaussian noises in
\eqref{eq:saddle-psi0-tau}--\eqref{eq:saddle-xi-t} have zero mean and covariances
\begin{align}
    \left\langle u^{\psi_0}(\tau)u^{\psi_0}(\tau')\right\rangle
    &=
    \frac{\sigma^2}{\alpha}C^{\psi_1}(\tau,\tau'),
    &
    \left\langle u^{\psi_1}(\tau)u^{\psi_1}(\tau')\right\rangle
    &=
    \sigma^2 C^{\psi_0}(\tau,\tau'),
    \\
    \left\langle u^\chi(t)u^\chi(s)\right\rangle
    &=
    \sigma^2 C^\phi(t,s),
    &
    \left\langle u^\xi(t)u^\xi(s)\right\rangle
    &=
    \frac{\sigma^2}{\alpha}C^g(t,s),
    \\
    \left\langle u^{\psi_0}(\tau)u^\xi(s)\right\rangle
    &=
    \frac{\sigma^2}{\alpha}c_1(\tau,s),
    &
    \left\langle u^{\psi_1}(\tau)u^\chi(s)\right\rangle
    &=
    \sigma^2 c_0(\tau,s).
\end{align}
The saddle-point values of the order parameters are deterministic and satisfy
\begin{align}
    C^{\psi_0}(\tau,\tau')
    &=
    \left\langle\psi_0(\tau)\psi_0(\tau')\right\rangle,
    &
    C^{\psi_1}(\tau,\tau')
    &=
    \left\langle\psi_1(\tau)\psi_1(\tau')\right\rangle,
    \\
    R^{\psi_0}(\tau,\tau')
    &=
    \left\langle
    \frac{\delta\psi_0(\tau)}
    {\delta u^{\psi_0}(\tau')}
    \right\rangle,
    &
    R^{\psi_1}(\tau,\tau')
    &=
    \left\langle
    \frac{\delta\psi_1(\tau)}
    {\delta u^{\psi_1}(\tau')}
    \right\rangle,
    \\
    C^\phi(t,s)
    &=
    \left\langle\phi(t)\phi(s)\right\rangle,
    &
    C^g(t,s)
    &=
    \left\langle g(t)g(s)\right\rangle,
    \\
    R^\phi(t,s)
    &=
    \left\langle
    \frac{\partial\phi(t)}
    {\partial u^\xi(s)}
    \right\rangle,
    &
    R^g(t,s)
    &=
    \left\langle
    \frac{\partial g(t)}
    {\partial u^\chi(s)}
    \right\rangle,
    \\
    c_0(\tau,t)
    &=
    \left\langle\psi_0(\tau)\phi(t)\right\rangle,
    &
    c_1(\tau,t)
    &=
    \left\langle\psi_1(\tau)g(t)\right\rangle,
    \\
    \rho_0(\tau,t)
    &=
    \left\langle
    \frac{\partial\psi_0(\tau)}
    {\partial u^\xi(t)}
    \right\rangle,
    &
    \rho_1(\tau,t)
    &=
    \left\langle
    \frac{\partial\psi_1(\tau)}
    {\partial u^\chi(t)}
    \right\rangle.
\end{align}

\paragraph{Projected saddle-point equations in \((\tau,t)\)-space.}
We now derive the closed equations for the mixed order parameters before taking any
Laplace transform. Multiplying \eqref{eq:saddle-psi0-tau} by \(\phi(t)\), averaging,
and using Stein's lemma gives
\begin{align}
    \partial_\tau c_0(\tau,t)
    &=
    \sigma^2
    \int d\tau'\,
    R^{\psi_1}(\tau,\tau')c_0(\tau',t)
    +
    \frac{\sigma^2}{\alpha}
    \sum_s R^\phi(t,s)c_1(\tau,s)
    \nonumber\\
    &\hspace{2cm}
    +
    \sigma^2
    \sum_s C^\phi(t,s)\rho_1(\tau,s)
    +
    \sum_s C^\phi(t,s)c_1(\tau,s).
    \label{eq:c0-saddle-tau}
\end{align}
Here we used
\[
    \left\langle u^{\psi_0}(\tau)\phi(t)\right\rangle
    =
    \frac{\sigma^2}{\alpha}
    \sum_s R^\phi(t,s)c_1(\tau,s).
\]
Differentiating \eqref{eq:saddle-psi0-tau} with respect to \(u^\xi(t)\) gives
\begin{align}
    \partial_\tau \rho_0(\tau,t)
    &=
    \sigma^2
    \int d\tau'\,
    R^{\psi_1}(\tau,\tau')\rho_0(\tau',t)
    +
    \sigma^2
    \sum_s R^\phi(s,t)\rho_1(\tau,s)
    +
    \sum_s R^\phi(s,t)c_1(\tau,s).
    \label{eq:rho0-saddle-tau}
\end{align}
Similarly, multiplying \eqref{eq:saddle-psi1-tau} by \(g(t)\), averaging, and using
\[
    \left\langle u^{\psi_1}(\tau)g(t)\right\rangle
    =
    \sigma^2
    \sum_s R^g(t,s)c_0(\tau,s),
\]
gives
\begin{align}
    c_1(\tau,t)
    &=
    \frac{\sigma^2}{\alpha}
    \int d\tau'\,
    R^{\psi_0}(\tau,\tau')c_1(\tau',t)
    +
    \sigma^2
    \sum_s R^g(t,s)c_0(\tau,s)
    \nonumber\\
    &\hspace{2cm}
    +
    \frac{\sigma^2}{\alpha}
    \sum_s C^g(t,s)\rho_0(\tau,s)
    +
    \sum_s C^g(t,s)c_0(\tau,s).
    \label{eq:c1-saddle-tau}
\end{align}
Finally, differentiating \eqref{eq:saddle-psi1-tau} with respect to \(u^\chi(t)\) gives
\begin{align}
    \rho_1(\tau,t)
    &=
    \frac{\sigma^2}{\alpha}
    \int d\tau'\,
    R^{\psi_0}(\tau,\tau')\rho_1(\tau',t)
    +
    \frac{\sigma^2}{\alpha}
    \sum_s R^g(s,t)\rho_0(\tau,s)
    +
    \sum_s R^g(s,t)c_0(\tau,s).
    \label{eq:rho1-saddle-tau}
\end{align}

\paragraph{Laplace transform in the auxiliary spectral time.}
Only after obtaining the saddle-point equations
\eqref{eq:c0-saddle-tau}--\eqref{eq:rho1-saddle-tau} do we Laplace transform in
\(\tau\). For any probe-time function \(f(\tau)\), define
\[
    f(z)
    =
    \int_0^\infty d\tau\,e^{-z\tau}f(\tau),
    \qquad
    \Re z>0.
\]
Boundary terms from the Laplace transform encode the initial probe condition and do not affect
the pole condition for outliers, so we suppress them below.

The pure probe responses obey, in \((\tau,\tau')\)-space,
\begin{align}
    \left[
    \partial_\tau
    -
    \sigma^2 R^{\psi_1}
    \right]
    \circ R^{\psi_0}
    =
    \delta,
    \qquad
    \left[
    I
    -
    \frac{\sigma^2}{\alpha}R^{\psi_0}
    \right]
    \circ R^{\psi_1}
    =
    \delta,
\end{align}
where \(\circ\) denotes convolution in the auxiliary time. After Laplace transformation,
\begin{align}
    R^{\psi_0}(z)
    &=
    \frac{1}{z-\sigma^2R^{\psi_1}(z)},
    \\
    R^{\psi_1}(z)
    &=
    \frac{1}{1-\frac{\sigma^2}{\alpha}R^{\psi_0}(z)}.
\end{align}
We identify
\[
    \mathcal G(z)
    :=
    R^{\psi_0}(z),
\]
the Stieltjes transform of the Marchenko--Pastur bulk. Hence
\[
    z-\sigma^2R^{\psi_1}(z)
    =
    \mathcal G(z)^{-1},
    \qquad
    1-\frac{\sigma^2}{\alpha}\mathcal G(z)
    =
    R^{\psi_1}(z)^{-1}.
\]

Laplace transforming
\eqref{eq:c0-saddle-tau}--\eqref{eq:rho1-saddle-tau} gives
\begin{align}
    \mathcal G(z)^{-1}c_0(t)
    &=
    \frac{\sigma^2}{\alpha}
    \sum_s R^\phi(t,s)c_1(s)
    +
    \sigma^2
    \sum_s C^\phi(t,s)\rho_1(s)
    +
    \sum_s C^\phi(t,s)c_1(s),
    \label{eq:c0-z-final}
    \\
    \mathcal G(z)^{-1}\rho_0(t)
    &=
    \sigma^2
    \sum_s R^\phi(s,t)\rho_1(s)
    +
    \sum_s R^\phi(s,t)c_1(s),
    \label{eq:rho0-z-final}
    \\
    \left(
    1-\frac{\sigma^2}{\alpha}\mathcal G(z)
    \right)c_1(t)
    &=
    \sigma^2
    \sum_s R^g(t,s)c_0(s)
    +
    \frac{\sigma^2}{\alpha}
    \sum_s C^g(t,s)\rho_0(s)
    +
    \sum_s C^g(t,s)c_0(s),
    \label{eq:c1-z-final}
    \\
    \left(
    1-\frac{\sigma^2}{\alpha}\mathcal G(z)
    \right)\rho_1(t)
    &=
    \frac{\sigma^2}{\alpha}
    \sum_s R^g(s,t)\rho_0(s)
    +
    \sum_s R^g(s,t)c_0(s).
    \label{eq:rho1-z-final}
\end{align}

\paragraph{Outlier determinant condition.}
Collecting the Laplace-domain mixed order parameters into
\[
    \bm q(z)
    =
    \begin{pmatrix}
        \bm c_0(z)\\
        \bm\rho_0(z)\\
        \bm c_1(z)\\
        \bm\rho_1(z)
    \end{pmatrix},
\]
equations
\eqref{eq:c0-z-final}--\eqref{eq:rho1-z-final} can be written as
\[
    \bm A(z)\bm q(z)= \bm q_0,
\]
where $\bm q_0$ is an initial condition and $\bm A$ has the form
\[
    \bm A(z)
    =
    \begin{bmatrix}
        \mathcal G(z)^{-1}\bm I
        & \bm 0
        & -\left(\sigma^2\alpha^{-1}\bm R^\phi+\bm C^\phi\right)
        & -\sigma^2\bm C^\phi
        \\
        \bm 0
        & \mathcal G(z)^{-1}\bm I
        & -(\bm R^\phi)^\top
        & -\sigma^2(\bm R^\phi)^\top
        \\
        -(\sigma^2\bm R^g+\bm C^g)
        & -\sigma^2\alpha^{-1}\bm C^g
        & \left(1-\frac{\sigma^2}{\alpha}\mathcal G(z)\right)\bm I
        & \bm 0
        \\
        -(\bm R^g)^\top
        & -\sigma^2\alpha^{-1}(\bm R^g)^\top
        & \bm 0
        & \left(1-\frac{\sigma^2}{\alpha}\mathcal G(z)\right)\bm I
    \end{bmatrix}.
\]
Nontrivial projected resolvents exist when
\[
    \det \bm A(z)=0.
\]
The isolated outlier eigenvalues are the solutions outside the Marchenko--Pastur support.

\section{Nonlinear Networks}\label{app:nonlinear_inner_dmft_eqn}

In this section, we summarize the results of Bordelon \& Pehlevan '22 \cite{bordelon2022self} which provides the evolution of hidden features and outputs of randomly initialized neural networks. 

\paragraph{Setting and Super-Wide Scaling Assumptions} We consider models in mean field parameterization (output scaling $\frac{1}{N\gamma}$ on the last layer) and consider SGD training dynamics from random initialization for all parameters $\bm \theta = \{ \bm W^\ell \}$. Let the hidden widths be $N$ and let depth $L$ be fixed but arbitrary. We consider SGD dynamics with an arbitrary mini-batch stream $\mathcal B_t$
\begin{align}
    \bm\theta(t+1) = \bm \theta(t) - \eta \gamma^2 N \ \mathbb{E}_{\bm x \in \mathcal B_t} \nabla_{\bm \theta} \mathcal L(\bm x, \bm\theta) 
\end{align}
These dynamics translate into each hidden weight matrix as
\begin{align}
    \bm W^\ell(t) = \bm W^\ell(0) + \frac{\eta \gamma}{ \sqrt N} \sum_{s<t} \mathbb{E}_{\bm x_\mu \in \mathcal B_s } \ \Delta_\mu(s) \bm g_\mu(s) \bm \phi_\mu(s)^\top 
\end{align}
where $\bm g^\ell_\mu(t) \equiv N \gamma \frac{\partial f_\mu(t)}{\partial \h^\ell_\mu(t)}$ and $\Delta_\mu(t) = - \frac{\partial}{\partial f_\mu(t)} \mathcal L(f_\mu(t))$. This leads to the following dynamics for variables $\bm h^{\ell}_\mu(t)$ and $\bm z^\ell_\mu(t)$
\begin{align}
    &\bm h^\ell_\mu(t) = \frac{1}{\sqrt N} \bm W^\ell(t) \bm\phi_\mu^{\ell-1}(t) = \bm\chi^\ell_\mu(t) + \eta\gamma \sum_{s<t} \mathbb{E}_{\bm x_\nu \in \mathcal B_s} \Delta_\nu(s)  C^{\phi^{\ell-1}}_{\mu\nu}(t,s) \bm g^\ell_\nu(s) \nonumber
    \\
    &\bm z^\ell_\mu(t) = \frac{1}{\sqrt N} \bm W^\ell(t)^\top \bm g_\mu^{\ell+1}(t) = \bm\xi^\ell_\mu(t) + \eta\gamma \sum_{s<t} \mathbb{E}_{\bm x_\nu \in \mathcal B_s} \Delta_\nu(s)  C^{g^{\ell+1}}_{\mu\nu}(t,s) \bm \phi_\mu^\ell(t)\nonumber \nonumber
    \\
    &\bm\phi_\mu^\ell(t) = \phi( \h^\ell_\mu(t))  \ , \ \bm g^\ell_\mu(t) = \bm z^\ell_\mu(t) \odot \dot\phi(\h^\ell_\mu(t)) 
\end{align}
where we introduced the variables $\bm\chi$ and $\bm\xi$ that depend on the random matrix 
\begin{align}
    \bm \chi^\ell_\mu(t) = \frac{1}{\sqrt N} \bm W^\ell(0) \bm\phi^{\ell-1}_\mu(t) \ , \ \bm \xi^\ell_\mu(t) = \frac{1}{\sqrt N} \bm W^\ell(0)^\top \bm g^{\ell+1}_\mu(t) .
\end{align}
The main claim to utilize our theory is that this dynamical system satisfies the requirements of the SVD result. We now will argue that this holds under the following assumptions
\begin{enumerate}
    \item \textbf{Random Init}: The weights $W^\ell_{ij}(0)$ are initialized with iid random entries. 
    \item \textbf{Super-wide Scaling}: The network width $N \to \infty$ with batch size $B$, number of update steps $T$, and input-dimension $D$ are held fixed.
\end{enumerate}
Under these assumptions, one can show using DMFT \cite{bordelon2022self} that the feature structural condition holds since we have
\begin{enumerate}
    \item \textbf{Concentration of Correlations and Errors}: The quantities $C^\phi$ and $C^g$ and $\Delta$ concentrate due to a law of large numbers effect (they are averages over neurons in each layer).
    \item \textbf{Decoupling of Neurons}: The neurons effectively decouple in their dynamics and become iid random processes throughout training. 
\end{enumerate}
As a consequence, we can indeed view the single-site equations as defining the features $\bm\phi$ and $\bm g$ elementwise in terms of $\bm \chi$ and $\bm\xi$
\begin{align}
    \bm\phi_\mu^\ell(t) = \phi_{\mu,t}^\ell\left( \{ \bm \chi^{\ell}, \bm \xi^\ell \} \right) \ , \ \bm g_\mu^{\ell+1}(t) = g_{\mu,t}^{\ell+1}\left( \{ \bm \chi^{\ell+1}, \bm \xi^{\ell+1}\} \right) ,
\end{align}
which verifies that our structural assumptions for Result 1 are satisfied. 

\paragraph{Connection to Result 1}
For each hidden layer \(\ell\), the trained weight matrix has the form
\[
    \bm W^\ell(T)
    =
    \bm W^\ell(0)
    +
    \frac{\eta\gamma}{\sqrt N}
    \sum_{s<T}
    \mathbb E_{\bm x_\nu\in\mathcal B_s}
    \bar\Delta_\nu(s)\,
    \bm g_\nu^{\ell+1}(s)
    \bm\phi_\nu^\ell(s)^\top .
\]
This is precisely a finite-rank deformation of the initialization. Moreover, the
spike vectors
\[
    \bm\phi_\nu^\ell(s),
    \qquad
    \bm g_\nu^{\ell+1}(s)
\]
have asymptotically iid entries whose single-site values are deterministic causal
functions of the Gaussian fields \((\chi,\xi)\) generated by the same initial matrix
\(\bm W^\ell(0)\). Hence the feature structural condition required by
Result~\ref{res:sing_value_thm} is satisfied. The matrices
\[
    \bm C^\phi,\quad \bm C^g,\quad \bm R^\phi,\quad \bm R^g
\]
appearing in the outlier determinant are the corresponding single-site correlations
and responses:
\[
    C^\phi_{ab}
    =
    \left\langle \phi_a\phi_b\right\rangle,
    \qquad
    C^g_{ab}
    =
    \left\langle g_a g_b\right\rangle,
\]
and
\[
    R^\phi_{ab}
    =
    \left\langle
    \frac{\partial \phi_a}{\partial \xi_b}
    \right\rangle,
    \qquad
    R^g_{ab}
    =
    \left\langle
    \frac{\partial g_a}{\partial \chi_b}
    \right\rangle,
\]
where the combined spike index \(a=(s,\nu)\) ranges over all training times and
mini-batch examples contributing to the finite-rank update.

\begin{figure}
    \centering
    \includegraphics[width=0.45\linewidth]{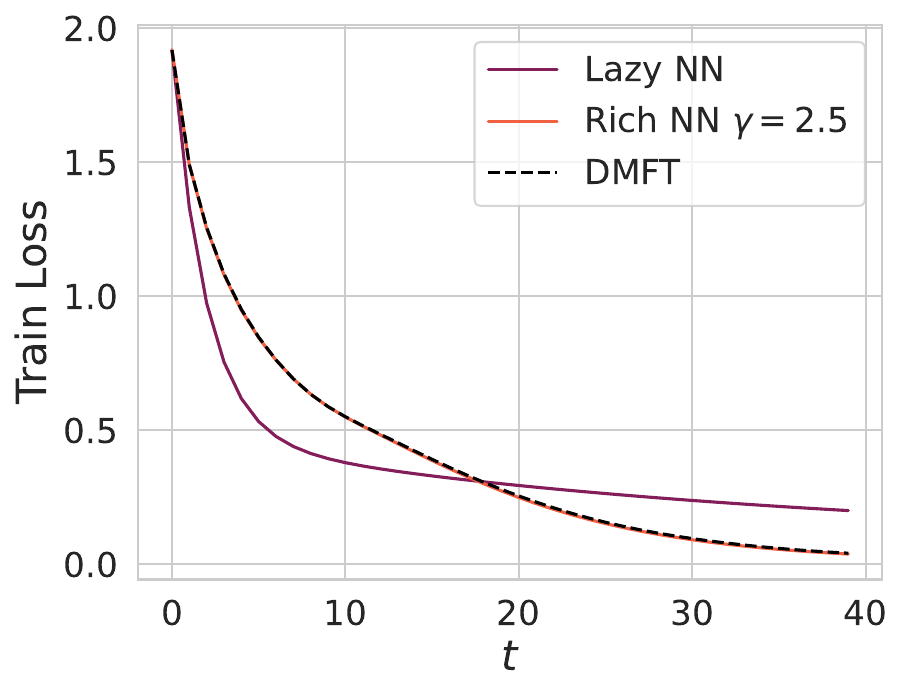}
    \caption{Loss Dynamics accurately predicted in the rich regime $\gamma > 0$. }
    \label{fig:placeholder}
\end{figure}

\begin{figure}
    \centering
    \includegraphics[width=0.45\linewidth]{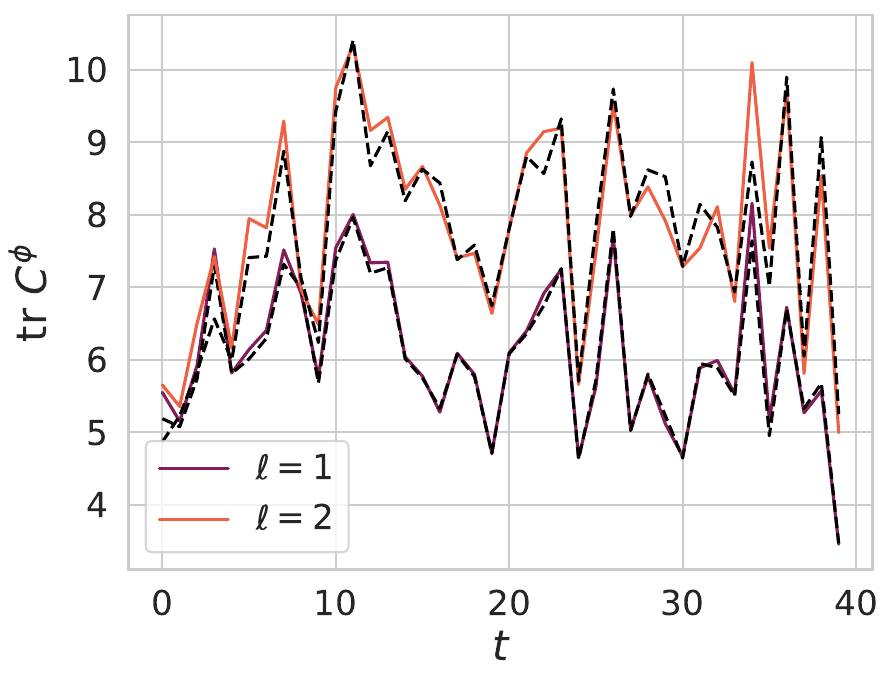}
    \includegraphics[width=0.45\linewidth]{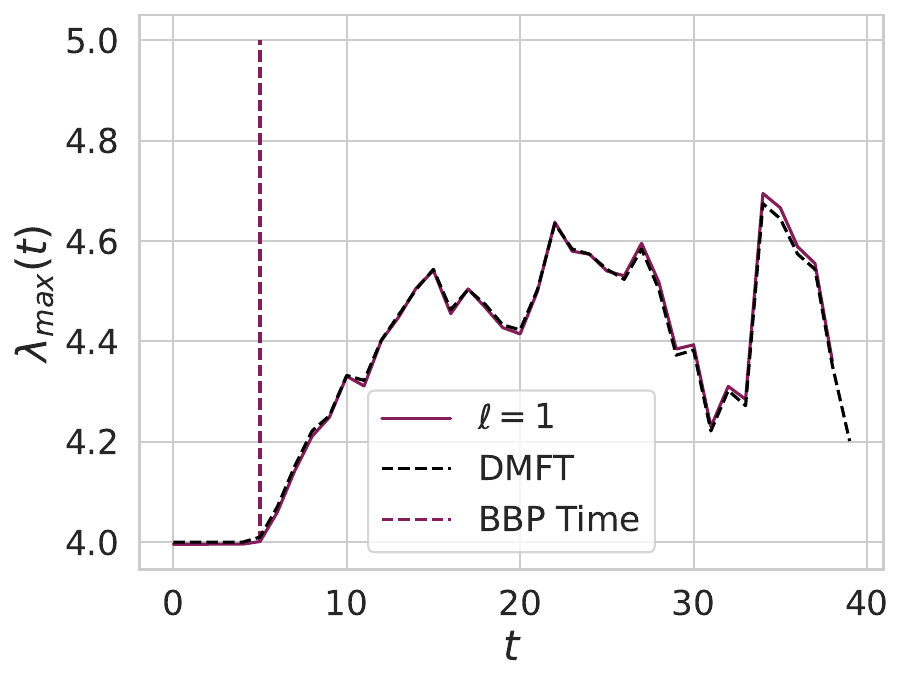}
    \caption{Our theory can also describe networks trained with minibatch SGD.  (a) Hidden correlation functions in a nonlinear $L=2$ hidden layer network trained with online SGD. (b) The maximum eigenvalue obeys a stochastic trajectory predicted by our outlier condition from Result 1.  }
    \label{fig:placeholder}
\end{figure}

\begin{figure}
    \centering
    \includegraphics[width=0.32\linewidth]{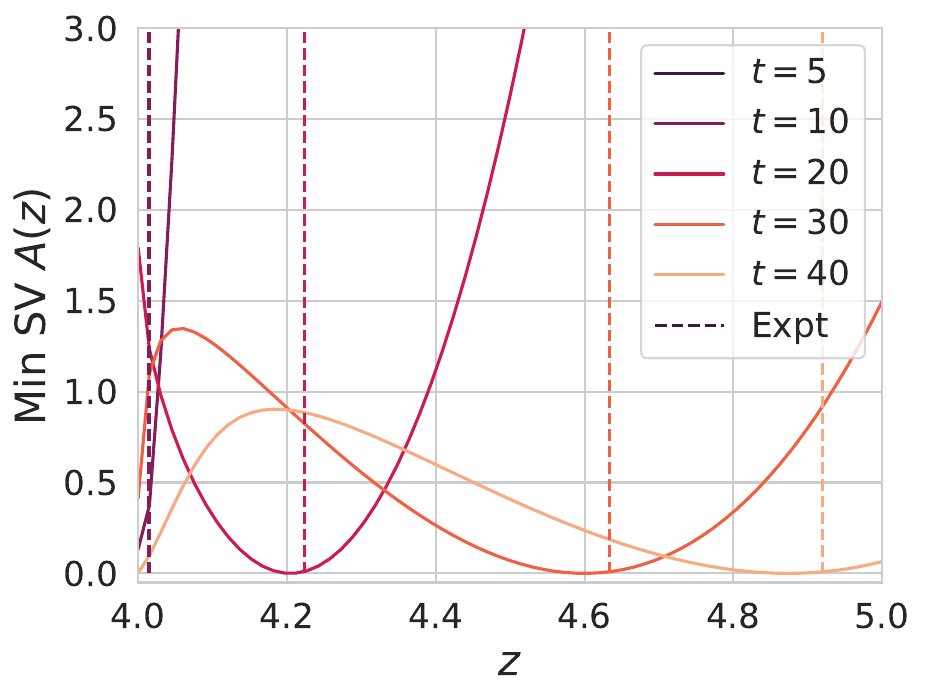}
    \includegraphics[width=0.32\linewidth]{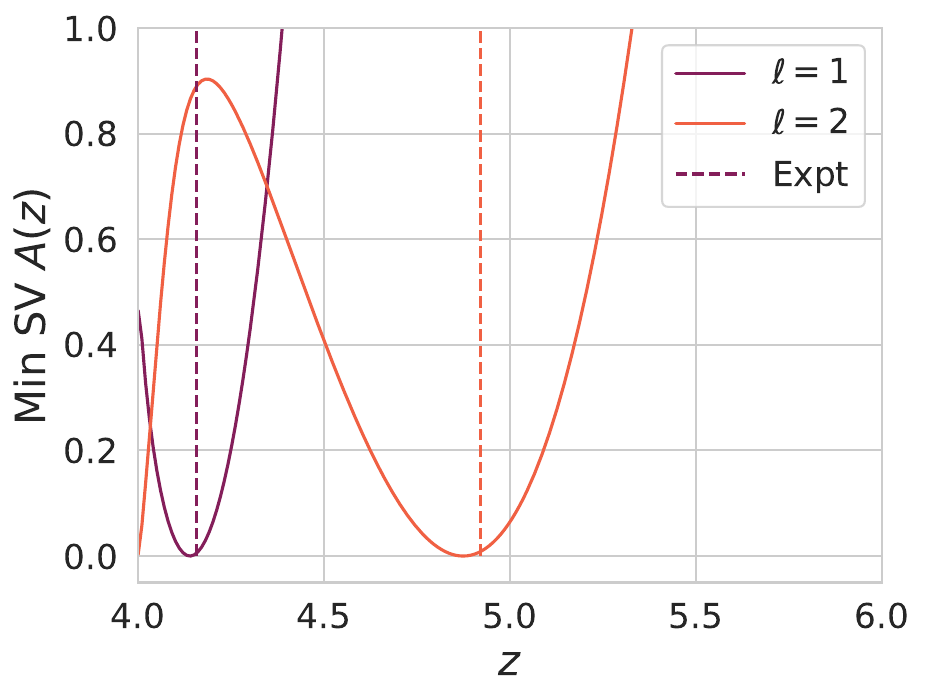}
    \includegraphics[width=0.32\linewidth]{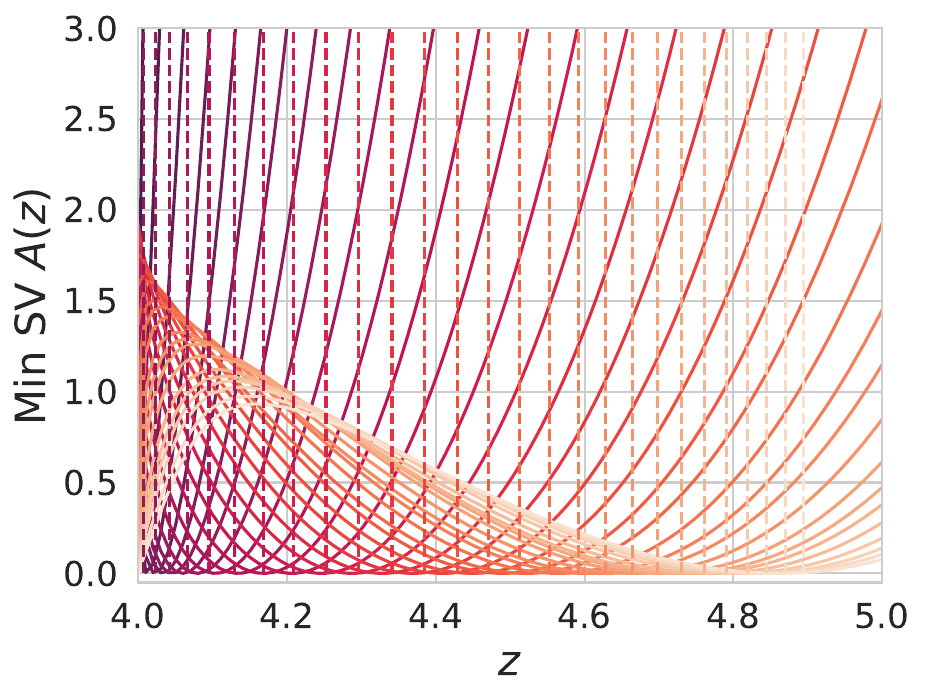}
    \caption{Demonstration of the root finding procedure for $\det \bm A(z) = 0$. In practice, we compute the minimum singular value for $\bm A(z)$. When the minimum singular value at $z$ is lower than a specified threshold, we declare that value of $z$ a solution in the set $O_S$. }
    \label{fig:placeholder}
\end{figure}

\section{Two-Level DMFT for Spectral Dynamics in Deep Linear Networks}
\label{app:linear_two_level_dmft}
In this section, we extend the proportional-limit DMFT for deep linear networks to compute the spectrum of trained weight matrices. As usual, the first level of the theory describes the training dynamics of the fields generated by gradient descent, while the second level introduces a linear spectral probe whose response gives the resolvent of the trained weight covariance. This two-level construction is still accurate because the directions generated by the gradient updates are both low-rank and statistically coupled to the random initial weights.

\subsection{Model definition and high dimensional limit}

We consider a depth-$L$ linear network with input dimension $D$, hidden width $N$, and scalar output
\begin{equation}
    f(\bm x)
    =
    \frac{\sqrt D}{N\gamma_0}
    (\bm w^L)^\top
    \left[
    \prod_{\ell=1}^{L-1}
    \left(
    \frac{1}{\sqrt N}\bm W^\ell
    \right)
    \right]
    \left(
    \frac{1}{\sqrt D}\bm W^0
    \right)
    \bm x .
\end{equation}
The first-layer weight matrix is $\bm W^0\in\mathbb R^{N\times D}$, the hidden-layer weights are $\bm W^\ell\in\mathbb R^{N\times N}$ for $\ell=1,\ldots,L-1$, and the readout is $\bm w^L\in\mathbb R^N$. The parameter $\gamma_0$ controls the feature-learning scale.

The training data are $\mathcal D=\{(\bm x_\mu,y_\mu)\}_{\mu=1}^P$, with isotropic random covariates and a linear teacher
\begin{equation}
    y_\mu
    =
    \frac{1}{\sqrt D}\bm w_\star\cdot \bm x_\mu.
\end{equation}
We study the proportional high-dimensional asymptotic limit
\begin{equation}
    P,N,D\to\infty,
    \qquad
    \alpha=\frac{P}{D},
    \qquad
    \nu=\frac{N}{D}
\end{equation}
and given a fixed number of training steps $T=\mathcal O(1)$.

Our goal is to characterize the spectrum of the trained weight covariances
\begin{equation}
    \bm M^\ell(t)
    =
    \frac{1}{N}
    \bm W^\ell(t)^\top \bm W^\ell(t)\in \mathbb{R}^{N\times N}
    \qquad
    \forall \ell=1,\ldots,L-1
\end{equation}
and similarly for the first layer
\begin{equation}
    \bm M^0(t)
    =
    \frac{1}{N}
    \bm W^0(t)^\top \bm W^0(t)
    \in\mathbb R^{D\times D}.
\end{equation}

\subsection{Training fields and finite-rank weight updates}

Let $\Delta_\mu(t)=y_\mu-f(\bm x_\mu,t)$ denote the training error at step $t$. We define the training-set averaged forward field
\begin{equation}
    \bm h^0(t)
    =
    \frac{\sqrt D}{P}
    \sum_{\mu=1}^P
    \Delta_\mu(t)\bm x_\mu,
\end{equation}
and the hidden forward and backward fields recursively by
\begin{align}
    \bm h^{\ell+1}(t)
    &=
    \frac{1}{\sqrt N}
    \bm W^\ell(t)\bm h^\ell(t)
    \qquad
    \ell=0,\ldots,L-1
    \\
    \bm g^L(t)
    &=
    \bm w^L(t)
    \\
    \bm g^\ell(t)
    &=
    \frac{1}{\sqrt N}
    \bm W^\ell(t)^\top \bm g^{\ell+1}(t),
    \qquad
    \ell=1,\ldots,L-1.
\end{align}
The gradient updates can be written as finite-rank deformations of the initial random Gaussian weights $W_{ij}^{\ell}(0)\sim \mathcal{N}(0,1)$
\begin{align}
    \bm W^0(t)
    &=
    \bm W^0(0)
    +
    \frac{\eta\gamma_0}{\sqrt D}
    \sum_{s<t}
    \bm g^1(s)\bm h^0(s)^\top
    \\
    \bm W^\ell(t)
    &=
    \bm W^\ell(0)
    +
    \frac{\eta\gamma_0}{\sqrt N}
    \sum_{s<t}
    \bm g^{\ell+1}(s)\bm h^\ell(s)^\top
    \qquad
    1\le \ell\le L-1
    \\
    \bm w^L(t)
    &=
    \bm w^L(0)
    +
    \eta\gamma_0
    \sum_{s<t}
    \bm h^L(s).
\end{align}
which at fixed training time $t$ is spanned by the training fields. To isolate the dependence on the initialized weights, we define
\begin{align}
    \bm \xi^0(t)
    &=
    \frac{\sqrt D}{N\gamma_0}
    \bm W^0(0)^\top \bm g^1(t)
    &
    \bm \chi^1(t)
    &=
    \frac{1}{\sqrt D}
    \bm W^0(0)\bm h^0(t)
    \\
    \bm \xi^\ell(t)
    &=
    \frac{1}{\sqrt N}
    \bm W^\ell(0)^\top \bm g^{\ell+1}(t),
    &
    \bm \chi^{\ell+1}(t)
    &=
    \frac{1}{\sqrt N}
    \bm W^\ell(0)\bm h^\ell(t)
    \qquad
    1\le \ell\le L-1.
\end{align}
Substituting the finite-rank weight updates into the forward and backward recursions gives
\begin{align}
    \bm v(t)
    &=
    \bm w_\star
    -
    \bm \xi^0(t)
    -
    \eta
    \sum_{s<t}
    C_g^1(t,s)\bm h^0(s)
    \\
    \bm h^{\ell+1}(t)
    &=
    \bm \chi^{\ell+1}(t)
    +
    \eta\gamma_0
    \sum_{s<t}
    C_h^\ell(t,s)\bm g^{\ell+1}(s)
    \\
    \bm g^\ell(t)
    &=
    \bm \xi^\ell(t)
    +
    \eta\gamma_0
    \sum_{s<t}
    C_g^{\ell+1}(t,s)\bm h^\ell(s).
\end{align}
Here
\begin{equation}
    C_h^\ell(t,s)
    =
    \frac{1}{N}
    \bm h^\ell(t)\cdot \bm h^\ell(s),
    \qquad
    C_g^\ell(t,s)
    =
    \frac{1}{N}
    \bm g^\ell(t)\cdot \bm g^\ell(s)
\end{equation}
with the convention that $C_h^0$ is normalized by $D$ when $\bm h^0\in\mathbb R^D$.

The first-level DMFT determines the joint law of the fields
\begin{equation}
    \bm v(t),\quad
    \Delta(t),\quad
    \bm h^\ell(t),\quad
    \bm g^\ell(t),
\end{equation}
through their correlation and response functions. These are the same training-level order parameters used to characterize the loss dynamics in the proportional limit (see~\cite{bordelon2025deep} for more details).

\subsection{Spectral probe for the trained weight covariance}

We now introduce a second-level probe to compute the spectrum of $\bm M^\ell(t)$. For a fixed layer $\ell$ and fixed training time $t$, consider the linear response dynamics
\begin{equation}
    \partial_\tau \bm \psi_0(\tau)
    =
    \bm M^\ell(t)\bm \psi_0(\tau)
    +
    \bm j(\tau).
\end{equation}
The response of $\bm \psi_0$ to the source $\bm j$ gives the resolvent of $\bm M^\ell(t)$
\begin{equation}
    \mathcal H^\ell(\omega)
    =
    \frac{1}{d_\ell}
    \operatorname{Tr}
    \left(
    i\omega \bm I
    -
    \bm M^\ell(t)
    \right)^{-1}
\end{equation}
where $d_0=D$ for the first layer and $d_\ell=N$ for the hidden layers.

The key step is to express the action of $\bm M^\ell(t)$ on the probe using the same initialized weights and training fields that appear in the first-level DMFT.

For the first layer, $\ell=0$, define
\begin{align}
    \bm{\psi}_{1}(\tau)&=\underbrace{\frac{1}{\sqrt{D}}\bm{W}^{0}(0)\bm{\psi}_{0}(\tau)}_{\zeta_{1}(\tau)}+\eta\gamma_{0}\sum_{s<t}\bm{g}^{1}(s)C^{h^{0}\psi_{0}}(s,\tau)
    \\
    \bm \psi_2(\tau)
    &=\underbrace{\frac{\sqrt{D}}{N}\bm{W}^{0}(0)^{\top}\bm{\psi}_{1}(\tau)}_{\zeta_{2}(\tau)}+\eta\gamma_{0}\sum_{s<t}\bm{h}^{0}(s)C^{g^{1}\psi_{1}}(s,\tau)
\end{align}
where
\begin{equation}
    C^{h^0\psi_0}(s,\tau)
    =
    \frac{1}{D}
    \bm h^0(s)\cdot \bm \psi_0(\tau),
    \qquad
    C^{g^1\psi_1}(s,\tau)
    =
    \frac{1}{N}
    \bm g^1(s)\cdot \bm \psi_1(\tau).
\end{equation}
By construction
\begin{equation}
    \bm \psi_2(\tau)
    =
    \bm M^0(t)\bm \psi_0(\tau).
\end{equation}

For hidden layers $1\le \ell\le L-1$, define instead
\begin{align}
    \bm{\psi}_{1}(\tau)&=\underbrace{\frac{1}{\sqrt{N}}\bm{W}^{\ell}(0)\bm{\psi}_{0}(\tau)}_{\zeta_{1}(\tau)}+\eta\gamma_{0}\sum_{s<t}\bm{g}^{\ell+1}(s)C^{h^{\ell}\psi_{0}}(s,\tau)\\
    \bm{\psi}_{2}(\tau)&=\underbrace{\frac{1}{\sqrt{N}}\bm{W}^{\ell}(0)^{\top}\bm{\psi}_{1}(\tau)}_{\zeta_{2}(\tau)}+\eta\gamma_{0}\sum_{s<t}\bm{h}^{\ell}(s)C^{g^{\ell+1}\psi_{1}}(s,\tau)
\end{align}
with
\begin{equation}
    C^{h^\ell\psi_0}(s,\tau)
    =
    \frac{1}{N}
    \bm h^\ell(s)\cdot \bm \psi_0(\tau),
    \qquad
    C^{g^{\ell+1}\psi_1}(s,\tau)
    =
    \frac{1}{N}
    \bm g^{\ell+1}(s)\cdot \bm \psi_1(\tau).
\end{equation}
Again
\begin{equation}
    \bm \psi_2(\tau)
    =
    \bm M^\ell(t)\bm \psi_0(\tau).
\end{equation}

The spectral problem therefore introduces new probe fields $\bm\psi_0,\bm\psi_1,\bm\psi_2$ and new mixed order parameters between the probe and the training fields. These mixed order parameters play the role of projected resolvents onto the finite-dimensional subspace generated by the training dynamics.
\subsection{Path-Integral (MGF)}
We now aim to characterize the joint distribution of our training fields and spectral probe $\{\bm v (t), \bm \Delta(t),\bm h^0 (t), \{\bm h^{\ell}(t),\bm g^{\ell}(t)\}_{\ell=1}^L, \bm \psi_0\}$ over draws of the quenched disorder $\mathcal{D}\equiv \{\bm X, \{\bm W^{\ell}(0)\}_{\ell=0}^L\}$ represented by data and initialization. To do so, we will start by calculating the moment generating function of these fields with a Martin-Siggia-Rose integral over the trajectories
\begin{equation}\label{eq:MGF_dln}
\begin{split}
     &Z \left[ \bm j_v, \bm j_{\Delta}, \{\bm j_{h^{\ell}}, \bm j_{\bm g^{\ell}}\}_{\ell \in \left[ L\right] },\bm j_{\psi}\right]\\
     &=\left< \exp\left(\sum_t (\bm j_v(t)\cdot \bm v(t)+\bm j_\Delta(t)\cdot\Delta(t) +\sum_{\ell} \bm j_{h^{\ell}}(t)\cdot \bm h^{\ell}(t)+\sum_{\ell} \bm j_{g^{\ell}}(t)\cdot \bm g^{\ell}(t) +\sum_{\tau} \bm j_{\psi} (\tau)\cdot \bm \psi_0 (\tau)\right)\right>.
\end{split}
\end{equation}
As in the previous section, imposing all the fields definitions that appear in~\eqref{eq:MGF_dln} leads to isolate each disorder contribution to then perform easily the averages. 
\subsection{Data average}
The relevant terms involving $\bm X$ are
\begin{equation}
    \left\langle
    \exp\left[
    -i
    \operatorname{Tr}
    \bm X^\top
    \sum_t
    \left(
    \frac{\sqrt D}{P}
    \bm\Delta(t)\hat{\bm h}^0(t)^\top
    +
    \frac{1}{\sqrt D}
    \hat{\bm\Delta}(t)\bm v(t)^\top
    \right)
    \right]
    \right\rangle_{\bm X}.
\end{equation}
Averaging over the Gaussian data gives
\begin{align}
    &\exp\left[
    -\frac{1}{2}
    \sum_{tt'}
    C_v(t,t')
    \hat{\bm\Delta}(t)\cdot \hat{\bm\Delta}(t')
    -
    \frac{1}{2\alpha}
    \sum_{tt'}
    C_\Delta(t,t')
    \hat{\bm h}^0(t)\cdot \hat{\bm h}^0(t')
    \right]
    \nonumber \\
    &\times
    \exp\left[
    D
    \sum_{tt'}
    R_\Delta(t',t)R_{vu}^0(t,t')
    \right]
\end{align}
where
\begin{align}
    C_v(t,t')
    &=
    \frac{1}{D}
    \bm v(t)\cdot \bm v(t')
    &
    C_\Delta(t,t')
    &=
    \frac{1}{P}
    \bm\Delta(t)\cdot \bm\Delta(t')
    \\
    R_\Delta(t,t')
    &=
    -\frac{i}{P}
    \bm\Delta(t)\cdot \hat{\bm\Delta}(t')
    &
    R_{vu}^0(t,t')
    &=
    -\frac{i}{D}
    \bm v(t)\cdot \hat{\bm h}^0(t').
\end{align}
To enforce these definitions, we use Fourier representation of Dirac delta function, for instance
\begin{align}
    1&\equiv\int\frac{dR_{\Delta}(t,t')d\hat{R}_{\Delta}(t,t')}{2\pi}\exp\Bigg(-P\sum_{tt'}R_{\Delta}(t,t')\hat{R}_{\Delta}(t,t')-i\sum_{tt'}\hat{R}_{\Delta}(t,t')\bm{\Delta}(t)\cdot\hat{\bm{\Delta}}(t')\Bigg)\\1&\equiv\int\frac{dR_{vu}^{0}(t,t')d\hat{R}_{vu}^{0}(t,t')}{2\pi}\exp\Bigg(-D\sum_{tt'}R_{vu}^{0}(t,t')\hat{R}_{vu}^{0}(t,t')-i\sum_{tt'}\hat{R}_{vu}^{0}(t,t')\bm{v}(t)\cdot\hat{\bm{h}}^{0}(t')\Bigg)
\end{align}
 At the saddle point, differentiating with respect to $R_{\Delta}$ and $R^0_{vu}$ gives
\begin{align}
    \hat{R}_{\Delta}(t',t)&=\frac{1}{\alpha}R_{vu}^{0}(t,t')\\\hat{R}_{vu}^{0}(t,t')&=R_{\Delta}(t',t).
\end{align}
These terms encode the memory induced by reusing the same random dataset in full-batch gradient descent.
\subsection{First-layer weight average}

We now average over the first-layer initialization $\bm W^0(0)$
\begin{equation}
    \begin{split}
        &\Bigg<\exp\left(-\frac{i}{\sqrt{D}}\text{Tr}\bm{W}^{0}(0)^{\top}\left(\sum_{t}\hat{\bm{\chi}}^{1}(t)\bm{h}^{0}(t)^{\top}+\frac{1}{\gamma_{0}\nu}\sum_{t}\bm{g}^{1}(t)\hat{\bm{\xi}}^{0}(t)^{\top}+\sum_{\tau}\hat{\bm{\zeta}}_{1}(\tau)\bm{\psi}_{0}(\tau)^{\top}+\frac{1}{\nu}\sum_{\tau}\bm{\psi}_{1}(\tau)\hat{\bm{\zeta}}_{2}(\tau)^{\top}\right)\right)\Bigg>\\
    & =\exp\Bigg(-\frac{1}{2}\sum_{tt'}C_{h}^{0}(t,t')\hat{\bm{\chi}}^{1}(t)\cdot\hat{\bm{\chi}}^{1}(t')-\frac{1}{2\gamma_{0}^{2}\nu}\sum_{tt'}C_{g}^{1}(t,t')\hat{\bm{\xi}}^{0}(t)\cdot\hat{\bm{\xi}}^{0}(t')\Bigg)\\
    &\times \exp\Bigg(-\frac{1}{2}\sum_{\tau\tau'}C^{\psi_{0}}(\tau,\tau')\hat{\bm{\zeta}}_{1}(\tau)\cdot\hat{\bm{\zeta}}_{1}(\tau')-\frac{1}{2\nu}\sum_{\tau\tau'}C^{\psi_{1}}(\tau,\tau')\hat{\bm{\zeta}}_{2}(\tau)\cdot\hat{\bm{\zeta}}_{2}(\tau')\Bigg)\\
    &\times\exp\Bigg(+\frac{D}{\gamma_{0}}\sum_{t,t'}R_{gu}^{1}(t',t)R_{hr}^{0}(t,t')-\sum_{t\tau}C^{h^{0}\psi_{0}}(t,\tau)\hat{\bm{\chi}}^{1}(t)\cdot\hat{\bm{\zeta}}_{1}(\tau)\Bigg)\\
    &\times\exp\Bigg(+D\sum_{t\tau}R^{\psi_{1}u^{1}}(\tau,t)R^{h_{0}\zeta_{2}}(t,\tau)+\frac{D}{\gamma_{0}}\sum_{t\tau}R^{\psi_{0}r^{0}}(\tau,t)R^{g^{1}\zeta_{1}}(t,\tau)\Bigg)\\
    &\times\exp\Bigg(-\frac{1}{\gamma_{0}\nu}\sum_{t\tau}C^{g^{1}\psi_{1}}(t,\tau)\hat{\bm{\xi}}^{0}(t)\cdot\hat{\bm{\zeta}}_{2}(\tau)+D\sum_{\tau\tau'}R^{\psi_{1}\zeta_{1}}(\tau,\tau')R^{\psi_{0}\zeta_{2}}(\tau',\tau)\Bigg).
    \end{split}
\end{equation}
This Gaussian average over $\bm W^0(0)$ generates covariance terms for each pair of fields, together with mixed training--probe response terms. 
 Define
\begin{align}
C_{h}^{0}(s,s')
&=
\frac{1}{D}\bm h^{0}(s)\cdot\bm h^{0}(s'),
&
C_{g}^{1}(s,s')
&=
\frac{1}{N}\bm g^{1}(s)\cdot\bm g^{1}(s'),
\\
C^{\psi_{0}}(\tau,\tau')
&=
\frac{1}{D}\bm\psi_{0}(\tau)\cdot\bm\psi_{0}(\tau'),
&
C^{\psi_{1}}(\tau,\tau')
&=
\frac{1}{N}\bm\psi_{1}(\tau)\cdot\bm\psi_{1}(\tau'),
\\
C^{h^{0}\psi_{0}}(s,\tau)
&=
\frac{1}{D}\bm h^{0}(s)\cdot\bm\psi_{0}(\tau),
&
C^{g^{1}\psi_{1}}(s,\tau)
&=
\frac{1}{N}\bm g^{1}(s)\cdot\bm\psi_{1}(\tau).
\end{align}
The response-type order parameters generated by the same average are
\begin{align}
R_{hr}^{0}(s,s')
&=
-\frac{i}{D}\bm h^{0}(s)\cdot\hat{\bm\xi}^{0}(s'),
&
R_{gu}^{1}(s,s')
&=
-\frac{i}{N}\bm g^{1}(s)\cdot\hat{\bm\chi}^{1}(s'),
\\
R^{h^{0}\zeta_{2}}(s,\tau)
&=
-\frac{i}{D}\bm h^{0}(s)\cdot\hat{\bm\zeta}_{2}(\tau),
&
R^{g^{1}\zeta_{1}}(s,\tau)
&=
-\frac{i}{N}\bm g^{1}(s)\cdot\hat{\bm\zeta}_{1}(\tau),
\\
R^{\psi_{0}r^{0}}(\tau,s)
&=
-\frac{i}{D}\bm\psi_{0}(\tau)\cdot\hat{\bm\xi}^{0}(s),
&
R^{\psi_{1}u^{1}}(\tau,s)
&=
-\frac{i}{N}\bm\psi_{1}(\tau)\cdot\hat{\bm\chi}^{1}(s),
\\
R^{\psi_{0}\zeta_{2}}(\tau,\tau')
&=
-\frac{i}{D}\bm\psi_{0}(\tau)\cdot\hat{\bm\zeta}_{2}(\tau'),
&
R^{\psi_{1}\zeta_{1}}(\tau,\tau')
&=
-\frac{i}{N}\bm\psi_{1}(\tau)\cdot\hat{\bm\zeta}_{1}(\tau').
\end{align}
\subsection{Hidden layers weight average}
For $1\leq \ell \leq L-1$ the weight matrix $\bm W^{\ell}$ is square. The Gaussian average is
\begin{align}
    &\Bigg<\exp\Bigg(-\frac{i}{\sqrt{N}}\text{Tr}\bm{W}^{\ell}(0)^{\top}\Bigg(\sum_{t}\hat{\bm{\chi}}^{\ell+1}(t)\bm{h}^{\ell}(t)^{\top}+\sum_{t}\bm{g}^{\ell+1}(t)\hat{\bm{\xi}}^{\ell}(t)^{\top}+\sum_{\tau}\hat{\bm{\zeta}}_{1}(\tau)\bm{\psi}_{0}(\tau)^{\top}+\sum_{\tau}\bm{\psi}_{1}(\tau)\hat{\bm{\zeta}}_{2}(\tau)^{\top}\Bigg)\Bigg)\Bigg>\\
    &=\exp\Bigg(-\frac{1}{2}\sum_{tt'}C_{h}^{\ell}(t,t')\hat{\bm{\chi}}^{\ell+1}(t)\cdot\hat{\bm{\chi}}^{\ell+1}(t')-\frac{1}{2}\sum_{tt'}C_{g}^{\ell+1}(t,t')\hat{\bm{\xi}}^{\ell}(t)\cdot\hat{\bm{\xi}}^{\ell}(t')\Bigg)\\
    &\times\exp\Bigg(-\frac{1}{2}\sum_{\tau\tau'}C^{\psi_{0}}(\tau,\tau')\hat{\bm{\zeta}}_{1}(\tau)\cdot\hat{\bm{\zeta}}_{1}(\tau')-\frac{1}{2}\sum_{\tau\tau'}C^{\psi_{1}}(\tau,\tau')\hat{\bm{\zeta}}_{2}(\tau)\cdot\hat{\bm{\zeta}}_{2}(\tau')\Bigg)\\&\times\exp\Bigg(N\sum_{tt'}R_{gu}^{\ell+1}(t',t)R_{hr}^{\ell}(t,t')-\sum_{t\tau}C^{h^{\ell}\psi_{0}}(t,\tau)\hat{\bm{\chi}}^{\ell+1}(t)\cdot\hat{\bm{\zeta}}_{1}(\tau)\Bigg)\\&\times\exp\Bigg(N\sum_{t\tau}R^{\psi_{1}u^{\ell+1}}(\tau,t)R^{h^{\ell}\zeta_{2}}(t,\tau)+N\sum_{t\tau}R^{\psi_{0}r^{\ell}}(\tau,t)R^{g^{\ell+1}\zeta_{1}}(t,\tau)\Bigg)\\&\times\exp\Bigg(-\sum_{t\tau}C^{g^{\ell+1}\psi_{1}}(t,\tau)\hat{\bm{\xi}}^{\ell}(t)\cdot\hat{\bm{\zeta}}_{2}(\tau)+N\sum_{\tau\tau'}R^{\psi_{1}\zeta_{1}}(\tau,\tau')R^{\psi_{0}\zeta_{2}}(\tau',\tau)\Bigg)
\end{align}
which yields the same structure as for the first layer, but without the aspect-ratio $\nu$ which comes from having a rectangular first layer $\bm W^0$. The hidden-layer weight average generates the following
correlation functions
\[
C_h^\ell(t,t')
=
\frac1N\bm h^\ell(t)\cdot \bm h^\ell(t'),
\qquad
C_g^{\ell+1}(t,t')
=
\frac1N\bm g^{\ell+1}(t)\cdot \bm g^{\ell+1}(t'),
\]
\[
C^{\psi_0}(\tau,\tau')
=
\frac1N\bm\psi_0(\tau)\cdot \bm\psi_0(\tau'),
\qquad
C^{\psi_1}(\tau,\tau')
=
\frac1N\bm\psi_1(\tau)\cdot \bm\psi_1(\tau'),
\]
and the mixed training-probe correlations
\[
C^{h^\ell\psi_0}(t,\tau)
=
\frac1N\bm h^\ell(t)\cdot \bm\psi_0(\tau),
\qquad
C^{g^{\ell+1}\psi_1}(t,\tau)
=
\frac1N\bm g^{\ell+1}(t)\cdot \bm\psi_1(\tau).
\]

The corresponding response functions are defined by contractions with
the MSR conjugate fields. The training response functions are
\[
R_{hr}^{\ell}(t,t')
=
-\frac{i}{N}\bm h^\ell(t)\cdot \widehat{\bm\xi}^{\ell}(t'),
\qquad
R_{gu}^{\ell+1}(t,t')
=
-\frac{i}{N}\bm g^{\ell+1}(t)\cdot \widehat{\bm\chi}^{\ell+1}(t').
\]
The mixed probe-training response functions are
\[
R^{\psi_0 r^\ell}(\tau,t)
=
-\frac{i}{N}\bm\psi_0(\tau)\cdot \widehat{\bm\xi}^{\ell}(t),
\qquad
R^{\psi_1 u^{\ell+1}}(\tau,t)
=
-\frac{i}{N}\bm\psi_1(\tau)\cdot \widehat{\bm\chi}^{\ell+1}(t),
\]
\[
R^{h^\ell\zeta_2}(t,\tau)
=
-\frac{i}{N}\bm h^\ell(t)\cdot \widehat{\bm\zeta}_2(\tau),
\qquad
R^{g^{\ell+1}\zeta_1}(t,\tau)
=
-\frac{i}{N}\bm g^{\ell+1}(t)\cdot \widehat{\bm\zeta}_1(\tau).
\]
Finally, the pure probe response functions are
\[
R^{\psi_1\zeta_1}(\tau,\tau')
=
-\frac{i}{N}\bm\psi_1(\tau)\cdot \widehat{\bm\zeta}_1(\tau'),
\qquad
R^{\psi_0\zeta_2}(\tau,\tau')
=
-\frac{i}{N}\bm\psi_0(\tau)\cdot \widehat{\bm\zeta}_2(\tau').
\]
\subsection{Last-layer readout average}

The readout initialization only appears in 
\begin{equation}
    \exp \Big[ -i \bm w^L (0)\cdot \sum_t \hat{\bm g}^L(t)\Big].
\end{equation}
Thus
\begin{equation}
    \left<\exp\left(-i \bm w^L (0)\cdot \sum_t \hat{\bm g}^L(t)\right) \right>_{\bm w^L (0)} = \exp \Big[ -\frac{1}{2}\sum_{tt'} \hat{\bm g}^L (t)\cdot \hat{g}^L (t')\Big].
\end{equation}
Equivalently, after Hubbard linearization, the readout contributes as a Gaussian process $r^L (t)$ with covariance
\begin{equation}
    \langle r^L (t) r^L (t')\rangle = 1.
\end{equation}
The covariance is one for all pairs of times because the same readout initialization is reused at every training step.
\subsection{From the averaged action to single-site processes}
All correlations and responses after the averages can be now enforced by delta-function identities. After these insertions, the joint generating functional has the schematic form
\begin{equation}
    Z[\mathcal J]
    =
    \int\dd\cQ\dd\widehat\cQ\,
    \exp\left[-D\cS(\cQ,\widehat\cQ;\mathcal J)\right]
\end{equation}
where $\cQ$ collects all training and probe order parameters.

The key factorization is as follows.  Once the disorder has been averaged, distinct data samples $\mu$, input coordinates $k$, and hidden neurons $i$ no longer interact directly.  They interact only through the macroscopic order parameters $\cQ$.  Therefore the microscopic path integral factorizes as
\begin{equation}
\begin{split}
    Z[\mathcal J]
    =
    \int\dd\cQ\dd\widehat\cQ\,
    e^{-D\bm \Phi(\cQ,\widehat\cQ)}
    \prod_{\mu=1}^{P}Z_\Delta^{(\mu)}[\cQ]
    \prod_{k=1}^{D}Z_0^{(k)}[\cQ]
    \prod_{\ell=1}^{L}\prod_{i=1}^{N}Z_\ell^{(i)}[\cQ]
    \times Z_{\rm pr}^{(a)}[\cQ].
\end{split}
\end{equation}
Here $Z_\Delta$ is the scalar process for one training sample, $Z_0$ is the scalar process for one input coordinate, and $Z_\ell$ is the scalar process for one hidden neuron at layer $\ell$.  The factor $Z_{\rm pr}^{(a)}$ denotes the same single-site factors augmented by probe variables on the probed layer.  At zero source the sites are exchangeable, so
\begin{equation}
    \frac1D\sum_{k=1}^D F_k\to\langle F\rangle_{Z_0},
    \qquad
    \frac1N\sum_{i=1}^N F_i\to\langle F\rangle_{Z_\ell},
    \qquad
    \frac1P\sum_{\mu=1}^P F_\mu\to\langle F\rangle_{Z_\Delta}.
\end{equation}

At the saddle point the conjugates of the correlation order parameters vanish in the zero-source limit, while the response conjugates produce the retarded memory kernels.  The remaining quadratic forms in the Fourier fields are linearized by Hubbard transformations, e.g.
\begin{align}
    \exp\left[-\frac12\sum_{tt'}\what\Delta(t)C_v(t,t')\what\Delta(t')\right]
    &=\left\langle e^{-i\sum_t\what\Delta(t)u_\Delta(t)}\right\rangle_{u_\Delta\sim GP(0,C_v)},
    \\
    \exp\left[-\frac1{2\alpha}\sum_{tt'}\what h^0(t)C_\Delta(t,t')\what h^0(t')\right]
    &=\left\langle e^{-i\sum_t\what h^0(t)u^0(t)}\right\rangle_{u^0\sim GP(0,\alpha^{-1}C_\Delta)},
    \\
    \exp\left[-\frac1{2\gamma_0^2\nu}\sum_{tt'}\what\xi^0(t)C_g^1(t,t')\what\xi^0(t')\right]
    &=\left\langle e^{-i\sum_t\what\xi^0(t)r^0(t)}\right\rangle_{r^0\sim GP(0,(\gamma_0^2\nu)^{-1}C_g^1)},
    \\
    \exp\left[-\frac12\sum_{tt'}\what\chi^\ell(t)C_h^{\ell-1}(t,t')\what\chi^\ell(t')\right]
    &=\left\langle e^{-i\sum_t\what\chi^\ell(t)u^\ell(t)}\right\rangle_{u^\ell\sim GP(0,C_h^{\ell-1})},
    \\
    \exp\left[-\frac12\sum_{tt'}\what\xi^\ell(t)C_g^{\ell+1}(t,t')\what\xi^\ell(t')\right]
    &=\left\langle e^{-i\sum_t\what\xi^\ell(t)r^\ell(t)}\right\rangle_{r^\ell\sim GP(0,C_g^{\ell+1})}.
\end{align}
Integrating over the Fourier variables then gives stochastic single-site equations.  The response functions can equivalently be read as functional derivatives with respect to the Gaussian noises.  For example,
\begin{equation}
    R_\Delta(t,t')=\left\langle\frac{\partial\Delta(t)}{\partial u_\Delta(t')}\right\rangle,
    \qquad
    R_{vu}^0(t,t')=\left\langle\frac{\partial v(t)}{\partial u^0(t')}\right\rangle.
\end{equation}

\subsection{Deterministic limiting order parameters}

For fixed $L$ and fixed training horizon, the action is proportional to $D$.  Hence the macroscopic order parameters concentrate at the saddle point.  In particular,
\begin{align}
    \frac1D\bv(t)\cdot\bv(t')&\xrightarrow[D\to\infty]{} C_v(t,t')=\langle v(t)v(t')\rangle,
    \\
    \frac1P\bD(t)\cdot\bD(t')&\xrightarrow[D\to\infty]{} C_\Delta(t,t')=\langle \Delta(t)\Delta(t')\rangle,
    \\
    \frac1D\bh^0(t)\cdot\bh^0(t')&\xrightarrow[D\to\infty]{} C_h^0(t,t')=\langle h^0(t)h^0(t')\rangle,
    \\
    \frac1N\bh^\ell(t)\cdot\bh^\ell(t')&\xrightarrow[D\to\infty]{} C_h^\ell(t,t')=\langle h^\ell(t)h^\ell(t')\rangle,
    \\
    \frac1N\bg^\ell(t)\cdot\bg^\ell(t')&\xrightarrow[D\to\infty]{} C_g^\ell(t,t')=\langle g^\ell(t)g^\ell(t')\rangle.
\end{align}
Similarly,
\begin{align}
    -\frac iP\bD(t)\cdot\what{\bD}(t')&\to R_\Delta(t,t')
    =\left\langle\frac{\partial\Delta(t)}{\partial u_\Delta(t')}\right\rangle,
    \\
    -\frac iD\bv(t)\cdot\what{\bh}^0(t')&\to R_{vu}^0(t,t')
    =\left\langle\frac{\partial v(t)}{\partial u^0(t')}\right\rangle,
    \\
    -\frac iD\bh^0(t)\cdot\what{\bxi}^0(t')&\to R_{hr}^0(t,t')
    =\left\langle\frac{\partial h^0(t)}{\partial r^0(t')}\right\rangle,
    \\
    -\frac iN\bh^\ell(t)\cdot\what{\bxi}^\ell(t')&\to R_{hr}^\ell(t,t')
    =\left\langle\frac{\partial h^\ell(t)}{\partial r^\ell(t')}\right\rangle,
    \\
    -\frac iN\bg^\ell(t)\cdot\what{\bchi}^\ell(t')&\to R_{gu}^\ell(t,t')
    =\left\langle\frac{\partial g^\ell(t)}{\partial u^\ell(t')}\right\rangle.
\end{align}
The mixed probe order parameters also converge to deterministic single-site averages.  For a hidden probed layer $a$,
\begin{align}
    C^{h^a\psi_0}(t,\tau)&\to\langle h^a(t)\psi_0(\tau)\rangle,
    &
    C^{g^{a+1}\psi_1}(t,\tau)&\to\langle g^{a+1}(t)\psi_1(\tau)\rangle,
    \\
    R^{\psi_0r^a}(\tau,t)&\to\left\langle\frac{\delta\psi_0(\tau)}{\delta r^a(t)}\right\rangle,
    &
    R^{\psi_1u^{a+1}}(\tau,t)&\to\left\langle\frac{\delta\psi_1(\tau)}{\delta u^{a+1}(t)}\right\rangle.
\end{align}
The first-layer version is identical except for the normalizations by $D,N$ and the factors of $\nu,\gamma_0$ already displayed in the first-layer average.

\subsection{Training-level DMFT}

The zero-source training DMFT is obtained from the joint saddle by setting the probe sources to zero after the disorder average has been performed.  This gives a closed scalar stochastic process for one data point, one input coordinate, and one representative hidden neuron in each layer.  The probe does not alter this training process; it only probes the solution obtained from it.

The data sector is
\begin{align}
    \Delta(t)
    &=u_\Delta(t)+\frac1\alpha\sum_{s<t}R_{vu}^0(t,s)\Delta(s)+\sigma\epsilon,
    &
    u_\Delta&\sim GP(0,C_v),
    \quad \epsilon\sim\cN(0,1),
    \\
    h^0(t)
    &=u^0(t)+\sum_{s<t}R_\Delta(t,s)v(s),
    &
    u^0&\sim GP(0,\frac{1}{\alpha}C_\Delta).
\end{align}
The first-layer boundary is
\begin{align}
    &h^1(t)
    =u^1(t)+\sum_{s<t}
      \left[\frac1{\gamma_0\nu}R_{hr}^0(t,s)+\eta\gamma_0 C_h^0(t,s)\right]g^1(s),\quad u^{1}(t)\sim GP(0,C_{h}^{0}),
    \\
    &v(t)
    =w_\star-r^0(t)-\sum_{s<t}
      \left[\frac1{\gamma_0}R_{gu}^1(t,s)+\eta C_g^1(t,s)\right]h^0(s),\quad r^{0}(t)\sim GP(0,\frac{1}{\gamma_{0}^{2}\nu}C_{g}^{1}).
\end{align}
For the remaining forward fields, \(2\leq\ell\leq L\),
\begin{align}
   h^\ell(t)
    =u^\ell(t)+\sum_{s<t}
      \left[R_{hr}^{\ell-1}(t,s)+\eta\gamma_0 C_h^{\ell-1}(t,s)\right]g^\ell(s),\quad u^\ell\sim GP(0,C_h^{\ell-1}).
\end{align}
For the backward fields below the readout, \(1\leq\ell\leq L-1\),
\begin{align}
   g^\ell(t)
    =r^\ell(t)+\sum_{s<t}
      \left[R_{gu}^{\ell+1}(t,s)+\eta\gamma_0 C_g^{\ell+1}(t,s)\right]h^\ell(s),\quad r^\ell&\sim GP(0,C_g^{\ell+1}).
\end{align}
The readout boundary is
\begin{equation}
    g^L(t)
    =r^L+\eta\gamma_0\sum_{s<t}
    \left[u^L(s)+\sum_{r<s}
      \left(R_{hr}^{L-1}(s,r)+\eta\gamma_0 C_h^{L-1}(s,r)\right)g^L(r)
    \right],\quad  \langle r^L(t)r^L(t')\rangle
    =1.
\end{equation}
The correlation fixed points are
\begin{align}
    C_v(t,t')&=\langle v(t)v(t')\rangle,
    &
    C_\Delta(t,t')&=\langle\Delta(t)\Delta(t')\rangle,
    \\
    C_h^\ell(t,t')&=\langle h^\ell(t)h^\ell(t')\rangle,
    &
    C_g^\ell(t,t')&=\langle g^\ell(t)g^\ell(t')\rangle.
\end{align}
The response fixed points are
\begin{align}
    R_\Delta(t,t')&=\left\langle\frac{\partial\Delta(t)}{\partial u_\Delta(t')}\right\rangle,
    &
    R_{vu}^0(t,t')&=\left\langle\frac{\partial v(t)}{\partial u^0(t')}\right\rangle,
    \\
    R_{hr}^0(t,t')&=\left\langle\frac{\partial h^0(t)}{\partial r^0(t')}\right\rangle,
    &
    R_{hr}^\ell(t,t')&=\left\langle\frac{\partial h^\ell(t)}{\partial r^\ell(t')}\right\rangle,
    \\
    R_{gu}^\ell(t,t')&=\left\langle\frac{\partial g^\ell(t)}{\partial u^\ell(t')}\right\rangle,
    &
    1\leq\ell&\leq L.
\end{align}
Finally,
\begin{equation}
    \mathcal L_{\rm test}(t)=C_v(t,t)+\sigma^2,
    \qquad
    \widehat{\mathcal L}_{\rm train}(t)=C_\Delta(t,t).
\end{equation}
\subsection{Probe-level DMFT fixed points}
The probe equations are solved after the training-level DMFT has produced $C_h^\ell,C_g^\ell,R_{hr}^\ell,R_{gu}^\ell$.  Thus the training process and the spectral probe are separated at the final stage: the training order parameters are inputs to the probe fixed point.
\subsubsection{First-layer probe}
For $\ell=0$, the effective first-layer probe process is
\begin{align}
    \psi_1(\tau)
    &=
    u_{\zeta_1}(\tau)
    +
    \frac{1}{\nu}
    \int d\tau'\,
    R^{\psi_0\zeta_2}(\tau,\tau')\psi_1(\tau')
    \nonumber \\
    &\quad
    +
    \sum_{s<t}
    \left[
    \eta\gamma_0 C^{h^0\psi_0}(s,\tau)
    +
    \frac{1}{\gamma_0\nu}
    R^{\psi_0 r^0}(\tau,s)
    \right]g^1(s)
    \\
    \psi_2(\tau)
    &=
    u_{\zeta_2}(\tau)
    +
    \int d\tau'\,
    R^{\psi_1\zeta_1}(\tau,\tau')\psi_0(\tau')
    \nonumber \\
    &\quad
    +
    \sum_{s<t}
    \left[
    \eta\gamma_0 C^{g^1\psi_1}(s,\tau)
    +
    R^{\psi_1u^1}(\tau,s)
    \right]h^0(s).
\end{align}
The Gaussian noises have covariances
\begin{align}
    \left\langle
    u_{\zeta_1}(\tau)u_{\zeta_1}(\tau')
    \right\rangle
    &=
    C^{\psi_0}(\tau,\tau'),
    &
    \left\langle
    u_{\zeta_2}(\tau)u_{\zeta_2}(\tau')
    \right\rangle
    &=
    \frac{1}{\nu}
    C^{\psi_1}(\tau,\tau'),
\end{align}
and mixed covariances
\begin{align}
    \left\langle
    u_{\zeta_1}(\tau)u^1(t)
    \right\rangle
    &=
    C^{h^0\psi_0}(t,\tau),
    &
    \left\langle
    u_{\zeta_2}(\tau)r^0(t)
    \right\rangle
    &=
    \frac{1}{\gamma_0\nu}
    C^{g^1\psi_1}(t,\tau).
\end{align}
These equations are the spectral-probe analogue of the first-layer single-site training equations.

To obtain the outlier condition, Fourier transform in the spectral time $\tau$ and set
\[
    z=i\omega,
    \qquad
    \mathcal H(z)
    =
    \left[
    z-R^{\psi_1\zeta_1}(z)
    \right]^{-1}.
\]
The $\psi_2$ equation gives
\begin{equation}
    \left[
    i\omega-R^{\psi_1\zeta_1}(\omega)
    \right]\psi_0(\omega)
    =
    u_{\zeta_2}(\omega)
    +
    \sum_{s<t}
    \left[
    \eta\gamma_0 C^{g^1\psi_1}(s,\omega)
    +
    R^{\psi_1u^1}(\omega,s)
    \right]h^0(s).
\end{equation}
Projecting onto $h^0$ once you stack the training time $T$ in $T$-dimensional vectors, and differentiating with respect to $r^0$ gives
\begin{align}
    \Big(i\omega-R^{\psi_{1},\zeta_{1}}(\omega)\Big)\bm{C}^{h^{0}\psi_{0}}(\omega)&=\langle\zeta_{2}(\omega)\bm{h}^{0}\rangle+\bm{C}_{h}^{0}\Big(\eta\gamma_{0}\bm{C}^{g^{1}\psi_{1}}(\omega)+\bm{R}^{\psi_{1},u^{1}}(\omega)\Big)\\\Big(i\omega-R^{\psi_{1},\zeta_{1}}(\omega)\Big)\bm{R}^{\psi_{0}r^{0}}(\omega)&=(\bm{R}_{hr}^{0})^{\top}\Big(\eta\gamma_{0}\bm{C}^{g^{1}\psi_{1}}(\omega)+\bm{R}^{\psi_{1},u^{1}}(\omega)\Big)
\end{align}
where we use Stein's lemma to write
\begin{equation}
    \left\langle
    u_{\zeta_2}(z)\bm h^0
    \right\rangle
    =
    \frac{1}{\gamma_0\nu}
    \bm R^0_{hr}\bm C^{g^1\psi_1}(z).
\end{equation}

Similarly, the $\psi_1$ equation gives
\begin{equation}
    \Big(1+\nu^{-1}\mathcal{H}(\omega)\Big)\psi_{1}(\omega)=u_{\zeta_{1}}(\omega)+\sum_{s<t}\Big(\eta\gamma_{0}C^{h^{0},\psi_{0}}(s,\tau)+\frac{1}{\gamma_{0}\nu}R^{\psi_{0},r^{0}}(\tau,s)\Big)g^{1}(s)
\end{equation}
where
\begin{equation}
    \mathcal{H}(\omega)=-R_{0,2}(\omega)=\Big(i\omega-R^{\psi_{1},\zeta_{1}}(\omega)\Big)^{-1}.
\end{equation}
Projecting onto $g^1$ and differentiating with respect to $u^1$ gives
\begin{align}
    \Big(1+\nu^{-1}\mathcal{H}(\omega)\Big)\bm{C}^{g^{1}\psi_{1}}(\omega)&=\langle u_{\zeta_{1}}(\omega)\bm{g}^{1}\rangle+\bm{C}_{g}^{1}\Big(\eta\gamma_{0}\bm{C}^{h^{0},\psi_{0}}(\omega)+\frac{1}{\gamma_{0}\nu}\bm{R}^{\psi_{0},r^{0}}(\omega)\Big)\\\Big(1+\nu^{-1}\mathcal{H}(\omega)\Big)\bm{R}^{\psi_{1}u^{1}}(\omega)&=(\bm{R}_{gu}^{1})^{\top}\Big(\eta\gamma_{0}\bm{C}^{h^{0},\psi_{0}}(\omega)+\frac{1}{\gamma_{0}\nu}\bm{R}^{\psi_{0},r^{0}}(\omega)\Big).
\end{align}
Here, again, we used Stein's lemma to express
\begin{equation}
    \left\langle
    u_{\zeta_1}(\omega)\bm g^1
    \right\rangle
    =
    \bm R^1_{gu}\bm C^{h^0\psi_0}(\omega).
\end{equation}

Collecting the four projected quantities and with the analytic continuation from $z = i\omega$ to the real spectral axis one gets
\begin{equation}
    \bm q(z)
    =
    \begin{pmatrix}
        \bm C^{h^0\psi_0}(z)\\
        \bm R^{\psi_0r^0}(z)\\
        \bm C^{g^1\psi_1}(z)\\
        \bm R^{\psi_1u^1}(z)
    \end{pmatrix}
\end{equation}
from which we obtain the finite-dimensional linear system
\begin{equation}
    \bm{\mathcal A}_0(z)\bm q(z)=0
\end{equation}
where
\begin{equation}
\bm{\mathcal A}_0(z)
=
\left(\begin{array}{cccc}
(z-R^{\psi_{1},\zeta_{1}}(z))\bm{I} & \bm{0} & -\frac{1}{\gamma_{0}\nu}\bm{R}_{hr}^{0}-\eta\gamma_{0}\bm{C}_{h}^{0} & -\bm{C}_{h}^{0}\\
\bm{0} & (z-R^{\psi_{1},\zeta_{1}}(z))\bm{I} & -\eta\gamma_{0}(\bm{R}_{hr}^{0})^{\top} & -(\bm{R}_{hr}^{0})^{\top}\\
-\bm{R}_{gu}^{1}-\eta\gamma_{0}\bm{C}_{g}^{1} & -\frac{1}{\gamma_{0}\nu}\bm{C}_{g}^{1} & (R^{\psi_{1},\zeta_{1}}(z))^{-1}\bm{I} & \bm{0}\\
-\eta\gamma_{0}(\bm{R}_{gu}^{1})^{\top} & -\frac{1}{\gamma_{0}\nu}(\bm{R}_{gu}^{1})^{\top} & \bm{0} & (R^{\psi_{1},\zeta_{1}}(z))^{-1}\bm{I}
\end{array}\right)
\end{equation}
where we remember
\begin{equation}
    R^{\psi_1\zeta_1}(z)
    =
    \left[
    1+\nu^{-1}\mathcal H(z)
    \right]^{-1}.
\end{equation}
The first-layer outliers are therefore obtained from the pole condition
\begin{equation}
    \det \bm{\mathcal A}_0(z)=0.
\end{equation}
\subsubsection{Hidden-layer probe}
As before, for $\ell=1,\ldots ,L-1$ the probe process generates 
\begin{align}
    \psi_{1}(\tau)&=u_{\zeta_{1}}(\tau)+\int R^{\psi_{0},\zeta_{2}}(\tau,\tau')\psi_{1}(\tau')+\sum_{s<t}\Big(\eta\gamma_{0}C^{h^{\ell}\psi_{0}}(s,\tau)+R^{\psi_{0},r^{\ell}}(\tau,s)\Big)g^{\ell+1}(s)\\\psi_{2}(\tau)&=u_{\zeta_{2}}(\tau)+\int R^{\psi_{1},\zeta_{1}}(\tau,\tau')\psi_{0}(\tau')+\sum_{s<t}\Big(\eta\gamma_{0}C^{g^{\ell+1}\psi_{1}}(s,\tau)+R^{\psi_{1},u^{\ell+1}}(\tau,s)\Big)h^{\ell}(s)
\end{align}
with 
\begin{align}
    \langle u_{\zeta_{1}}(\tau)u_{\zeta_{1}}(\tau')\rangle&=C^{\psi_{0}}(\tau,\tau'),\quad\langle\zeta_{1}(\tau)u^{\ell+1}(t)\rangle=C^{h^{\ell}\psi_{0}}(t,\tau)\\\langle u_{\zeta_{2}}(\tau)u_{\zeta_{2}}(\tau')\rangle&=C^{\psi_{1}}(\tau,\tau'),\quad\langle\zeta_{2}(\tau)r^{\ell}(t)\rangle=C^{g^{\ell+1}\psi_{1}}(t,\tau).
\end{align}
Thus, the variable of interest has the form
\begin{equation}
    \Big(i\omega-R^{\psi_{1},\zeta_{1}}(\omega)\Big)\psi_{0}(\omega)=u_{\zeta_{2}}(\omega)+\sum_{s<t}\Big(\eta\gamma_{0}C^{g^{\ell+1}\psi_{1}}(s,\omega)+R^{\psi_{1},u^{\ell+1}}(\omega,s)\Big)h^{\ell}(s)
\end{equation}
from which by projecting onto $h^{\ell}$ and differentiating with respect to $r^{\ell}$ one gets
\begin{align}
    \Big(i\omega-R^{\psi_{1},\zeta_{1}}(\omega)\Big)\bm{C}^{h^{\ell}\psi_{0}}(\omega)&=\langle\zeta_{2}(\omega)\bm{h}^{\ell}\rangle+\bm{C}_{h}^{\ell}\Big(\eta\gamma_{0}\bm{C}^{g^{\ell+1}\psi_{1}}(\omega)+\bm{R}^{\psi_{1},u^{\ell+1}}(\omega)\Big)\\\Big(i\omega-R^{\psi_{1},\zeta_{1}}(\omega)\Big)\bm{R}^{\psi_{0}r^{\ell}}(\omega)&=(\bm{R}_{hr}^{\ell})^{\top}\Big(\eta\gamma_{0}\bm{C}^{g^{\ell+1}\psi_{1}}(\omega)+\bm{R}^{\psi_{1},u^{\ell+1}}(\omega)\Big).
\end{align}
We can again apply Stein's Lemma for
\begin{equation}
    \langle\zeta_{2}(\omega)\bm{h}^{\ell}\rangle=\bm{R}_{hr}^{\ell}\bm{C}^{g^{\ell+1}\psi_{1}}(\omega).
\end{equation}
Similarly for $\psi_1 (\omega)$
\begin{equation}
    \Big(1+\mathcal{H}(\omega)\Big)\psi_{1}(\omega)=u_{\zeta_{1}}(\omega)+\sum_{s<t}\Big(\eta\gamma_{0}C^{h^{\ell},\psi_{0}}(s,\omega)+R^{\psi_{0},r^{\ell}}(\omega,s)\Big)g^{\ell+1}(s)
\end{equation}
with $\mathcal{H}(\omega)=-R_{0,2}(\omega)=\Big(i\omega-R^{\psi_{1},\zeta_{1}}(\omega)\Big)^{-1}$. From this, 
\begin{align}
    \Big(1+\mathcal{H}(\omega)\Big)\bm{C}^{g^{\ell+1}\psi_{1}}(\omega)&=\langle u_{\zeta_{1}}(\omega)\bm{g}^{\ell+1}\rangle+\bm{C}_{g}^{\ell+1}\Big(\eta\gamma_{0}\bm{C}^{h^{\ell},\psi_{0}}(\omega)+\bm{R}^{\psi_{0},r^{\ell}}(\omega)\Big)\\\Big(1+\mathcal{H}(\omega)\Big)\bm{R}^{\psi_{1}u^{\ell+1}}(\omega)&=(\bm{R}_{gu}^{\ell+1})^{\top}\Big(\eta\gamma_{0}\bm{C}^{h^{\ell},\psi_{0}}(\omega)+\bm{R}^{\psi_{0},r^{\ell}}(\omega)\Big).
\end{align}
By collecting all the four projected quantities as for the first layer and setting $z = i\omega$, then
\begin{equation}
    \bm q^{\ell} (z) = \left(\begin{array}{c}
\bm{C}^{h^{\ell}\psi_{0}}(z)\\
\bm{R}^{\psi_{0}r^{\ell}}(z)\\
\bm{C}^{g^{\ell+1}\psi_{1}}(z)\\
\bm{R}^{\psi_{1}u^{\ell+1}}(z)
\end{array}\right)
\end{equation}
by giving the poles condition of
\begin{equation}
    \det \bm A^{\ell} (z) = 0
\end{equation}
with
\begin{equation}
    \bm A^{\ell} (z) =\left(\begin{array}{cccc}
(z-R^{\psi_{1},\zeta_{1}}(z))\bm{I} & \bm{0} & -\bm{R}_{hr}^{\ell}-\eta\gamma_{0}\bm{C}_{h}^{\ell} & -\bm{C}_{h}^{\ell}\\
\bm{0} & (z-R^{\psi_{1},\zeta_{1}}(z))\bm{I} & -\eta\gamma_{0}(\bm{R}_{hr}^{\ell})^{\top} & -(\bm{R}_{hr}^{\ell})^{\top}\\
-\bm{R}_{gu}^{\ell+1}-\eta\gamma_{0}\bm{C}_{g}^{\ell+1} & -\bm{C}_{g}^{\ell+1} & (R^{\psi_{1},\zeta_{1}}(z))^{-1}\bm{I} & \bm{0}\\
-\eta\gamma_{0}(\bm{R}_{gu}^{\ell+1})^{\top} & -(\bm{R}_{gu}^{\ell+1})^{\top} & \bm{0} & (R^{\psi_{1},\zeta_{1}}(z))^{-1}\bm{I}
\end{array}\right).
\end{equation}
\subsection{Multi-Class setting}

The extension to multiple output classes is straightforward when the number
of classes \(C\) remains \(\mathcal{O}(1)\) as
\[
    D,N,P\to\infty,
    \qquad
    \alpha=\frac PD,
    \qquad
    \nu=\frac ND .
\]
The important point is that the class index is not averaged away in the
proportional limit. Instead, every training-time order parameter becomes a
block kernel indexed by both time and class. Thus the scalar-output DMFT is
lifted to a class-time DMFT by replacing
\[
    t
    \quad\longrightarrow\quad
    (c,t),
    \qquad
    c\in\{1,\ldots,C\}.
\]
Equivalently, every \(T\times T\) correlation or response kernel becomes an
\((CT)\times(CT)\) block kernel.

For each class \(c\in\{1,\ldots,C\}\), introduce a teacher vector
\(\bm w_{\star,c}\), a readout vector \(\bm w_c^L\), an error vector
\(\bm\Delta_c(t)\), and class-resolved training fields
\[
    \bm v_c(t),
    \qquad
    \bm h_c^\ell(t),
    \qquad
    \bm g_c^\ell(t).
\]
The hidden weights are shared across classes, while the readout vectors are
class-specific. Keeping the same normalization as in the scalar-output case,
the finite-rank weight updates become
\begin{align}
    \bm W^0(t)
    &=
    \bm W^0(0)
    +
    \frac{\eta\gamma_0}{\sqrt D}
    \sum_{s<t}
    \sum_{c=1}^{C}
    \bm g_c^1(s)\bm h_c^0(s)^\top,
    \\
    \bm W^\ell(t)
    &=
    \bm W^\ell(0)
    +
    \frac{\eta\gamma_0}{\sqrt N}
    \sum_{s<t}
    \sum_{c=1}^{C}
    \bm g_c^{\ell+1}(s)\bm h_c^\ell(s)^\top,
    \qquad
    1\leq \ell\leq L-1,
    \\
    \bm w_c^L(t)
    &=
    \bm w_c^L(0)
    +
    \eta\gamma_0
    \sum_{s<t}\bm h_c^L(s).
\end{align}
The initialized-weight fields are now class-dependent as well
\begin{align}
    \bm \xi_c^0(t)
    &=
    \frac{\sqrt D}{N\gamma_0}
    \bm W^0(0)^\top \bm g_c^1(t),
    &
    \bm \chi_c^1(t)
    &=
    \frac1{\sqrt D}
    \bm W^0(0)\bm h_c^0(t),
    \\
    \bm \xi_c^\ell(t)
    &=
    \frac1{\sqrt N}
    \bm W^\ell(0)^\top \bm g_c^{\ell+1}(t),
    &
    \bm \chi_c^{\ell+1}(t)
    &=
    \frac1{\sqrt N}
    \bm W^\ell(0)\bm h_c^\ell(t),
    \qquad
    1\leq \ell\leq L-1.
\end{align}
Expanding the finite-rank updates gives
\begin{align}
    \bm v_c(t)
    &=
    \bm w_{\star,c}
    -
    \bm \xi_c^0(t)
    -
    \eta
    \sum_{s<t}
    \sum_{c'=1}^{C}
    C_{g,cc'}^1(t,s)\bm h_{c'}^0(s),
    \\
    \bm h_c^{\ell+1}(t)
    &=
    \bm \chi_c^{\ell+1}(t)
    +
    \eta\gamma_0
    \sum_{s<t}
    \sum_{c'=1}^{C}
    C_{h,cc'}^\ell(t,s)\bm g_{c'}^{\ell+1}(s),
    \\
    \bm g_c^\ell(t)
    &=
    \bm \xi_c^\ell(t)
    +
    \eta\gamma_0
    \sum_{s<t}
    \sum_{c'=1}^{C}
    C_{g,cc'}^{\ell+1}(t,s)\bm h_{c'}^\ell(s).
\end{align}

The only structural change in the full-batch DMFT is that every correlation
and response function acquires two class indices. The class-resolved
correlations are
\begin{align}
    C_{v,cc'}(t,t')
    &=
    \frac1D
    \bm v_c(t)\cdot \bm v_{c'}(t'),
    &
    C_{\Delta,cc'}(t,t')
    &=
    \frac1P
    \bm\Delta_c(t)\cdot \bm\Delta_{c'}(t'),
    \\
    C_{h,cc'}^0(t,t')
    &=
    \frac1D
    \bm h_c^0(t)\cdot \bm h_{c'}^0(t'),
    &
    C_{h,cc'}^\ell(t,t')
    &=
    \frac1N
    \bm h_c^\ell(t)\cdot \bm h_{c'}^\ell(t'),
    \\
    C_{g,cc'}^\ell(t,t')
    &=
    \frac1N
    \bm g_c^\ell(t)\cdot \bm g_{c'}^\ell(t').
\end{align}
The corresponding response functions are
\begin{align}
    R_{\Delta,cc'}(t,t')
    &=
    -\frac{i}{P}
    \bm\Delta_c(t)\cdot\widehat{\bm\Delta}_{c'}(t'),
    &
    R_{vu,cc'}^0(t,t')
    &=
    -\frac{i}{D}
    \bm v_c(t)\cdot\widehat{\bm h}_{c'}^0(t'),
    \\
    R_{hr,cc'}^0(t,t')
    &=
    -\frac{i}{D}
    \bm h_c^0(t)\cdot\widehat{\bm\xi}_{c'}^0(t'),
    &
    R_{gu,cc'}^1(t,t')
    &=
    -\frac{i}{N}
    \bm g_c^1(t)\cdot\widehat{\bm\chi}_{c'}^1(t'),
    \\
    R_{hr,cc'}^\ell(t,t')
    &=
    -\frac{i}{N}
    \bm h_c^\ell(t)\cdot\widehat{\bm\xi}_{c'}^\ell(t'),
    &
    R_{gu,cc'}^{\ell+1}(t,t')
    &=
    -\frac{i}{N}
    \bm g_c^{\ell+1}(t)\cdot\widehat{\bm\chi}_{c'}^{\ell+1}(t').
\end{align}

Equivalently, in the single-site description, the scalar Gaussian noises
become \(C\)-dimensional Gaussian processes over class index. For instance
\begin{align}
    \left\langle
    u_{\Delta,c}(t)u_{\Delta,c'}(t')
    \right\rangle
    &=
    C_{v,cc'}(t,t'),
    &
    \left\langle
    u_c^0(t)u_{c'}^0(t')
    \right\rangle
    &=
    \frac1\alpha C_{\Delta,cc'}(t,t'),
    \\
    \left\langle
    u_c^\ell(t)u_{c'}^\ell(t')
    \right\rangle
    &=
    C_{h,cc'}^{\ell-1}(t,t'),
    &
    \left\langle
    r_c^\ell(t)r_{c'}^\ell(t')
    \right\rangle
    &=
    C_{g,cc'}^{\ell+1}(t,t').
\end{align}
The first-layer backward noise has covariance
\[
    \left\langle
    r_c^0(t)r_{c'}^0(t')
    \right\rangle
    =
    \frac{1}{\nu\gamma_0^2}
    C_{g,cc'}^1(t,t').
\]
The readout boundary becomes
\[
    g_c^L(t)
    =
    r_c^L
    +
    \eta\gamma_0
    \sum_{s<t}h_c^L(s),
    \qquad
    \left\langle r_c^L r_{c'}^L\right\rangle
    =
    \delta_{cc'}.
\]

Thus the training DMFT equations are the scalar-output equations with the
replacement \(t\mapsto(c,t)\), together with sums over repeated class
indices. Explicitly
\begin{align}
    \Delta_c(t)
    &=
    u_{\Delta,c}(t)
    +
    \frac1\alpha
    \sum_{s<t}
    \sum_{c'=1}^{C}
    R_{vu,cc'}^0(t,s)\Delta_{c'}(s)
    +
    \sigma\epsilon_c(t),
    \\
    h_c^0(t)
    &=
    u_c^0(t)
    +
    \sum_{s<t}
    \sum_{c'=1}^{C}
    R_{\Delta,cc'}(t,s)v_{c'}(s),
    \\
    v_c(t)
    &=
    w_{\star,c}
    -
    r_c^0(t)
    -
    \sum_{s<t}
    \sum_{c'=1}^{C}
    \left[
        \frac1{\gamma_0}R_{gu,cc'}^1(t,s)
        +
        \eta C_{g,cc'}^1(t,s)
    \right]
    h_{c'}^0(s),
    \\
    h_c^1(t)
    &=
    u_c^1(t)
    +
    \sum_{s<t}
    \sum_{c'=1}^{C}
    \left[
        \frac1{\nu\gamma_0}R_{hr,cc'}^0(t,s)
        +
        \eta\gamma_0 C_{h,cc'}^0(t,s)
    \right]g_{c'}^1(s),
    \\
    h_c^\ell(t)
    &=
    u_c^\ell(t)
    +
    \sum_{s<t}
    \sum_{c'=1}^{C}
    \left[
        R_{hr,cc'}^{\ell-1}(t,s)
        +
        \eta\gamma_0 C_{h,cc'}^{\ell-1}(t,s)
    \right]g_{c'}^\ell(s),
    \qquad 2\leq \ell\leq L,
    \\
    g_c^\ell(t)
    &=
    r_c^\ell(t)
    +
    \sum_{s<t}
    \sum_{c'=1}^{C}
    \left[
        R_{gu,cc'}^{\ell+1}(t,s)
        +
        \eta\gamma_0 C_{g,cc'}^{\ell+1}(t,s)
    \right]h_{c'}^\ell(s),
    \qquad 1\leq \ell\leq L-1.
\end{align}
The responses are again functional derivatives of the single-site process
\[
    R_{\Delta,cc'}(t,t')
    =
    \left\langle
    \frac{\partial \Delta_c(t)}
    {\partial u_{\Delta,c'}(t')}
    \right\rangle,
    \qquad
    R_{vu,cc'}^0(t,t')
    =
    \left\langle
    \frac{\partial v_c(t)}
    {\partial u_{c'}^0(t')}
    \right\rangle,
\]
\[
    R_{hr,cc'}^\ell(t,t')
    =
    \left\langle
    \frac{\partial h_c^\ell(t)}
    {\partial r_{c'}^\ell(t')}
    \right\rangle,
    \qquad
    R_{gu,cc'}^\ell(t,t')
    =
    \left\langle
    \frac{\partial g_c^\ell(t)}
    {\partial u_{c'}^\ell(t')}
    \right\rangle.
\]
The deterministic self-consistency conditions are
\begin{align}
    C_{v,cc'}(t,t')
    &=
    \left\langle
    v_c(t)v_{c'}(t')
    \right\rangle,
    &
    C_{\Delta,cc'}(t,t')
    &=
    \left\langle
    \Delta_c(t)\Delta_{c'}(t')
    \right\rangle,
    \\
    C_{h,cc'}^\ell(t,t')
    &=
    \left\langle
    h_c^\ell(t)h_{c'}^\ell(t')
    \right\rangle,
    &
    C_{g,cc'}^\ell(t,t')
    &=
    \left\langle
    g_c^\ell(t)g_{c'}^\ell(t')
    \right\rangle.
\end{align}

The spectral calculation changes in the same way. The probe vector itself has
no class index because it probes the shared weight matrix. However, the
finite-rank update directions are now the collection
\[
    \left\{
    \bm h_c^\ell(s),\bm g_c^{\ell+1}(s)
    \right\}_{c=1}^{C,\,s<t}.
\]
For example, for a hidden layer $\ell$
\begin{align}
    \bm\psi_1(\tau)
    &=
    \bm\zeta_1(\tau)
    +
    \eta\gamma_0
    \sum_{s<t}
    \sum_{c=1}^{C}
    \bm g_c^{\ell+1}(s)
    C^{h_c^\ell\psi_0}(s,\tau),
    \\
    \bm\psi_2(\tau)
    &=
    \bm\zeta_2(\tau)
    +
    \eta\gamma_0
    \sum_{s<t}
    \sum_{c=1}^{C}
    \bm h_c^\ell(s)
    C^{g_c^{\ell+1}\psi_1}(s,\tau).
\end{align}
Therefore the pole matrix derived above is unchanged in structure, but its
blocks are enlarged from \(T\times T\) matrices to
\((CT)\times(CT)\) block matrices. If
\[
    I=(c,t),
    \qquad
    J=(c',t'),
\]
we define
\[
    [\mathsf C_h^\ell]_{I,J}
    =
    C_{h,cc'}^\ell(t,t'),
    \qquad
    [\mathsf C_g^{\ell+1}]_{I,J}
    =
    C_{g,cc'}^{\ell+1}(t,t'),
\]
and similarly for
\[
    \mathsf R_{hr}^\ell,
    \qquad
    \mathsf R_{gu}^{\ell+1}.
\]
Then the hidden-layer spectral condition is obtained by the substitutions
\[
    \bm C_h^\ell\to \mathsf C_h^\ell,
    \qquad
    \bm C_g^{\ell+1}\to \mathsf C_g^{\ell+1},
    \qquad
    \bm R_{hr}^{\ell}\to \mathsf R_{hr}^{\ell},
    \qquad
    \bm R_{gu}^{\ell+1}\to \mathsf R_{gu}^{\ell+1},
    \qquad
    \bm I_T\to \bm I_{CT}.
\]
Thus
\begin{equation}
    \det \bm{\mathcal A}_\ell^{\rm multi-class}(z)=0
\end{equation}
has exactly the same form as in the scalar-output case.

If the teacher/readout initialization is class-symmetric and the targets are
orthogonal or independent across classes, the saddle can often be restricted
to a class-diagonal or permutation-symmetric ansatz. In the most general case,
however, one should keep all cross-class order parameters.
\subsection{Online Dynamics}
The full-batch calculation above assumes that the same dataset
\(\bm X\in\mathbb R^{P\times D}\) is reused at every training step. In online
SGD, this data block is replaced by a fresh mini-batch
\(\bm X(t)\in\mathbb R^{B\times D}\) at each step \(t\), with
\[
    B=\alpha_B D,
    \qquad
    D,N,B\to\infty,
    \qquad
    \nu=\frac ND,
    \qquad
    \alpha_B=\frac BD .
\]
The input-layer data fields are now
\begin{equation}
    \bm h^0(t)
    =
    \frac{\sqrt D}{B}\bm X(t)^\top \bm\Delta(t),
    \qquad
    \bm\Delta(t)
    =
    \frac1{\sqrt D}\bm X(t)\bm v(t)
    +
    \sigma \bm\epsilon(t),
\end{equation}
where \(\bm X(t)\) and \(\bm\epsilon(t)\) are resampled independently at
each training step. All weight-dependent fields and all finite-rank weight
updates are unchanged.

The only modification to the DMFT is therefore in the data average. Since
the data are fresh at each time step, the average over \(\bm X(t)\) factorizes
over \(t\). Thus the nonlocal data-response sector of the full-batch theory is
removed. Equivalently, in the full-batch data block one makes the replacement
\begin{equation}
    R_\Delta(t,t')
    \longrightarrow
    \delta_{tt'},
\end{equation}
and the error covariance becomes
\begin{equation}
    C_\Delta(t,t')
    =
    C_v(t,t')
    +
    \sigma^2\delta_{tt'}.
\end{equation}
The stochastic input-layer process is therefore
\begin{equation}
    \Delta(t)
    =
    u_\Delta(t)
    +
    \sigma\epsilon(t),
    \qquad
    u_\Delta\sim \mathrm{GP}(0,C_v),
    \qquad
    \langle \epsilon(t)\epsilon(t')\rangle=\delta_{tt'},
\end{equation}
and
\begin{equation}
    h^0(t)
    =
    v(t)
    +
    u^0(t),
    \qquad
    u^0\sim
    \mathrm{GP}
    \left(
        0,
        \frac1{\alpha_B}
        C_\Delta(t,t')\delta_{tt'}
    \right).
\end{equation}
In particular, the full-batch memory term
\[
    \frac1\alpha
    \sum_{t'<t}
    R_{vu}^0(t,t')\Delta(t')
\]
in the \(\Delta\)-equation is absent in online SGD. The finite-batch effect
instead appears as a time-local stochastic-gradient noise in \(h^0\). On the
diagonal,
\begin{equation}
    \langle h^0(t)^2\rangle
    =
    C_v(t,t)
    +
    \frac1{\alpha_B}C_\Delta(t,t),
\end{equation}
so smaller batches increase the effective input noise.

The remaining single-site training equations are unchanged, except that they
are driven by the online-SGD input process above. Thus
\begin{equation}
    v(t)
    =
    w_\star
    -
    r^0(t)
    -
    \sum_{s<t}
    \left[
        \frac1{\gamma_0}R_{gu}^{1}(t,s)
        +
        \eta C_g^1(t,s)
    \right]
    h^0(s),
    \qquad
    r^0\sim
    \mathrm{GP}
    \left(
        0,
        \frac1{\nu\gamma_0^2}C_g^1
    \right),
\end{equation}
\begin{equation}
    h^1(t)
    =
    u^1(t)
    +
    \sum_{s<t}
    \left[
        \frac1{\nu\gamma_0}R_{hr}^{0}(t,s)
        +
        \eta\gamma_0 C_h^0(t,s)
    \right]
    g^1(s),
    \qquad
    u^1\sim \mathrm{GP}(0,C_h^0),
\end{equation}
and, for \(1<\ell\leq L\),
\begin{equation}
    h^\ell(t)
    =
    u^\ell(t)
    +
    \sum_{s<t}
    \left[
        R_{hr}^{\ell-1}(t,s)
        +
        \eta\gamma_0 C_h^{\ell-1}(t,s)
    \right]
    g^\ell(s),
    \qquad
    u^\ell\sim \mathrm{GP}(0,C_h^{\ell-1}).
\end{equation}
Similarly, for \(1\leq \ell\leq L-1\),
\begin{equation}
    g^\ell(t)
    =
    r^\ell(t)
    +
    \sum_{s<t}
    \left[
        R_{gu}^{\ell+1}(t,s)
        +
        \eta\gamma_0 C_g^{\ell+1}(t,s)
    \right]
    h^\ell(s),
    \qquad
    r^\ell\sim \mathrm{GP}(0,C_g^{\ell+1}),
\end{equation}
with the same readout boundary condition
\begin{equation}
    g^L(t)
    =
    r^L
    +
    \eta\gamma_0
    \sum_{s<t}h^L(s),
    \qquad
    r^L\sim\mathcal N(0,1).
\end{equation}

The self-consistency equations for the correlations and responses are also
unchanged in form:
\begin{align}
    C_v(t,t')
    &=
    \langle v(t)v(t')\rangle,
    &
    C_\Delta(t,t')
    &=
    \langle \Delta(t)\Delta(t')\rangle,
    \\
    C_h^\ell(t,t')
    &=
    \langle h^\ell(t)h^\ell(t')\rangle,
    &
    C_g^\ell(t,t')
    &=
    \langle g^\ell(t)g^\ell(t')\rangle,
\end{align}
and
\begin{align}
    R_{vu}^0(t,t')
    &=
    \left\langle
    \frac{\partial v(t)}{\partial u^0(t')}
    \right\rangle,
    &
    R_{hr}^\ell(t,t')
    &=
    \left\langle
    \frac{\partial h^\ell(t)}{\partial r^\ell(t')}
    \right\rangle,
    \\
    R_{gu}^\ell(t,t')
    &=
    \left\langle
    \frac{\partial g^\ell(t)}{\partial u^\ell(t')}
    \right\rangle.
\end{align}
Finally, the spectral-probe DMFT derived above is not modified at the level of
the weight averages. The probe equations and the pole conditions
\[
    \det \bm{\mathcal A}_0(z)=0,
    \qquad
    \det \bm{\mathcal A}_\ell(z)=0
\]
retain the same form. The only change is that the training order parameters
\[
    C_h^\ell,\quad C_g^\ell,\quad R_{hr}^\ell,\quad R_{gu}^\ell
\]
that enter the probe fixed-point equations must now be computed from the
online-SGD training DMFT rather than from the full-batch DMFT.
\subsection{NTK spectrum}

\begin{figure}
    \centering
    \includegraphics[width=0.45\linewidth]{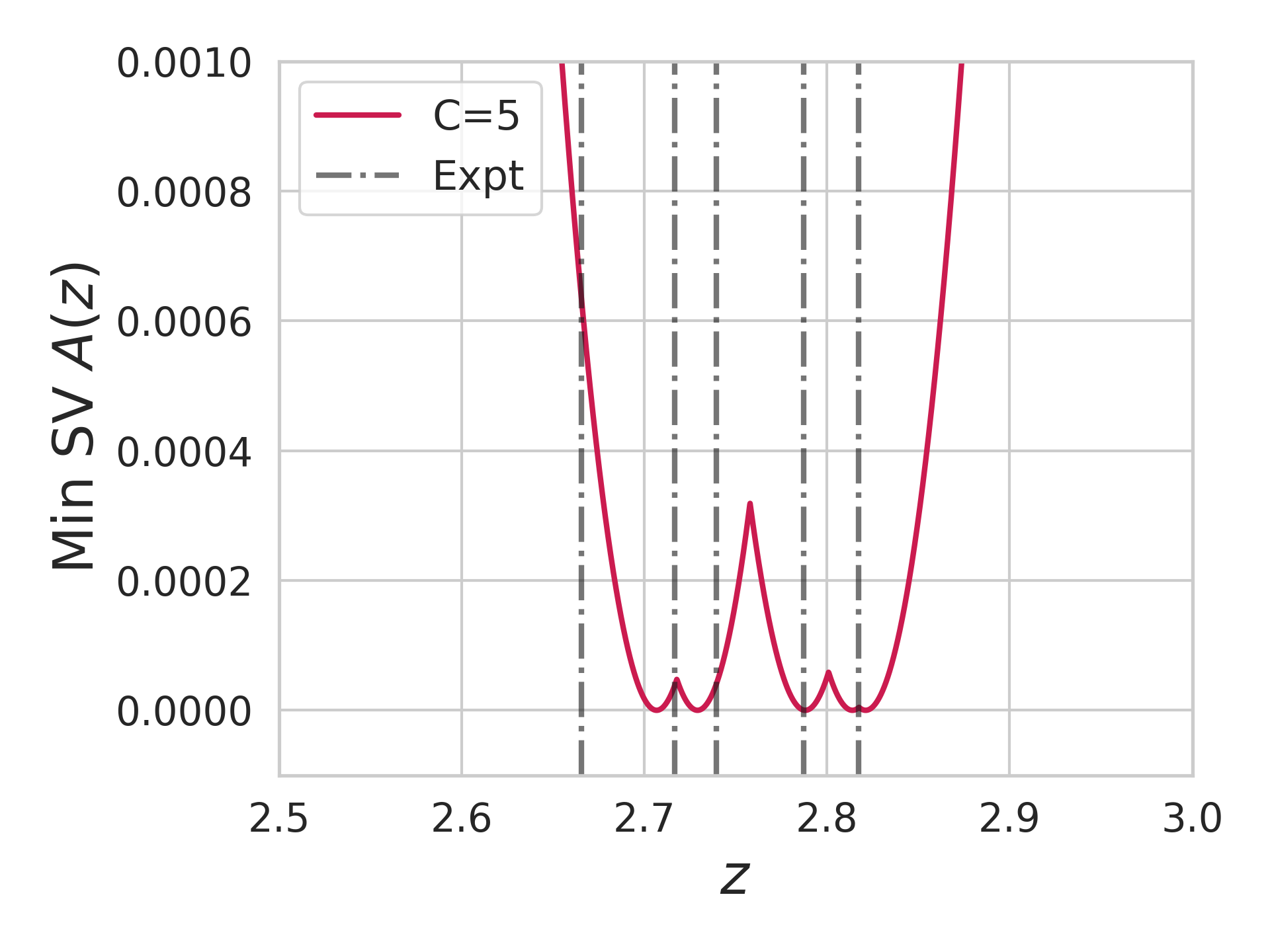}
    \includegraphics[width=0.45\linewidth]{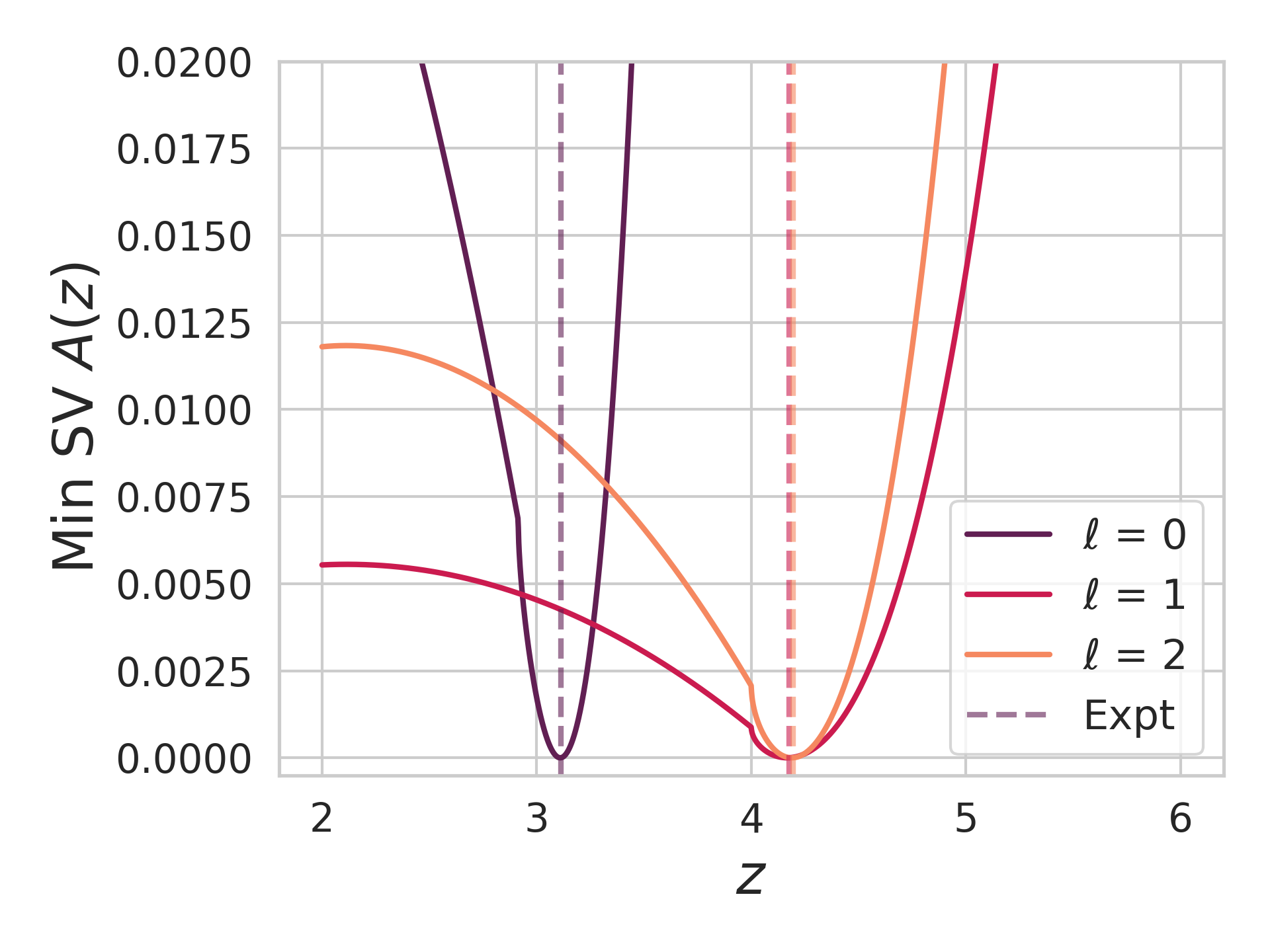}
    \caption{Root-finding procedure for $\det \bm A(z)=0$ in a multiclass model ($C=5$). We compute the minimum singular value of $\bm A(z)$ as a function of $z$; values where it approaches zero identify predicted spectral outliers. The presence of multiple minima reflects the rank-$C$ structure of the problem. Dashed lines indicate empirical outlier eigenvalues of $\bm W^\top \bm W/N$.}
    \label{fig:placeholder}
\end{figure}

For a two-layer linear network, i.e. \(L=1\), the model is
\begin{equation}
    f_t(\bm x)
    =
    \frac{1}{N\gamma_0}
    \bm w(t)^\top \bm W(t)\bm x ,
\end{equation}
where \(\bm W(t)\equiv \bm W^0(t)\in\mathbb R^{N\times D}\) and
\(\bm w(t)\equiv \bm w^1(t)\in\mathbb R^N\). In this case the backward
field is simply
\begin{equation}
    \bm g^1(t)=\bm w(t).
\end{equation}
The population NTK can be written as a kernel on input space. Indeed,
\begin{equation}
    \frac{\partial f_t(\bm x)}{\partial \bm w}
    =
    \frac{1}{N\gamma_0}\bm W(t)\bm x,
    \qquad
    \frac{\partial f_t(\bm x)}{\partial W_{ij}}
    =
    \frac{1}{N\gamma_0}w_i(t)x_j .
\end{equation}
Therefore, up to the overall convention-dependent prefactor coming from the
learning-rate normalization, the NTK has the \begin{equation}
    \Theta_t(\bm x,\bm x')
    =
    \frac{1}{D}
    \bm x^\top \bm B_t \bm x',
\end{equation}
where the \(D\times D\) population NTK core is
\begin{equation}
    \bm B_t
    =
    \frac{1}{N}\bm W(t)^\top\bm W(t)
    +
    q_g(t)\bm I_D
\end{equation}
with
\begin{equation}
     q_g(t)
    =
    \frac{1}{N}\|\bm w(t)\|^2
    =
    C_g^1(t,t)
\end{equation}
in the usual notation. Thus, in the two-layer case, the NTK core is exactly the first-layer weight covariance plus an isotropic readout-gradient contribution. The scalar \(q_g(t)\) is already one of the training DMFT order
parameters, since \(C_g^1(t,t)=N^{-1}\bm g^1(t)\cdot \bm g^1(t)\).

Consequently, the spectrum of the population NTK core is just a shift of the
spectrum of the first-layer weight covariance
\begin{equation}
    \bm M^0(t)
    =
    \frac{1}{N}\bm W(t)^\top\bm W(t).
\end{equation}
If
\begin{equation}
    \lambda_1^M(t),\ldots,\lambda_D^M(t)
\end{equation}
are the eigenvalues of \(\bm M^0(t)\), then the eigenvalues of \(\bm B_t\) are
\begin{equation}
    \lambda_a^{\mathrm{NTK}}(t)
    =
    \lambda_a^M(t)+q_g(t),
    \qquad a=1,\ldots,D.
\end{equation}
Equivalently, if \(m_M(z;t)\) is the Stieltjes transform of the spectral
measure of \(\bm M^0(t)\), then the NTK-core resolvent is
\begin{equation}
    m_{\mathrm{NTK}}(z;t)
    =
    m_M(z-q_g(t);t).
\end{equation}
Thus the NTK spectral density satisfies
\begin{equation}
    \rho_{\mathrm{NTK}}(\lambda;t)
    =
    \rho_M(\lambda-q_g(t);t).
\end{equation}
Notice that computing the spectrum of \(\bm K_t\) would require in principle
an additional probe involving the data matrix \(\bm X\). The simple shift
relation above applies to the population limit only.
\begin{figure}
    \centering
    \includegraphics[width=0.5\linewidth]{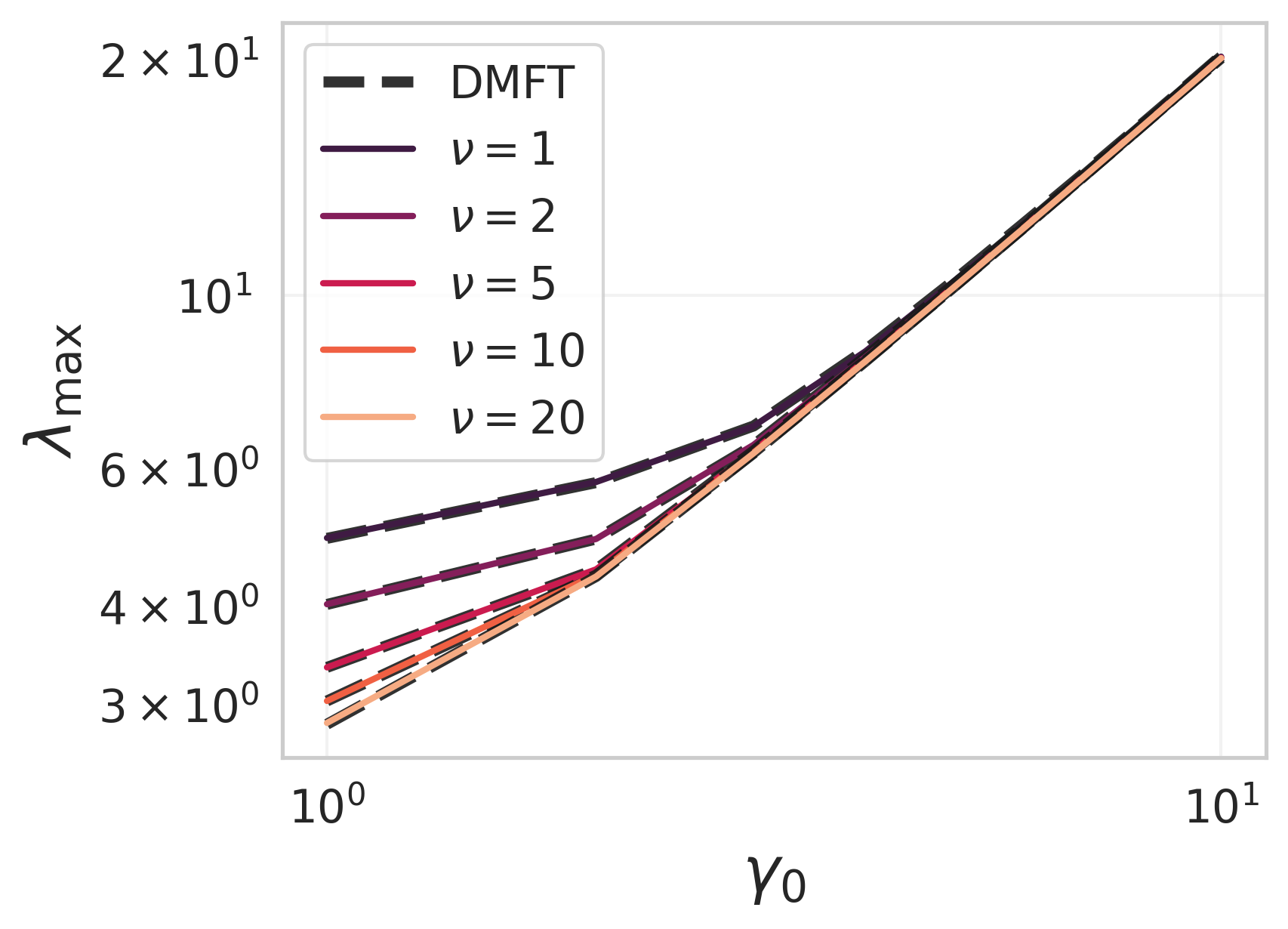}
    \caption{Leading NTK eigenvalue $\lambda_{\text{max}}$ as a function of richness $\gamma_0$ for different aspect rations $\nu$. As $\gamma_0 $ increases, $\lambda_{\text{max}}$ rapidly grows and the curves collapse, indicating fast convergence to a width-stable sharpening regime.}
    \label{fig:placeholder}
\end{figure}
\section{Large Output Channels}\label{app:extensive_outputs}
Consider a linear network with input dimension $D$, width $N$, and $C$ output channels. Let
$\bm X\in\mathbb R^{D\times P}$ be the data matrix, $\bm Y\in\mathbb R^{C\times P}$ the labels,
$\bm W\in\mathbb R^{N\times D}$ the hidden weight matrix, and $\bm A\in\mathbb R^{N\times C}$ the readout. We study the spectrum of the weight covariance
\begin{equation}
    \bm M
    =
    \frac{1}{N}\bm W^\top \bm W
    \in \mathbb R^{D\times D}
\end{equation}
in a proportional limit where $D,N,C\to\infty$ with fixed ratios $\nu = \frac{N}{D}$ and $\chi = \frac{C}{D}$. The network output and teacher are respectively
\begin{equation}
    \bm F
    =
    \frac{1}{\gamma_0 N}
    \bm A^\top \bm W \bm X,
    \qquad
    \bm Y
    =
    \frac{1}{\sqrt D}
    \bm B^\star \bm X,
\end{equation}
being $\bm B^\star\in\mathbb R^{C\times D}$ the target matrix. We are interested in the quadratic loss $ \mathcal L= \frac{1}{2CP}\|\bm F-\bm Y\|_F^2$ in the population limit $P\to\infty$ where $\frac{1}{P}\bm X\bm X^\top\to \bm I_D$. One gradient step from random initialization produces an update of the form
\begin{equation}
    \bm W(1)
    =
    \bm W(0)
    +
    \frac{\eta\gamma_0}{\sqrt{C}}
    \bm A\bm B^\star .
\end{equation}
Thus
\begin{equation}
    \bm M
    =
    \frac{1}{N}
    \left(
    \bm W(0)
    +
    \frac{\eta\gamma_0}{\sqrt{C}}
    \bm A\bm B^\star
    \right)^\top
    \left(
    \bm W(0)
    +
    \frac{\eta\gamma_0}{\sqrt{C}}
    \bm A\bm B^\star
    \right).
\end{equation}
At initialization, $\bm W(0)$, $\bm A$, and $\bm B^\star$ are independent random matrices, which makes the model tractable for the extensive output regime. 

To study the spectrum of $\bm M$, we introduce an auxiliary linear response problem driven by a source $\bm j(t)$
\begin{equation}
    \frac{d\bm v(t)}{dt}
    =
    -\bm M \bm v(t)
    +
    \bm j(t).
\end{equation}
The reason for introducing this dynamics is that its response function is exactly the resolvent of $\bm M$. Indeed, the causal solution is $\bm v(t)=\int_{-\infty}^{t} ds\,e^{-\bm M(t-s)}\bm j(s)$, so that $ \frac{\delta \bm v(t+\tau)}{\delta \bm j(t)}= \Theta(\tau)e^{-\bm M\tau}.$
Taking the Fourier transform of the normalized trace of this response gives
\begin{align}
    \mathcal H(\omega)
    &=
    \frac{1}{D}
    \text{Tr}
    \int_{0}^{\infty} d\tau\,
    e^{-(i\omega \bm I_D+\bm M)\tau}
    \nonumber \\
    &=
    \frac{1}{D}
    \text{Tr}
    \left(
    i\omega \bm I_D+\bm M
    \right)^{-1}.
\end{align}
Thus the spectral density of $\bm M$ can be recovered from the analytic continuation of $\mathcal H(\omega)$.

We implement this response calculation using a Martin--Siggia--Rose generating functional (MGF)
\begin{equation}
    \mathcal Z[\bm j]
    =
    \int D\bm v^0 D\hat{\bm v}^{0}
    \exp\left[
    i\int dt\,
    \hat{\bm v}^{0}(t)^\top
    \left(
    \partial_t \bm v^0(t)
    +
    \bm M\bm v^0(t)
    -
    \bm j(t)
    \right)
    \right].
\end{equation}
The remaining task is to average this generating functional over the random matrices entering $\bm M$. To expose the random-matrix structure, we represent the action of $\bm M$ on $\bm v^0(t)$ through the auxiliary fields
\begin{align}
    \bm v^1(t)
    &=
    \frac{1}{\sqrt D}\bm B\bm v^0(t),
    \\
    \bm v^2(t)
    &=
    \frac{1}{\sqrt D}\bm W(0)\bm v^0(t)
    +
    \frac{\gamma}{\sqrt C}\bm A\bm v^1(t),
    \\
    \bm v^3(t)
    &=
    \frac{\sqrt C}{N}\bm A^\top\bm v^2(t),
    \\
    \bm v^4(t)
    &=
    \frac{\sqrt D}{N}\bm W(0)^\top\bm v^2(t)
    +
    \gamma\frac{\sqrt D}{C}\bm B^\top\bm v^3(t).
\end{align}
With these definitions,
\begin{equation}
    \bm v^4(t)
    =
    \bm M\bm v^0(t)
\end{equation}
and resolvent is obtained from the response of the linear system
\begin{equation}
    \partial_t \bm v^0(t)
    =
    -\bm v^4(t)
    +
    \bm j(t).
\end{equation}

We enforce the above using the response fields $\hat{\bm v}^i$ through the integral representation of the Dirac delta function, e.g.
\begin{equation}
    1
    \equiv
    \int D\bm v^1 D\hat{\bm v}^1
    \exp\left[
    i\int dt\,
    \hat{\bm v}^1(t)^\top
    \left(
    \bm v^1(t)
    -
    \frac{1}{\sqrt D}\bm B\bm v^0(t)
    \right)
    \right].
\end{equation}
Since $\bm W(0)$, $\bm A$, and $\bm B$ are independent Gaussian matrices, their averages generate only quadratic contractions between the dynamical fields. The resulting effective theory depends on the correlation order parameters
\begin{align}
    C_{00}(t,t')
    &=
    \frac{1}{D}
    \bm v^0(t)\cdot \bm v^0(t'),
    &
    C_{11}(t,t')
    &=
    \frac{1}{C}
    \bm v^1(t)\cdot \bm v^1(t'),
    \\
    C_{22}(t,t')
    &=
    \frac{1}{N}
    \bm v^2(t)\cdot \bm v^2(t'),
    &
    C_{33}(t,t')
    &=
    \frac{1}{C}
    \bm v^3(t)\cdot \bm v^3(t'),
\end{align}
and response order parameters
\begin{align}
    R_{04}(t,t')
    &=
    -\frac{i}{D}
    \bm v^0(t)\cdot \hat{\bm v}^4(t'),
    &
    R_{13}(t,t')
    &=
    -\frac{i}{C}
    \bm v^1(t)\cdot \hat{\bm v}^3(t'),
    \\
    R_{22}(t,t')
    &=
    -\frac{i}{N}
    \bm v^2(t)\cdot \hat{\bm v}^2(t'),
    &
    R_{31}(t,t')
    &=
    -\frac{i}{C}
    \bm v^3(t)\cdot \hat{\bm v}^1(t')
\end{align}
which converge to deterministic quantities in the high-dimensional limit.
\subsection{DMFT action}
At large $D,N,C$, with $\nu=N/D$ and $\chi=C/D$ fixed, the disorder-averaged generating functional factorizes into independent single-site problems for the $D$-dimensional sector $(\bm v^0,\bm v^4)$, the $N$-dimensional sector $\bm v^2$, and the $C$-dimensional sector $(\bm v^1,\bm v^3)$
\begin{equation}
    \mathcal Z[\bm j]
    =
    \int d\mathcal Q\,
    \exp\left\{
    D S_{\rm eff}[\mathcal Q;\bm j]
    \right\},
\end{equation}
where
\begin{equation}
    \mathcal Q
    =
    \{C_{00},C_{11},C_{22},C_{33},R_{04},R_{13},R_{22},R_{31}\},
\end{equation}
while the DMFT action can be compressed as
\begin{equation}
    S_{\rm eff}
    =
    S_{\rm op}[\mathcal Q]
    +
    \log Z_{04}[\bm j]
    +
    \nu \log Z_2
    +
    \chi \log Z_{13}.
\end{equation}
Here, $S_{\rm op}$ collects the order-parameter terms generated by the Gaussian averages, while $Z_{04}$, $Z_2$, and $Z_{13}$ are the corresponding single-site path integrals. The response-sector saddle point equations from $\frac{\partial S_{\rm eff}}{\partial R}=0$ are
\begin{align}
    \hat R_{04}
    &=
    R_{22}
    +
    \gamma R_{31}
    &
    \hat R_{13}
    &=
    \gamma R_{22}
    \\
    \hat R_{22}
    &=
    \frac{1}{\nu}R_{04}
    +
    \gamma\frac{\chi}{\nu}R_{13}
    &
    \hat R_{31}
    &=
    \frac{\gamma}{\chi}R_{04}.
\end{align}
Substituting these back into the effective action gives three coupled single-site theories. Specifically, the $D$-sector, which contains the resolvent response, is
\begin{equation}
    Z_{04}[\bm j]
    =
    \int
    Dv^0D\hat v^0Dv^4D\hat v^4
    \exp\left\{
    S_{04}[\bm j]
    \right\},
\end{equation}
with
\begin{align}
    S_{04}[\bm j]
    &=
    i\int dt\,
    \hat v^0(t)
    \left(
    \partial_t v^0(t)
    +
    v^4(t)
    -
    j(t)
    \right)
    -
    \frac{1}{2}
    \int dtdt'\,
    \hat C_{00}(t,t')v^0(t)v^0(t')
    \nonumber \\
    &\quad
    -
    \frac{1}{2}
    \int dtdt'\,
    \left[
    \frac{1}{\nu}C_{22}(t,t')
    +
    \frac{\gamma^2}{\chi}C_{33}(t,t')
    \right]
    \hat v^4(t)\hat v^4(t')
    \nonumber \\
    &\quad
    -
    i\int dtdt'\,
    \left[
    R_{22}(t,t')
    +
    \gamma R_{31}(t,t')
    \right]
    v^0(t)\hat v^4(t') +i \int dt v_{4}(t)\hat{v}_{4}(t).
\end{align}

The $N$-sector is
\begin{equation}
    Z_2
    =
    \int
    Dv^2D\hat v^2
    \exp\left\{
    S_2
    \right\},
\end{equation}
with
\begin{align}
    S_2
    &=
    -
    \frac{1}{2}
    \int dtdt'\,
    \hat C_{22}(t,t')v^2(t)v^2(t')
    -
    \frac{1}{2}
    \int dtdt'\,
    \left[
    C_{00}(t,t')
    +
    \gamma^2 C_{11}(t,t')
    \right]
    \hat v^2(t)\hat v^2(t')
    \nonumber \\
    &\quad
    -
    i\int dtdt'\,
    \left[
    \frac{1}{\nu}R_{04}(t,t')
    +
    \gamma\frac{\chi}{\nu}R_{13}(t,t')
    \right]
    v^2(t)\hat v^2(t')
    +
    i\int dt\,v^2(t)\hat v^2(t).
\end{align}

Finally, the $C$-sector is
\begin{equation}
    Z_{13}
    =
    \int
    Dv^1D\hat v^1Dv^3D\hat v^3
    \exp\left\{
    S_{13}
    \right\},
\end{equation}
with
\begin{align}
    S_{13}
    &=
    -
    \frac{1}{2}
    \int dtdt'\,
    \hat C_{11}(t,t')v^1(t)v^1(t')
    -
    \frac{1}{2}
    \int dtdt'\,
    \hat C_{33}(t,t')v^3(t)v^3(t')
    \nonumber \\
    &\quad
    -
    \frac{1}{2}
    \frac{\chi}{\nu}
    \int dtdt'\,
    C_{22}(t,t')\hat v^3(t)\hat v^3(t')
    -
    \frac{1}{2}
    \int dtdt'\,
    C_{00}(t,t')\hat v^1(t)\hat v^1(t')
    \nonumber \\
    &\quad
    -
    i\gamma
    \int dtdt'\,
    R_{22}(t,t')v^1(t)\hat v^3(t')
    -
    i\frac{\gamma}{\chi}
    \int dtdt'\,
    R_{04}(t,t')v^3(t)\hat v^1(t')
    \nonumber \\
    &\quad
    +
    i\int dt\,v^1(t)\hat v^1(t)
    +
    i\int dt\,v^3(t)\hat v^3(t).
\end{align}
\subsubsection{Saddle point equations}
The saddle point of the effective action identifies the order parameters with the corresponding single-site correlations and responses. We start by differentiating with respect to the conjugate order parameters for the correlation functions which give
\begin{align}
    C_{00}(t,t')
    &=
    \left\langle
    v^0(t)v^0(t')
    \right\rangle_{04}
    &
    C_{11}(t,t')
    &=
    \left\langle
    v^1(t)v^1(t')
    \right\rangle_{13}
    \\
    C_{22}(t,t')
    &=
    \left\langle
    v^2(t)v^2(t')
    \right\rangle_{2}
    &
    C_{33}(t,t')
    &=
    \left\langle
    v^3(t)v^3(t')
    \right\rangle_{13}.
\end{align}
while the response functions are
\begin{align}
    R_{04}(t,t')
    &=
    \left\langle
    \frac{\delta v^0(t)}{\delta u_4(t')}
    \right\rangle_{04}
    &
    R_{22}(t,t')
    &=
    \left\langle
    \frac{\delta v^2(t)}{\delta u_2(t')}
    \right\rangle_{2}
    \\
    R_{13}(t,t')
    &=
    \left\langle
    \frac{\delta v^1(t)}{\delta u_3(t')}
    \right\rangle_{13}
    &
    R_{31}(t,t')
    &=
    \left\langle
    \frac{\delta v^3(t)}{\delta u_1(t')}
    \right\rangle_{13}.
\end{align}
For causality conditions, the conjugate correlation fields vanish at the saddle
\begin{equation}
    \hat C_{00}
    =
    \hat C_{11}
    =
    \hat C_{22}
    =
    \hat C_{33}
    =
    0.
\end{equation}

The final DMFT equations are then 
\begin{equation}
    v^2(t)
    =
    u_2(t)
    +
    \sum_{t'}
    \left[
    \frac{1}{\nu}R_{04}(t,t')
    +
    \gamma\frac{\chi}{\nu}R_{13}(t,t')
    \right]
    v^2(t')
\end{equation}
where
\begin{equation}
    u_2
    \sim
    \mathcal N
    \left(
    0,\,
    C_{00}
    +
    \gamma^2 C_{11}
    \right).
\end{equation}
The $C$-sector gives the coupled equations
\begin{align}
    v^1(t)
    &=
    u_1(t)
    +
    \frac{\gamma}{\chi}
    \sum_{t'}
    R_{04}(t,t')v^3(t')
    \\
    v^3(t)
    &=
    u_3(t)
    +
    \gamma
    \sum_{t'}
    R_{22}(t,t')v^1(t')
\end{align}
where
\begin{equation}
    u_1
    \sim
    \mathcal N(0,C_{00})
    \qquad
    u_3
    \sim
    \mathcal N
    \left(
    0,\,
    \frac{\chi}{\nu}C_{22}
    \right)
\end{equation}
and the $D$-sector gives the resolvent probe
\begin{align}
    v^4(t)
    &=
    u_4(t)
    +
    \sum_{t'}
    \left[
    R_{22}(t,t')
    +
    \gamma R_{31}(t,t')
    \right]v^0(t')
    \\
    \partial_t v^0(t)
    &=
    -
    u_4(t)
    -
    \sum_{t'}
    \left[
    R_{22}(t,t')
    +
    \gamma R_{31}(t,t')
    \right]v^0(t')
    +
    j(t)
\end{align}
where
\begin{equation}
    u_4
    \sim
    \mathcal N
    \left(
    0,\,
    \frac{1}{\nu}C_{22}
    +
    \frac{\gamma^2}{\chi}C_{33}
    \right).
\end{equation}
These equations form the DMFT closure: the correlation functions are computed from the stochastic processes above, while the response functions are obtained by differentiating the same processes with respect to their Gaussian noises.

Since the linear system is time-translation-invariant (T.T.I.), in Fourier space 
\begin{align}
    &\int dt\,e^{-i\omega t}\Bigg(\frac{\partial}{\partial t'}v^{0}(t')+u_{4}(t)+\sum_{t'}\Big(R_{2,2}(t,t')+\gamma R_{3,1}(t,t')\Big)v^{0}(t')-j(t)\Bigg)
    \\&\Rightarrow\Big(i\omega+R_{2,2}(\omega)+\gamma R_{3,1}(\omega)\Big)v_{0}(\omega)+u_{4}(\omega)-j(\omega)=0\\&\Rightarrow\mathcal{H}(\omega)=\frac{1}{i\omega+R_{2,2}(\omega)+\gamma R_{3,1}(\omega)}
\end{align}
and since $R_{04}$ is the response of $v^0$ to $u_4$, one has
\begin{equation}
    R_{04}(\omega)
    =
    -
    \mathcal H(\omega).
\end{equation}
The remaining response functions can be solved algebraically from the single-site equations. Defining
\begin{equation}
    \Delta(\omega)
    =
    1
    +
    \frac{\gamma^2}{\chi}
    \mathcal H(\omega)R_{22}(\omega)
\end{equation}
we obtain
\begin{align}
    R_{31}(\omega)
    &=
    \frac{
    \gamma R_{22}(\omega)
    }{
    \Delta(\omega)
    }
    \\
    R_{13}(\omega)
    &=
    -
    \frac{\gamma}{\chi}
    \frac{
    \mathcal H(\omega)
    }{
    \Delta(\omega)
    }
    \\
    R_{22}(\omega)
    &=
    \left[
    1
    +
    \frac{1}{\nu}
    \mathcal H(\omega)
    +
    \frac{\gamma^2}{\nu}
    \frac{
    \mathcal H(\omega)
    }{
    \Delta(\omega)
    }
    \right]^{-1}.
\end{align}
Together with
\begin{equation}
    \mathcal H(\omega)
    =
    \left[
    i\omega
    +
    R_{22}(\omega)
    +
    \frac{
    \gamma^2R_{22}(\omega)
    }{
    \Delta(\omega)
    }
    \right]^{-1}
\end{equation}
these equations give a closed self-consistency problem for the resolvent.

The spectral density of $\bm M$ is obtained by analytic continuation,
\begin{equation}
    \rho(\lambda)
    =
    \frac{1}{\pi}
    \lim_{\epsilon\to0^+}
    \operatorname{Im}
    \mathcal H(\omega=i\lambda-\epsilon).
\end{equation}
This set of equations shows explicitly why the extensive-output regime differs from the finite-rank spike setting: when $\chi=\mathcal{O}(1)$, the readout-induced update contributes at the level of the full single-site, modifying the resolvent self-consistently and restructuring the bulk spectrum rather than only producing isolated outliers.
\subsection{Edge of the Bulk}
Spectral edges correspond to poles of the Stieltjes transform. Suppose we have $G(z) = \text{tr}\left[ z- \bm M\right]^{-1} = \frac{1}{z-R(G)}$. We can re-express this as
\begin{equation}
    z = \frac{1}{G}+ R(G).
\end{equation}
The derivative $\frac{dz}{dG} = \Big(\frac{dG}{dz}\Big)^{-1}$ should be zero at the spectral edge since $\frac{dG}{dz}$ diverges there. 
In our notation, the resolvent is
\begin{equation}
    \mathcal H(z)
    =
    \left[
    z
    +
    R_{22}(z)
    +
    \frac{\gamma^2 R_{22}(z)}
    {
    1+\frac{\gamma^2}{\chi}\mathcal H(z)R_{22}(z)
    }
    \right]^{-1}
\end{equation}
together with
\begin{equation}
    R_{22}(z)
    =
    \left[
    1
    +
    \frac{1}{\nu}\mathcal H(z)
    +
    \frac{\gamma^2}{\nu}
    \frac{\mathcal H(z)}
    {
    1+\frac{\gamma^2}{\chi}\mathcal H(z)R_{22}(z)
    }
    \right]^{-1}.
\end{equation}
In general, the bulk edge is obtained by solving these equations together with the condition that the Jacobian of the map from $(\mathcal H,R_{22})$ to $z$ becomes singular. A simple closed form is obtained in the limit $\nu\to\infty$. In this case $ R_{22}(z)\to 1$,
and the resolvent equation reduces to
\begin{equation}
    \mathcal H(z)
    =
    \left[
    z
    +
    1
    +
    \frac{\gamma^2}
    {
    1+\frac{\gamma^2}{\chi}\mathcal H(z)
    }
    \right]^{-1}.
\end{equation}
Equivalently, using the physical spectral variable $\lambda=-z$,
\begin{equation}
    \lambda(\mathcal H)
    =
    -\frac{1}{\mathcal H}
    +
    1
    +
    \frac{\gamma^2}
    {
    1+\frac{\gamma^2}{\chi}\mathcal H
    }.
\end{equation}
The bulk edges are therefore determined by
\begin{equation}
    \frac{d\lambda}{d\mathcal H}
    =
    \frac{1}{\mathcal H^2}
    -
    \frac{\gamma^4/\chi}
    {
    \left(
    1+\frac{\gamma^2}{\chi}\mathcal H
    \right)^2
    }
    =
    0.
\end{equation}
Solving this equation and substituting back into $\lambda(\mathcal H)$ gives a shifted Marchenko--Pastur bulk with upper edge
\begin{equation}
    \lambda_{\rm edge}
    =
    1
    +
    \gamma^2
    \left(
    1
    +
    \chi^{-1/2}
    \right)^2 .
\end{equation}
This is consistent with the intuition that, when $N/D\to\infty$, the initialization covariance concentrates as
\begin{equation}
    \frac{1}{N}\bm W(0)^\top \bm W(0)\to \bm I_D,
\end{equation}
while the learned extensive-output term contributes a Wishart-like component proportional to $\bm B^\top \bm B/C$.
\subsubsection{Expansion in the $\nu \to \infty$ limit}
\begin{figure}
    \centering
    \includegraphics[width=0.45\linewidth]{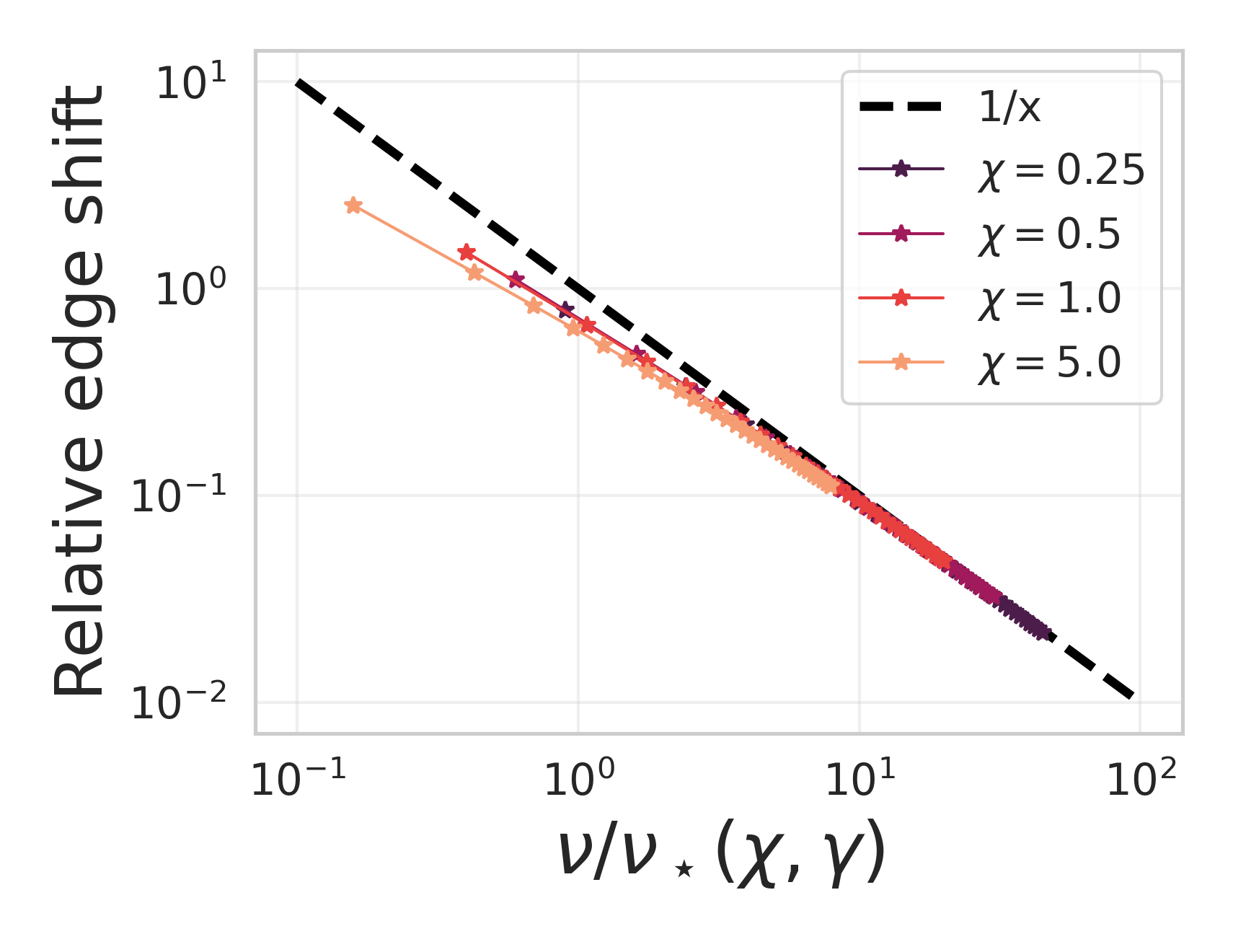}
    \caption{Scaling collapse of the finite-width correction to the upper bulk edge. The relative edge shift $(\lambda_+(\nu)-\lambda_+^\infty)/\lambda_+^\infty$ is plotted against the rescaled width $\nu/\nu_\star(\chi,\gamma)$.
Curves for different $\chi$ collapse at large $\nu/\nu_\star$ and follow the predicted $\nu_\star/\nu$ scaling, shown as the dashed $1/x$ line.}
    \label{fig:edge_collapse}
\end{figure}

In the large $\nu$ limit, we expect the corrections term to the response function to be
\begin{equation}
    R_{22} = 1+ \frac{1}{\nu}R^{(1)}+ \mathcal{O}(\nu^{-2})
\end{equation}
with \begin{equation}
    R^{(1)} = - \mathcal{H} \Big[1+\frac{\gamma^2}{1+\frac{\gamma^2}{\chi}\mathcal{H}}\Big].
\end{equation}
By expanding the edge to first order in $1/\nu$ as well, one gets 
\begin{equation}
    \lambda(\mathcal{H}) = -\frac{1}{\mathcal{H}}+1+\frac{\gamma^2}{1+\frac{\gamma^2}{\chi}\mathcal{H}}+ \frac{1}{\nu}R^{(1)}\Big[1+ \frac{\gamma^2}{(1+\frac{\gamma^2}{\chi}\mathcal{H})^2} \Big] + \mathcal{O}(\nu^{-2})
\end{equation}
and therefore $\lambda(\mathcal{H}) = \lambda_0 (\mathcal{H}) + \frac{1}{\nu}\lambda_1(\mathcal{H})+\mathcal{O}(\nu^{-2})$ with
\begin{equation}
    \lambda_1(\mathcal{H}) = -\mathcal{H} \Big[1+\frac{\gamma^2}{1+\frac{\gamma^2}{\chi}\mathcal{H}}\Big]\Big[1+ \frac{\gamma^2}{(1+\frac{\gamma^2}{\chi}\mathcal{H})^2} \Big].
\end{equation}
The edge is stable across $\nu$ when the relative correction to the bulk edge is small
\begin{equation}
    \frac{1}{\nu}\frac{\lambda_1(\mathcal{H})}{\lambda_{\rm edge}}\ll 1
\end{equation}
therefore the condition for the upper edge to have stabilized is
\begin{equation}
    \nu \gg  \nu_{\star}= \sqrt{\chi} + \frac{\chi}{\gamma^2(1+\sqrt{\chi})}.
\end{equation}
Figure~\ref{fig:edge_collapse} verifies this scaling collapse: when the relative edge shift is plotted against $\nu/\nu_\star(\chi,\gamma)$, curves at different $\chi$ collapse onto the predicted $\nu_\star/\nu$ behavior at large rescaled width.
\begin{figure}[!t]
    \centering
\includegraphics[width=1\linewidth]{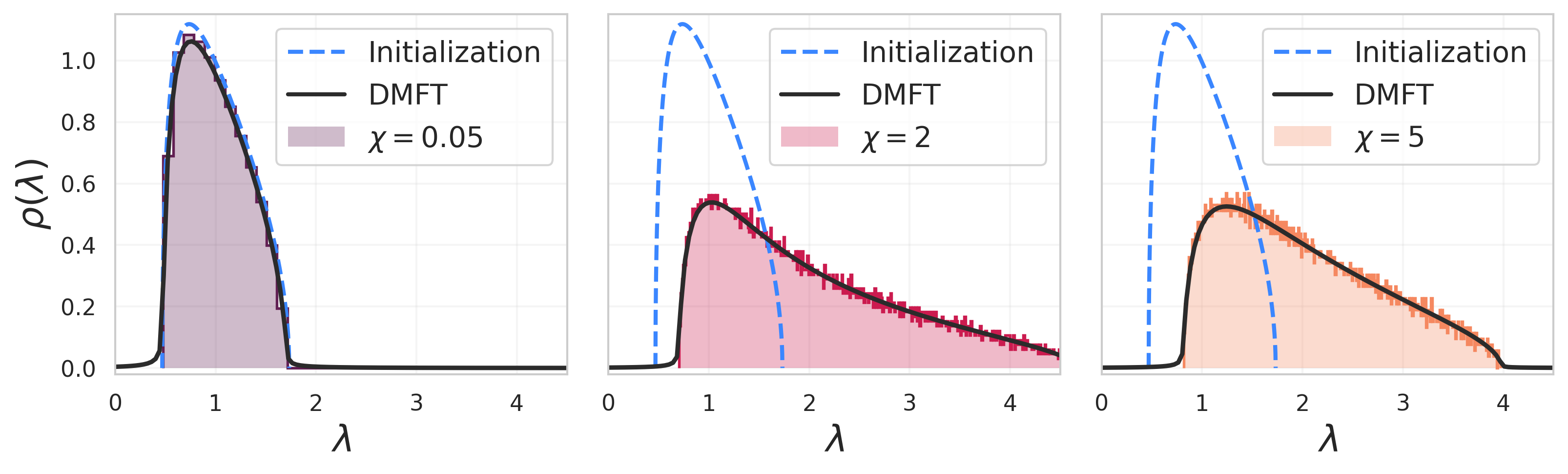}
    \caption{Weight-covariance spectrum after one gradient step for a deep ($L=2$) linear network in the proportional limit $D, N, C \to \infty$ with fixed $\nu = N/D$ and varying $\chi = C/D$. For small $\chi$, the bulk remains close to the MP initialization spectrum; for larger $\chi$, the bulk shifts and broadens, qualitatively resembling spectra of trained language models. }
    \label{fig:vary_chi}
\end{figure}
\begin{figure}
    \centering
\includegraphics[width=1\linewidth]{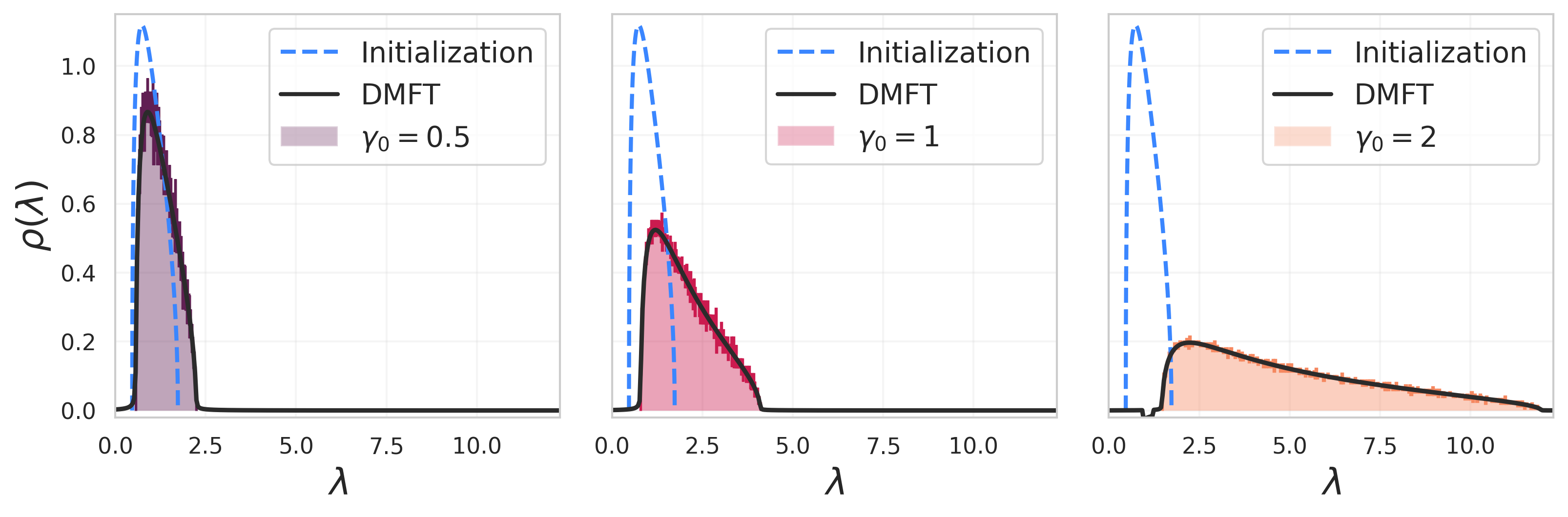}
    \caption{Weight-covariance spectrum of a deep ($L=2$) linear neural network after one gradient step in the proportional limit at fixed $\nu = 5$, $\chi = 4$ and varying richness $\gamma_0$. As $\gamma_0$ increases, the bulk shifts and broadens substantially, indicating an extensive learning-induced deformation of the weight covariance. }
    \label{fig:large_C_vary_gamma}
\end{figure}

\begin{figure}[!t]
    \centering

    \begin{subfigure}[t]{0.32\linewidth}
        \centering
    \includegraphics[width=\linewidth]{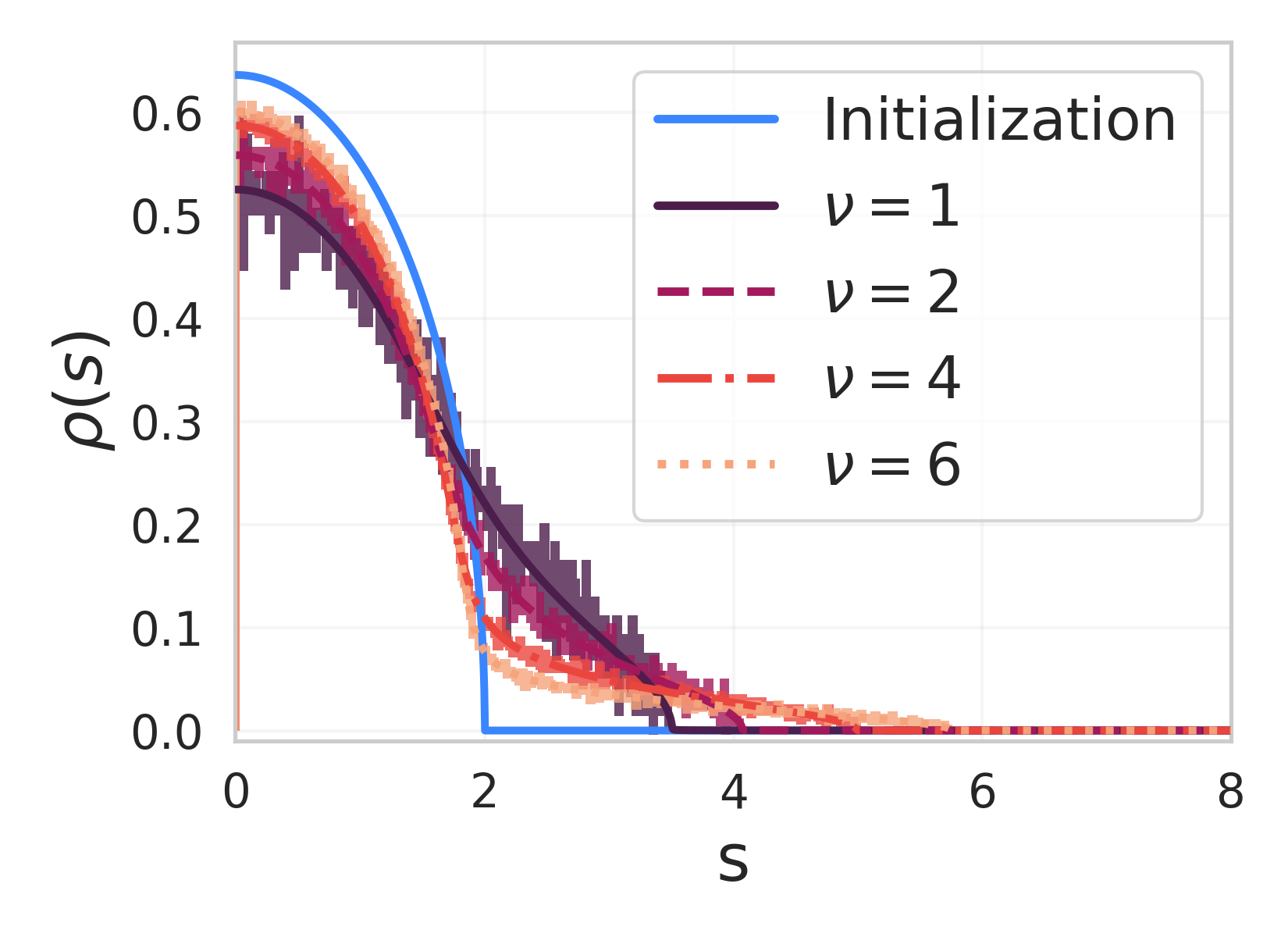}
        \caption{$\gamma_0 = 1$}
    \end{subfigure}
    \hfill
    \begin{subfigure}[t]{0.32\linewidth}
        \centering
        \includegraphics[width=\linewidth]{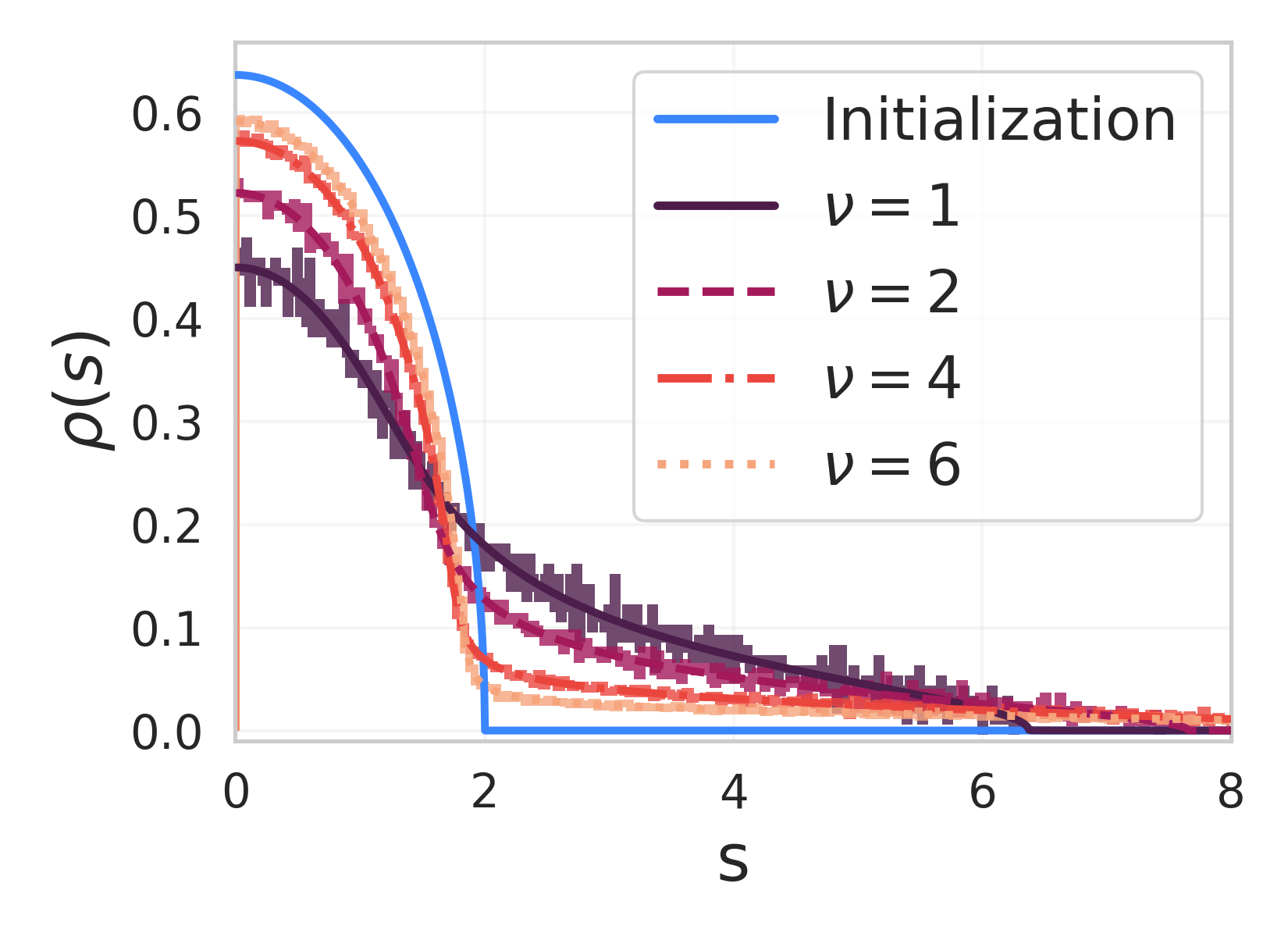}
        \caption{$\gamma_0 = 2$}
    \end{subfigure}
    \hfill
    \begin{subfigure}[t]{0.32\linewidth}
        \centering
        \includegraphics[width=\linewidth]{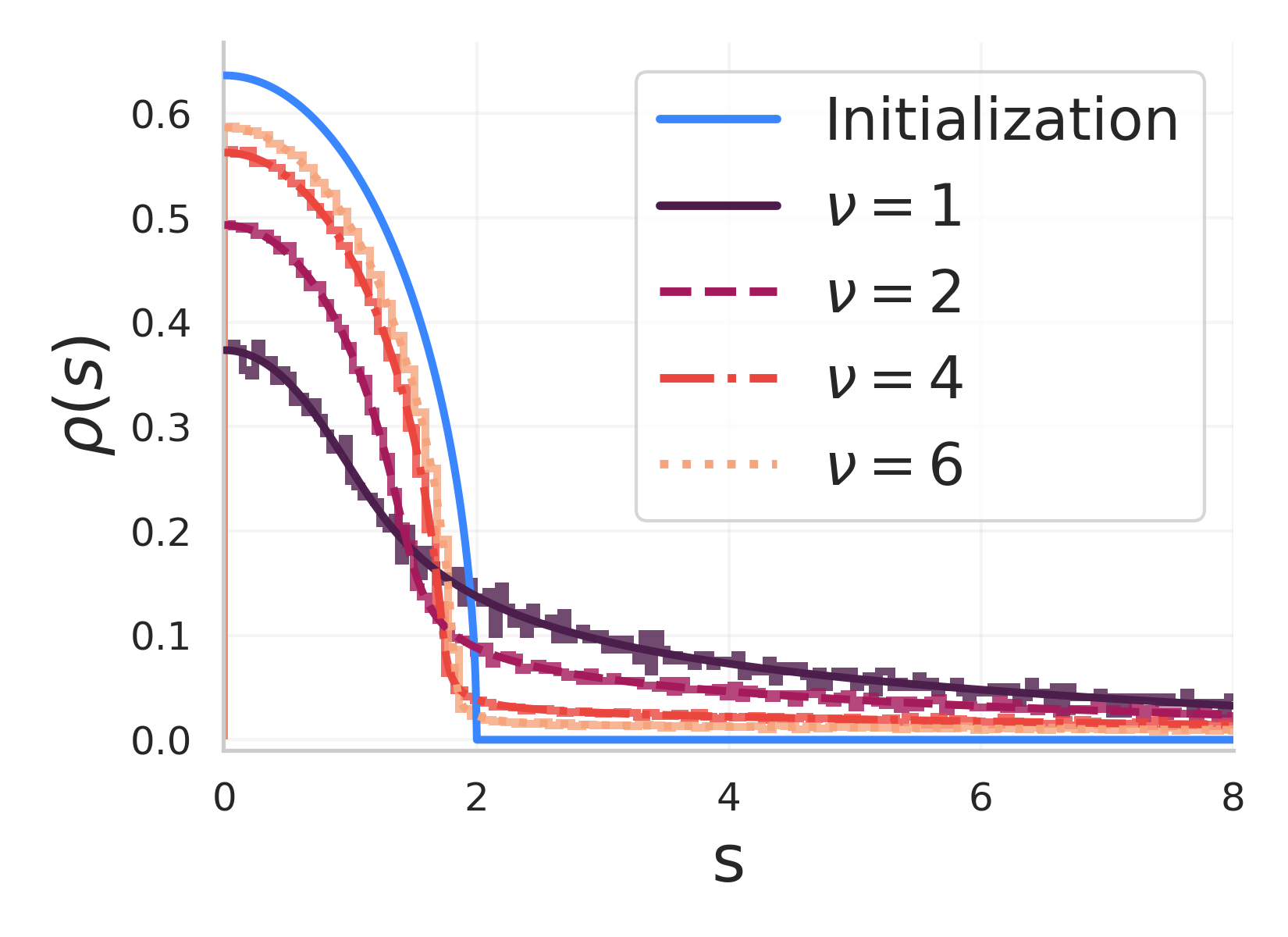}
        \caption{$\gamma_0 = 4$}
    \end{subfigure}

    \caption{Depth $L=3$ linear networks with extensive $C 
    = \mathcal O(N)$ classes. Singular value density plots after one gradient step at different $\nu_s$ and for different level of richness $\gamma_0$.}
    \label{fig:3hl_nu}
\end{figure}
\subsection{Multi-Layer setting}
It is easy to extend the calculation to a multi-layer setting. Let the forward pass be
\begin{equation}
    \bm H^{\ell} = \Big(\frac{1}{\sqrt{N}}\bm W^{\ell-1}(0)\Big) \ldots \Big(\frac{1}{\sqrt{D}}\bm W^{0}(0)\Big)\in \mathbb{R}^{N\times D}
\end{equation}
and backwards 
\begin{equation}
    \bm G^{\ell+1} = \Big(\frac{1}{\sqrt{N}}\bm W^{\ell+1}(0)^{\top}\Big)\ldots \Big(\frac{1}{\sqrt{N}}\bm W^{L-1}(0)^{\top}\Big)\bm A\in \mathbb{R}^{N\times C}.
\end{equation}
After one gradient step, the weight update for a generic hidden layer $\ell $ is
\begin{equation}
    \bm W^{\ell}(1) = \bm W^{\ell}(0)+\frac{\gamma_0}{\sqrt{CD}}\bm G^{\ell+1}\bm B^{\star}(\bm H^{\ell})^\top.
\end{equation}
We wish to study the spectrum of 
\begin{equation}
    \bm M^{\ell} = \frac{1}{N}\bm W^{\ell}(1)^{\top} \bm W^{\ell}(1). 
\end{equation}
In order to do that, we can define the following fields
\begin{align}
    \bm{v}^{1}(t)&=\frac{1}{\sqrt{N}}\bm{H}^{\ell\top}\bm{v}^{0}(t)\in\mathbb{R}^{D}\\\bm{v}^{2}(t)&=\frac{1}{\sqrt{D}}\bm{B}^{\star}\bm{v}^{1}(t)\in\mathbb{R}^{C}\\\bm{v}^{3}(t)&=\frac{\gamma_{0}}{\sqrt{C}}\bm{G}^{\ell+1}\bm{v}^{2}(t)+\frac{1}{\sqrt{N}}\bm{W}^{\ell}(0)\bm{v}^{0}(t)\in\mathbb{R}^{N}\\\bm{v}^{4}(t)&=\frac{\sqrt{C}}{N}\bm{G}^{\ell+1\,\top}\bm{v}_{3}(t)\in\mathbb{R}^{C}\\\bm{v}^{5}(t)&=\frac{\sqrt{D}}{C}\bm{B}^{\star \top}\bm{v}^{4}(t)\in\mathbb{R}^{D}\\\bm{v}^{6}(t)&=\frac{\gamma_{0}\sqrt{N}}{D}\bm{H}^{\ell}\bm{v}^{5}(t)+\frac{1}{\sqrt{N}}\bm{W}^{\ell}(0)^{\top}\bm{v}^{3}(t)\in\mathbb{R}^{N}.
\end{align}
Then
\begin{equation}
    \bm v^6 (t) = \bm M^{\ell} \bm v^0 (t)
\end{equation}
and the probe dynamics is simply 
\begin{equation}
    \partial_t \bm v^0 (t) = - \bm v^6 (t).
\end{equation}
And the calculation proceeds by just computing averages over the random independent Gaussian weights. For instance, for three hidden layers linear network, if we to study the spectrum of the hidden layer $\ell=1$, the moment generating function can be expressed as
\begin{align}
    \mathcal{Z}[j]&=\int\prod_{a=0}^{6}D\bm{v}^{a}(t)D\hat{\bm{v}}^{a}(t)\:\exp\Bigg[i\int dt\hat{\bm{v}}^{0}(t)\cdot\Big(\partial_{t}\bm{v}^{0}(t)+\bm{v}^{6}(t)-\bm{j}(t)\Big)+i\sum_{a=0}^{6}\int dt\bm{v}^{a}(t)\cdot\hat{\bm{v}}^{a}\Bigg]\\&\times\exp\Bigg[-i\int dt\Bigg(\frac{1}{\sqrt{N}}\hat{\bm{v}}^{1}(t)\cdot\bm{W}^{0\top}\bm{v}^{0}(t)+\frac{1}{\sqrt{D}}\hat{\bm{v}}^{2}(t)\cdot\bm{B}\bm{v}^{1}(t)\Bigg)\Bigg]\\&\times\exp\Bigg[-i\int dt\Bigg(\frac{\sqrt{C}}{N}\hat{\bm{v}}^{4}(t)\cdot\bm{A}^{\top}\bm{v}_{3}(t)+\frac{\sqrt{D}}{C}\hat{\bm{v}}^{5}(t)\cdot\bm{B}^{\top}\bm{v}^{4}(t)+\frac{\gamma_{0}\sqrt{N}}{D}\hat{\bm{v}}^{6}(t)\cdot\bm{W}^{0}\bm{v}^{5}(t)\Bigg]\\
    &\times \exp \Bigg[-i \int dt\frac{1}{\sqrt{N}}\hat{\bm{v}}^{6}(t)\cdot\bm{W}^{1}(0)^{\top}\bm{v}^{3}(t)+\frac{\gamma_{0}}{\sqrt{C}}\hat{\bm{v}}^{3}(t)\cdot\bm{A}\bm{v}^{2}(t)+\frac{1}{\sqrt{N}}\hat{\bm{v}}^{3}(t)\cdot\bm{W}^{1}(0)\bm{v}^{0}(t)\Bigg)\Bigg]
\end{align}
\subsubsection{Gaussian averages over the weights}
From $\bm W^0 (0)$
\begin{align}
    \exp\Bigg[-i\text{Tr}\bm{W}^{0}(0)^{\top}\int dt\Big(\frac{1}{\sqrt{N}}\bm{v}^{0}(t)\hat{\bm{v}}^{1}(t)^{\top}+\frac{\gamma_{0}\sqrt{N}}{D}\hat{\bm{v}}^{6}(t)\bm{v}^{5}(t)^{\top}\Big)\Bigg]\\=\exp\Bigg(-\frac{1}{2}\int dtdt'C_{v}^{0}(t,t')\hat{\bm{v}}^{1}(t)\cdot\hat{\bm{v}}^{1}(t')-\frac{\gamma_{0}^{2}\nu}{2}\int dtdt'C_{v}^{5}(t,t')\hat{\bm{v}}^{6}(t)\cdot\hat{\bm{v}}^{6}(t')\Bigg)\\\times\exp\Bigg(\gamma_{0}N\int dtdt'R_{06}(t,t')R_{51}(t',t)\Bigg).
\end{align}
From $\bm W^1 (0)$
\begin{align}
    \exp\Bigg[-i\text{Tr}\bm{W}^{1}(0)^{\top}\int dt\Big(\frac{1}{\sqrt{N}}\hat{\bm{v}}^{3}(t)\bm{v}^{0}(t)^{\top}+\frac{1}{\sqrt{N}}\bm{v}^{3}(t)\hat{\bm{v}}^{6}(t)^{\top}\Big)\Bigg]\\=\exp\Bigg(-\frac{1}{2}\int dtdt'C_{v}^{0}(t,t')\hat{\bm{v}}^{3}(t)\cdot\hat{\bm{v}}^{3}(t')-\frac{1}{2}\int dtdt'C_{v}^{3}(t,t')\hat{\bm{v}}^{6}(t)\cdot\hat{\bm{v}}^{6}(t')\Bigg)\\\times\exp\Bigg(N\int dtdt'R_{06}(t,t')R_{33}(t',t)\Bigg).
\end{align}
From $\bm A(0)$
\begin{align}
    \exp\Bigg[-i\text{Tr}\bm{A}(0)^{\top}\int dt\Big(\frac{\gamma_{0}}{\sqrt{C}}\hat{\bm{v}}^{3}(t)\bm{v}^{2}(t)^{\top}+\frac{\sqrt{C}}{N}\bm{v}_{3}(t)\hat{\bm{v}}^{4}(t)^{\top}\Big)\Bigg]\\=\exp\Bigg(-\frac{\gamma_{0}^{2}}{2}\int dtdt'C_{v}^{2}(t,t')\hat{\bm{v}}^{3}(t)\cdot\hat{\bm{v}}^{3}(t')-\frac{1}{2}\frac{\chi}{\nu}\int dtdt'C_{v}^{3}(t,t')\hat{\bm{v}}^{4}(t)\cdot\hat{\bm{v}}^{4}(t')\Bigg)\\\times\exp\Bigg(\gamma_{0}C\int dtdt'R_{33}(t,t')R_{24}(t',t)\Bigg).
\end{align}
From $\bm B^{\star} = \bm B$
\begin{align}
    \exp\Bigg[-i\text{Tr}\bm{B}^{\top}\int dt\Big(\frac{1}{\sqrt{D}}\hat{\bm{v}}^{2}(t)\bm{v}^{1}(t)^{\top}+\frac{\sqrt{D}}{C}\bm{v}^{4}(t)\hat{\bm{v}}^{5}(t)^{\top}\Big)\Bigg]\\=\exp\Bigg(-\frac{1}{2}\int dtdt'C_{v}^{1}(t,t')\hat{\bm{v}}^{2}(t)\cdot\hat{\bm{v}}^{2}(t')-\frac{1}{2\chi}\int dtdt'C_{v}^{4}(t,t')\hat{\bm{v}}^{5}(t)\cdot\hat{\bm{v}}^{5}(t')\Bigg)\\\times\exp\Bigg(D\int dtdt'R_{42}(t,t')R_{15}(t',t)\Bigg).
\end{align}
By enforcing these definitions
\begin{align}
    R_{06}(t,t')&=-\frac{i}{N}\bm{v}^{0}(t)\cdot\hat{\bm{v}}^{6}(t')\\R_{51}(t',t)&=-\frac{i}{D}\bm{v}^{5}(t')\cdot\hat{\bm{v}}^{1}(t)\\R_{33}(t',t)&=-\frac{i}{N}\bm{v}^{3}(t')\cdot\hat{\bm{v}}^{3}(t)\\R_{24}(t',t)&=-\frac{i}{C}\bm{v}^{2}(t')\cdot\hat{\bm{v}}^{4}(t)\\R_{42}(t,t')&=-\frac{i}{C}\bm{v}^{4}(t)\cdot\hat{\bm{v}}^{2}(t')\\R_{15}(t',t)&=-\frac{i}{D}\bm{v}^{1}(t')\cdot\hat{\bm{v}}^{5}(t)
\end{align}
one gets at the saddle point
\begin{align}
    \hat{R}_{06}&=\gamma_{0}R_{51}+R_{33}\\\hat{R}_{51}&=\gamma_{0}\nu R_{06}\\\hat{R}_{33}&=R_{06}+\gamma_{0}\frac{\chi}{\nu}R_{24}\\\hat{R}_{24}&=\gamma_{0}R_{33}\\\hat{R}_{42}&=\frac{1}{\chi}R_{15}\\\hat{R}_{15}&=R_{42}
\end{align}
giving the DMFT equations
\begin{align}
    v_{1}(t)&=u_{1}(t)+\gamma_{0}\nu\int dsR_{06}(t,s)v_{5}(s),\quad u_{1}\sim\mathcal{N}(0,C_{v}^{0})\\v_{2}(t)&=u_{2}(t)+\frac{1}{\chi}\int dsR_{15}(t,s)v_{4}(s),\quad\mathcal{N}(0,C_{v}^{1})\\v_{3}(t)&=u_{3}(t)+\int ds\Big(R_{06}+\gamma_{0}\frac{\chi}{\nu}R_{24}\Big)v_{3}(s),\quad u_{3}\sim\mathcal{N}(0,C_{v}^{0}+\gamma_{0}^{2}C_{v}^{2})\\v_{4}(t)&=u_{4}(t)+\gamma_{0}\int dsR_{33}(t,s)v_{2}(s),\quad u_{4}\sim\mathcal{N}(0,\frac{\chi}{\nu}C_{v}^{3})\\v_{5}(t)&=u_{5}(t)+\int dsR_{42}(t,s)v_{1}(s)\quad u_{5}\sim\mathcal{N}(0,\frac{1}{\chi}C_{v}^{4})\\v_{6}(t)&=u_{6}(t)+\int ds\Big(\gamma_{0}R_{51}(t,s)+R_{33}(t,s)\Big)v_{0}(s),\quad u_{6}\sim\mathcal{N}(0,C_{v}^{3}+\gamma_{0}^{2}\nu C_{v}^{5})
\end{align}
and finally
\begin{equation}
    \partial_{t}v_{0}(t)=-u_{6}(t)-\int ds\Big(\gamma_{0}R_{51}(t,s)+R_{33}(t,s)\Big)v_{0}(s)+j(t).
\end{equation}
In Fourier space
\begin{align}
    R_{06}(\omega)&=-\frac{1}{i\omega+R_{33}(\omega)+\gamma_{0}R_{51}(\omega)}=-\mathcal{H}(\omega)\\
    R_{15}(\omega)&=-\frac{\gamma_{0}\nu\mathcal{H}(\omega)}{1+\gamma_{0}\nu\mathcal{H}(\omega)R_{42}(\omega)}\\
    R_{51}(\omega)&=\frac{R_{42}(\omega)}{1+\gamma_{0}\nu R_{42}(\omega)\mathcal{H}(\omega)}\\
    R_{24}(\omega)&=\frac{\frac{1}{\chi}R_{15}(\omega)}{1-\frac{\gamma_{0}}{\chi}R_{15}(\omega)R_{33}(\omega)}\\
     R_{42}(\omega)&=\frac{\gamma_{0}R_{33}(\omega)}{1-\frac{\gamma_{0}}{\chi}R_{33}(\omega)R_{15}(\omega)}\\
    R_{33}(\omega)&=\frac{1}{1+\mathcal{H}(\omega)-\gamma_{0}\frac{\chi}{\nu}R_{24}(\omega)}.
\end{align}
The equations above give the closed response-sector DMFT for the spectrum, which can then be obtained from
\begin{equation}
    \rho(\lambda) = \frac{1}{\pi} \lim_{\epsilon\to 0^+} \text{Im}\mathcal{H}(i\lambda -\epsilon). 
\end{equation}

\section{Experimental Details and Additional Figures}

\subsection{CIFAR}
Models are trained on CIFAR-10~\cite{Krizhevsky09learningmultiple} with SGD in mean-field/$\mu$P scaling. The architecture is a residual CNN with $5$ residual blocks and constant hidden width $N$. Training uses batch size $128$ for $T=1000$ gradient steps, without momentum or learning-rate scheduling. We sweep width $N$ and output multiplier $\gamma_0$. For spectral plots, a convolutional tensor
\[
W \in \mathbb{R}^{C_{\rm out}\times C_{\rm in}\times k_1\times k_2}
\]
is square-unfolded as
\begin{equation}
W_{\rm sq}
\in
\mathbb{R}^{k_1 C_{\rm out}\times k_2 C_{\rm in}}
\end{equation}
and singular values are computed from
\[
\frac{W_{\rm sq}}{\sqrt{k_2 C_{\rm in}}}
\]
with aspect ratio
\begin{equation}
\phi
=
\frac{k_1 C_{\rm out}}{k_2 C_{\rm in}} .
\end{equation}

\subsection{ImageNet}
We train a width-$256$ $\mu$P ResNet18 on ImageNet~\cite{ILSVRC15} using the public Flax/JAX implementation accompanying Vyas et al.~\cite{vyas2023feature}. The model is trained with Adam in the $\mu$P parameterization, using minibatches of $1024$ images split into microbatches of size $128$. We track training time by the number of images seen and save model checkpoints at logarithmically spaced targets, including $10^3,10^4,10^5,$ and $10^6$ images seen. The reported loss is the validation cross-entropy evaluated on a held-out subset of ImageNet.

To probe the evolution of the weight spectrum, we extract convolutional kernels from saved checkpoints and compute their singular-value distributions. In the main plot we use the hidden $256$-channel convolutional layer, which does not directly touch either the input pixels or the final readout. For a convolutional kernel $\bm W\in\mathbb R^{k_1\times k_2\times C{\rm in}\times C_{\rm out}}$, we form the square unfolding
\[
\bm W_{\rm flat}\in\mathbb R^{k_1 C_{\rm out}\times k_2 C_{\rm in}}
\]
obtained by grouping one spatial index with output channels and the other spatial index with input channels. We then normalize, so that an unstructured random matrix with the same aspect ratio has Marchenko--Pastur singular-value edge
\[
s_{\rm edge}=1+\sqrt{\frac{k_1 C_{\rm out}}{k_2 C_{\rm in}}}.
\]
We plot the empirical singular-value density together with singular values exceeding this MP edge.
\subsection{GPT Style Pretraining}

Models are trained in $\mu$P with Adam \cite{yang2021tuning} on the C4 corpus \cite{raffel2020exploring}. The models are based on the OLMo repository models and training scripts \cite{olmo20242olmo2furious}, but modified to be in $\mu$P width scaling and CompleteP depth scaling \cite{yang2021tuning, dey2025don}. Scaling attention layers is achieved by scaling up the number of attention heads keeping $d_{\text{head}} = 64$ \cite{bordelon2024infinite}. The base model is $D_{\text{model}} = 768$ with $H=12$ heads and $L=6$ hidden layers. The C4 data are tokenized with the GPT-2 tokenizer whose vocabulary is $\approx 50$k. Weight decay is $0.1$ and a cosine annealing schedule for the $T=4$B run is used across all the models.  
\begin{figure}
    \centering
    \includegraphics[width=\linewidth]{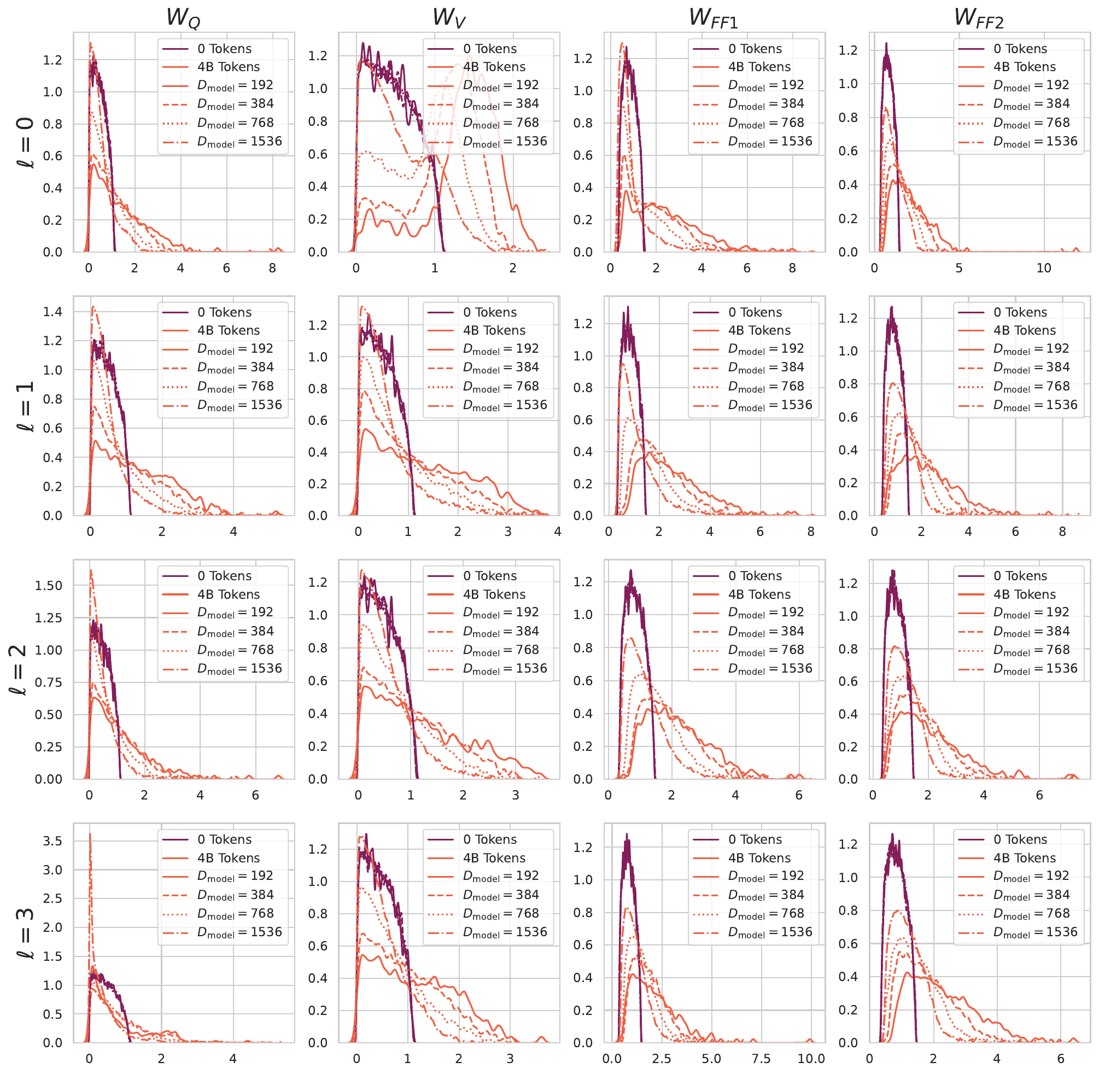}
    \caption{Weight spectra across the first $4$ hidden layers of the transformer after pretraining on $T=4$B tokens.}
    \label{fig:placeholder}
\end{figure}

\subsection{Compute Resources and Estimate of Compute Time}

\paragraph{Theory Plots} The theory curves are computed with a single H100 in a jupyter notebook (attached). Most figures can be generated within a few minutes, less than $\sim 20$ minutes. These were run sequentially and different settings were executed in separate notebooks. 
\\
\paragraph{CIFAR-10 Experiments} For CIFAR-10, we also use a single H100 on jupyter notebook. Training takes around 10 minutes. 
\\
\paragraph{ImageNet} We train on ImageNet for one epoch using a single H100 GPU. This takes about 90 minutes.  
\\
\paragraph{GPT experiments} For the language model pretraining experiments, we use 3x Nvidia A100 per run for 4B tokens varying the widths $D_{model}$. The runtime ranges between 1-6 hours depending on model size. 

\newpage


\end{document}